\documentclass{article}

\usepackage{float}
\usepackage{amsmath}
\usepackage{arxiv}
\usepackage{varwidth}
\usepackage[utf8]{inputenc} 
\usepackage[T1]{fontenc}    
\usepackage{hyperref}       
\usepackage{url}            
\usepackage{amsfonts}       
\usepackage{nicefrac}       
\usepackage{microtype}      
\usepackage{lipsum}		
\usepackage{graphicx}
\usepackage[
    backend=biber,
    style=nature,
]{biblatex}
\input{V3,97}
\usepackage{amssymb}
\usepackage{stmaryrd}
\usepackage{wasysym}
\usepackage{nccmath}
\usepackage{MnSymbol}
\input{V3,97}

\usepackage[dvipsnames]{xcolor}
\usepackage[ruled,vlined]{algorithm2e}
\usepackage{placeins}
\usepackage{subcaption}
\usepackage{svg}
\usepackage{longtable}
\usepackage{relsize}

\usepackage{graphicx}
\usepackage{tikz}
\newcommand*\halfcirc[1][1ex]{%
  \begin{tikzpicture}
  \draw[fill] (0,0)-- (90:#1) arc (90:270:#1) -- cycle ;
  \draw (0,0) circle (#1);
  \end{tikzpicture}}

\title{Synaptic spine head morphodynamics from graph grammar rules for actin dynamics}

\author{\hspace{1mm} Matthew Hur \\
	Program in Mathematical, Computational, and Systems Biology\\
	Univeristy of California Irvine\\
	Irvine, CA, USA \\
	\And
	\hspace{1mm}Thomas M. Bartol \\
	Computational Neurobiology Laboratory\\
	Salk Institute for Biological Studies\\
	La Jolla, CA, USA \\
	\And
	\hspace{1mm}Padmini Rangamani \\
	Departments of Pharmacology and Mechanical and Aerospace Engineering\\
	University of California San Diego\\
	La Jolla, CA, USA \\
	\And
	\hspace{1mm}Terrence J. Sejnowski \\
	Computational Neurobiology Laboratory\\
	Salk Institute for Biological Studies\\
	La Jolla, CA, USA \\
    Deparment of Neurobiology\\
    University of California San Diego\\
    La Jolla, CA, USA
    \And
	\hspace{1mm}Eric Mjolsness \\
	Departments of Computer Science and Mathematics\\
	Univeristy of California Irvine\\
	Irvine, CA, USA \\
    \textit{emj@uci.edu}
}

\date{}

\renewcommand{\headeright}{}
\renewcommand{\undertitle}{}
\renewcommand{\shorttitle}{Synaptic spine head morphodynamics from actin graph grammar dynamics}

\usepackage{fancyhdr}
\hypersetup{
pdftitle={A template for the arxiv style},
pdfsubject={q-bio.NC, q-bio.QM},
pdfauthor={David S.~Hippocampus, Elias D.~Striatum},
pdfkeywords={First keyword, Second keyword, More},
}

\begin{document}
\maketitle

\begin{abstract}
There is a morphodynamic component to synaptic learning by which changes in dendritic (postsynaptic) spine head size are associated with the strengthening or weakening of the synaptic connection between two neurons. The membrane shape and size dynamics is sculpted by the growth dynamics of the enclosed actin cytoskeleton. We use Dynamical Graph Grammars (DGGs) governing dynamic labelled graphs embedded in two dimensions to model networks of actin filaments and the enclosing membrane in spine head morphology. We demonstrate the flexibility and extensibility of the framework by encoding detailed biophysical as well as biochemical models, obeying constraints of invariance and conservation, in DGG rule sets. From graph-local energy functions for cytoskeleton actin interacting and membrane, we specialize dissipative stochastic dynamics to an exhaustive collection of graph-local neighborhood types for the rule left hand sides. Extensively simulating the resulting model delineates effects of four actin-binding proteins, and their epistatic relationships, on morphology.
\end{abstract}

\section{Introduction}

\label{section::intro}
Actin cytoskeletal dynamics 
leads
to changes in the shape of cells through 
action on the membranes of signal-receiving neuronal compartments called 
dendritic
spine heads (\cite{Hotulainen2010}). Synaptic spine heads influence learning because their size influences the strength of the synaptic connection between two neurons (\cite{Araya2014}). Repeated firing across the same synapse can lead to a long-lasting connection in the form of memory, as in Hebbian learning (\cite{Hebb2002}). 
Models of the synaptic spine head hold potential to model behaviors such as addiction and have been posited to be explanatory for the engram hypothesis in how memory is stored in the brain (\cite{Quintana2022,Lisman2017,Lee2024}). Actin's importance in synaptic 
{biophysical kinetics}
can lead to understanding of biological memory and learning, as well as to potential therapeutics (\cite{Haseena2025}).

Actin filament
biophysics can be modeled from principles that 
incorporate
mechanical memory of signaling from neuron to neuron. A system dependent on a global, anharmonic energy potential can lead to many energy minima consistently arising. The wells of the energy potentials are positioned nearby a filament's axis. Spatial transitions between them create a memory system that involves a dynamic system of moving wells of energy.
The CaMKII$\beta$ molecule is a bundling protein that strengthens individual chains of actins by 
cross-linking them
(\cite{Okamoto2007}). Its binding to the filament form of actin (F-actin) 
thus 
leads to stronger filaments (\cite{Yasuda2022}). 
CaMKII$\beta$'s binding along actin filaments allows the intermolecular potentials to 
exert a force on
the connected nodes of the bundling link.


Prior actin models have been 
implemented in
niche simulation packages such as Cytosim (\cite{Nedelec2007, Abekhoukh2014}), but they make assumptions about biophysical filament systems that can be 
fully accounted for using a declarative and highly expressive 
simulation package based on DGGs (\cite{Mjolsness2019}). 
Moving boundaries have been approached as part of hybrid models, e.g. partial differential equations (\cite{Quintana-Rangamani2024}) with finite element methods (\cite{Hernandez-Aristizabal2024}). Yet,
these approaches may
lack the ability to impose 
complex biophysical constraints through simple local rules,
or they lack fine-granularity in either the internal cytoskeleton or the forces onto the membrane mesh. 
What is needed is a graph-based
approach that can downscale
the granularity of the internal cytoskeleton representation and 
that enables easy expression of biophysical complexities
such as membrane mesh 
dynamics including for example the
Newtonian reaction forces between the cytoskeleton and membrane.

In order to simulate actin taking into account connectivity of an actin network, 
we build a model
as a Dynamical Graph Grammar (DGG)
(\cite{Mjolsness2019})
implemented in a computer algebra system in which expressive rules are written for pattern matching in the underlying package. 
DGGs  are theoretically grounded (\cite{Mjolsness2022}) and
originate as a graph-based 
notational
extension of Stochastic Parameterized Grammars (SPGs) which, highly expressive by nature, 
is
applied to node-labeled multisets instead of graphs, and  
incorporates probability distributions over rule firings and their outcomes (\cite{Yosiphon2006}). 
DGGs also incorporate differential equation dynamics. 

As we apply biophysical rules that move objects parameterized with spatial positions, we explain the relevance of simulating in an expressive and declarative theory-based modeling language that is grounded in particle physics theory (\cite{Mjolsness2022}). The creation and annihilation paradigm, which correspond to graph operations, implemented computationally, can apply to dynamical graphs. This is similar to how the operators of the paradigm apply to the statistical mechanics of particles such as the general set of fermion particles. Relaxing constraints -- as the DGG derives from application to node-labeled multisets and from creation/annihilation particle operators of studied fermions -- allows a wide range of criteria for DGG use-cases. Important to this study, spatial simulations and high-performance computing (HPC) are included in those criteria (\cite{Mjolsness2022}).

Here, hundreds
of mathematical rules expressed in Plenum,
a Mathematica package that implements DGGs (\cite{Yosiphon2009}),
model the behavior of the actin network remodeling, biophysical kinetics, 
and interaction with membrane, while the membrane has its own rules for curvature updates.
We show that we are able to simulate a dynamically changing cellular protrusion within reasonable 
time scales
using the highly expressive 
{DGG}
package Plenum in the 
Mathematica computer algebra system 
(\cite{Yosiphon2009}).

Our work complements prior modeling of synaptic spine heads based on differential equations that model the actin cytoskeleton. Previously, there has been work on modeling actin cytoskeleton forces exerted onto synaptic spine membranes using differential equations (\cite{Quintana2020, Quintana2021,Quintana-Rangamani2024}). In our work, we implement the Brownian ratchet (BR) 
hypothesis
which posits that fluctuations in both the lipid bilayer and ends of actin filaments lead to large enough distances between them for actin to polymerize (\cite{Mogilner2003}). This polymerization then provides force onto the membrane. 
Previous models,
with a membrane, coarse-grain membrane-filament interaction as a spring attachment without BR (\cite{Ni2021}) and others, along with the coarse inter-module interaction, use a partial membrane section in the simulations for endocytosis (\cite{Akamatsu2020, Serwas2023}). Thus, the cell's morphology 
is treated as
a plane with mainly barbed end mechanics in comparison to our fully enclosed circular compartment of the dendritic spine head. To accurately capture biophysical memory effects, we implement anharmonic 
potentials (\cite{Bader1996}) that can have 
multiple, spatially 
dependent energy wells and are designed, through DGG rules, 
to follow biophysical constraints and laws. The anharmonic potential thermal movement, e.g. generated by friction within the highly viscous spine head medium, exhibits memory friction (\cite{Bader1996, Tuckerman1993}) similar to in this paper implicitly.
Implementation within the well-founded DGG framework may also, in future, allow 
theoretical advances (\cite{Mjolsness2013, Mjolsness2019})
such as model reduction to be applied to the dendritic spine head rule system.

In this paper, we show that we can build a model for simulation of physical interaction between the actin cytoskeleton and the membrane using 
DGGs,
and
within a plausible model we can functionally characterize actin binding proteins (ABPs) 
that bind to and modulate individual actin monomers, 
affecting membrane shape. This framework can be used to grow an entire spine head from a membrane-enclosed 
area, though all in two dimensions
for simplicity and computational tractability.

While in the past DGGs have been used in non-biophysical simulation of microtubules (\cite{Medwedeff2023}), 
using it with 
the biophysical kinetics of fibers and membranes in an agent-based simulation 
as presented in this paper is novel.
(By ``biophysical kinetics'', hereinafter just ``biophysics'', we refer to the
dynamics of spatial positions of particles and elements of extended objects 
such as fibers and membranes, all due to position-dependent forces.)
We consider the propulsive force of a cytoskeleton onto a membrane polygon and the respective Newtonian reaction force of the membrane onto the cytoskeleton polymers in determining the inter-module dynamics of the system. In simulating the spatial movement of the cytoskeleton, we hope to gain a better and more accurate understanding of how the actin cytoskeleton determines shape and size of a cellular protrusion.

\section{Results}\label{section::results}

\subsection{DGG simulations represent growth of synaptic spine heads}
\label{section::growth}
We implement a DGG model inside a simulator in Plenum that minimizes the biophysical energy of a dynamically changing network. The total energy function is shown in Eq.~\ref{eq::membrane}

\begin{align}
\begin{split}
    \label{eq::membrane}
    E_{\text{tot}} &= 
    \underbrace{\sum_{i, j \neq i}^N G_{i j}
    U_{\text{sep}}(x_i,x_j)}_{\text{Anisotropic Buckling}} 
    + \underbrace{\sum_{i \neq j \neq k \neq i}^N  G_{i j} G_{j k} U_{\text{ang}}(x_i,x_j,x_k,\theta_\text{target})}_{\text{Angular Bending Energy}} \\
    &+\underbrace{2\kappa\sum_{a \neq b \neq c \neq a}^Mg_{a b}g_{b c} H_{\Gamma(x_a,x_b,x_c)}^2}_{\text{Membrane Mean Helfrich Energy}}.
\end{split}
\end{align}
In this formula, $G$ denotes the adjacency matrix of an actin cytoskeleton network's graph representation, and $g$ denotes that for an enclosing membrane mesh. The membrane term is
mediated by a directed graph with adjacency matrix $g_{b c}$
which consists of the counterclockwise cycle of edges
around the 1D membrane embedded in 2D space. 

\begin{figure}[h]
\centering
\includegraphics[width=\textwidth]{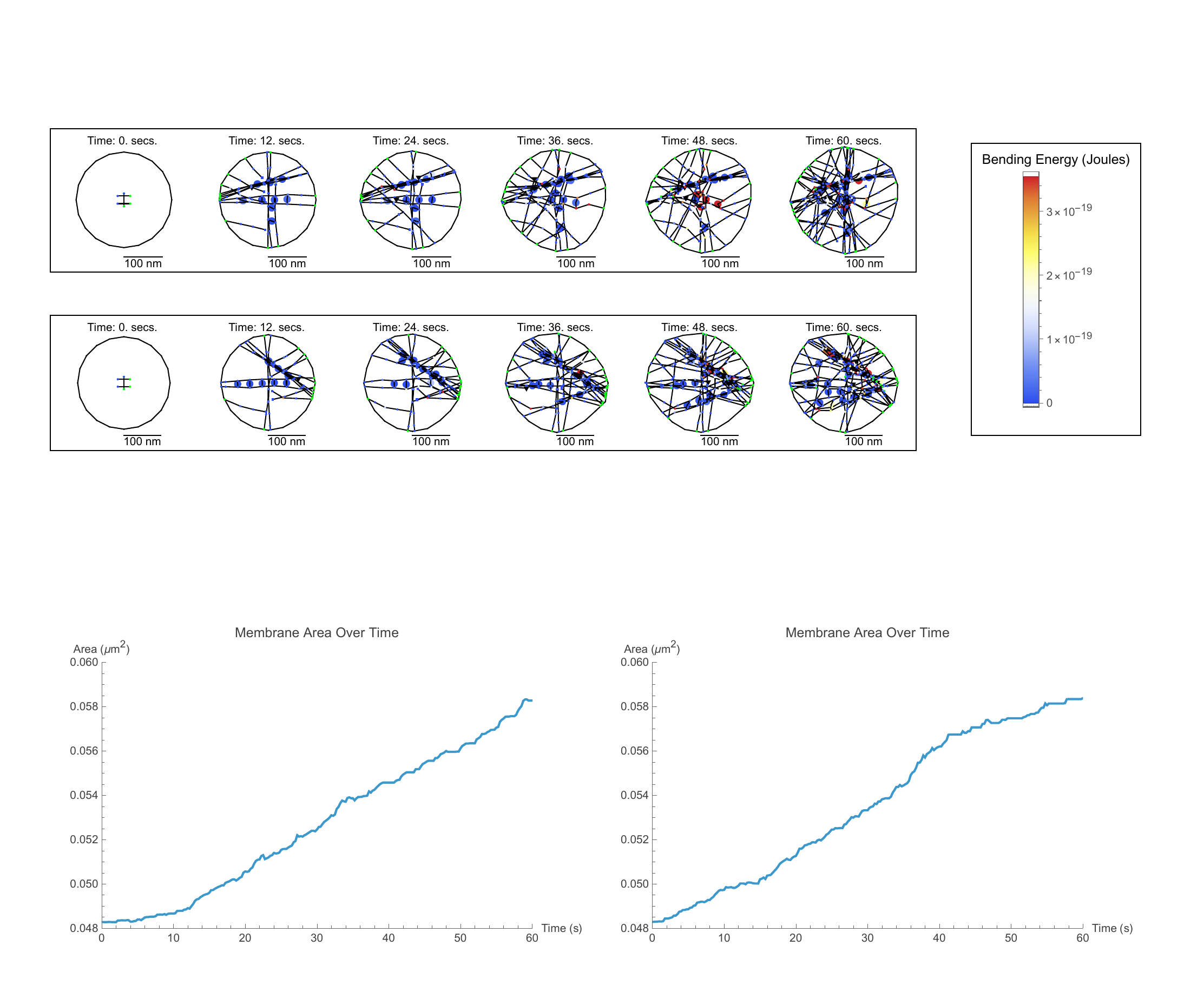}
\caption{\\
Top: Snapshots of two stochastic simulations of the stubby synaptic spine head. Simulation snapshots evolve from the same initial condition from stochastic variability. Model includes Aip1. Color indicates the degree of angle bending energy (or green for a barbed end) for each object, circles represent coarse-grained actin objects ($N_{\text{CG}}=12$), triangles represent caps or colored triangles represent junction objects, and dodecagons represent CaMKII$\beta$ bundling complexes. \\
Bottom: Membrane-enclosed area over time. Plots arranged from left-to-right correspond to simulations arranged top-to-bottom.}
\label{fig:exsim}
\end{figure}

Simulations run up to 60 seconds of biological time achieve compartment growth up to 20\% in Fig. \ref{fig:exsim}. An area of 0.048 $\mu m^2$ roughly corresponds to the transverse cross-sectional area of a biological synaptic spine head (\cite{Schünemann2024}). This indicates the possibility of simulating the effects of long-term potentiation (LTP) from repeated electrical signals that mediate Hebbian learning as in \cite{Quintana-Rangamani2024}.

Dendritic spines belong to a few types based on their shapes (i.e. stubby, mushroom, thin, filopodia, branched) (\cite{Pchitskaya2020}). In this paper, we investigate the spine head at an actin synthesis rate specific to the stubby type (\cite{Helm2021}). Of the types, the stubby spine type is in a more expansive growth phase, able to readily develop into more mature spines. Mushroom spines are known as mature ``memory spines'' because they are considered the developed type, with markers of past growth stored in the internal actin cytoskeleton (\cite{Shonesy2014}). We characterize the stubby spine type further through parameter sweeps, including spine type specific effects for the four ABPs (Arp2/3, CaMKII$\beta$, cofilin, Aip1), by varying their synthesis rates.

In all our simulations in this paper, synthesis rates are altered from their estimated basal values from \cite{Bosch2014} that are from differential equations in \cite{Quintana-Rangamani2024}, by adding and subtracting additional rates estimated from the same ODEs. The set of estimated synthesis rates is pursuant to the glutamate uncaging model in the \cite{Bosch2014} study. As glutamate stimulates the synaptic spine in this model, it grows to a larger stable size for spine head size in a process known as long-term potentiation (LTP) (\cite{Matsuki2004}). LTP is part of learning and memory (\cite{Voronin1983}). 

Biologically, neurotransmitters reaching ligand-gated ion channels on the membrane of the postsynaptic terminal can elevate the concentration of proteins related to the actin cytoskeleton (\cite{Gentile2022}). Synthesis rates can increase or decrease from influx due to electrical stimuli, which trigger the release of neurotransmitters, responding to different neuronal firing rates. 

We include four ABPs in our simulation model. Here, we list their individual, putative functions, which may interact with each other such as by competitive binding and will be explored in computer simulations. Arp2/3 creates other F-actin filament branches off of F-actin filaments, CaMKII$\beta$ bundles separate F-actin filaments, and cofilin and Aip1 sever F-actin filaments in a joint mechanism. 

\begin{figure}[h]
\centering
\includegraphics[width=\textwidth]
{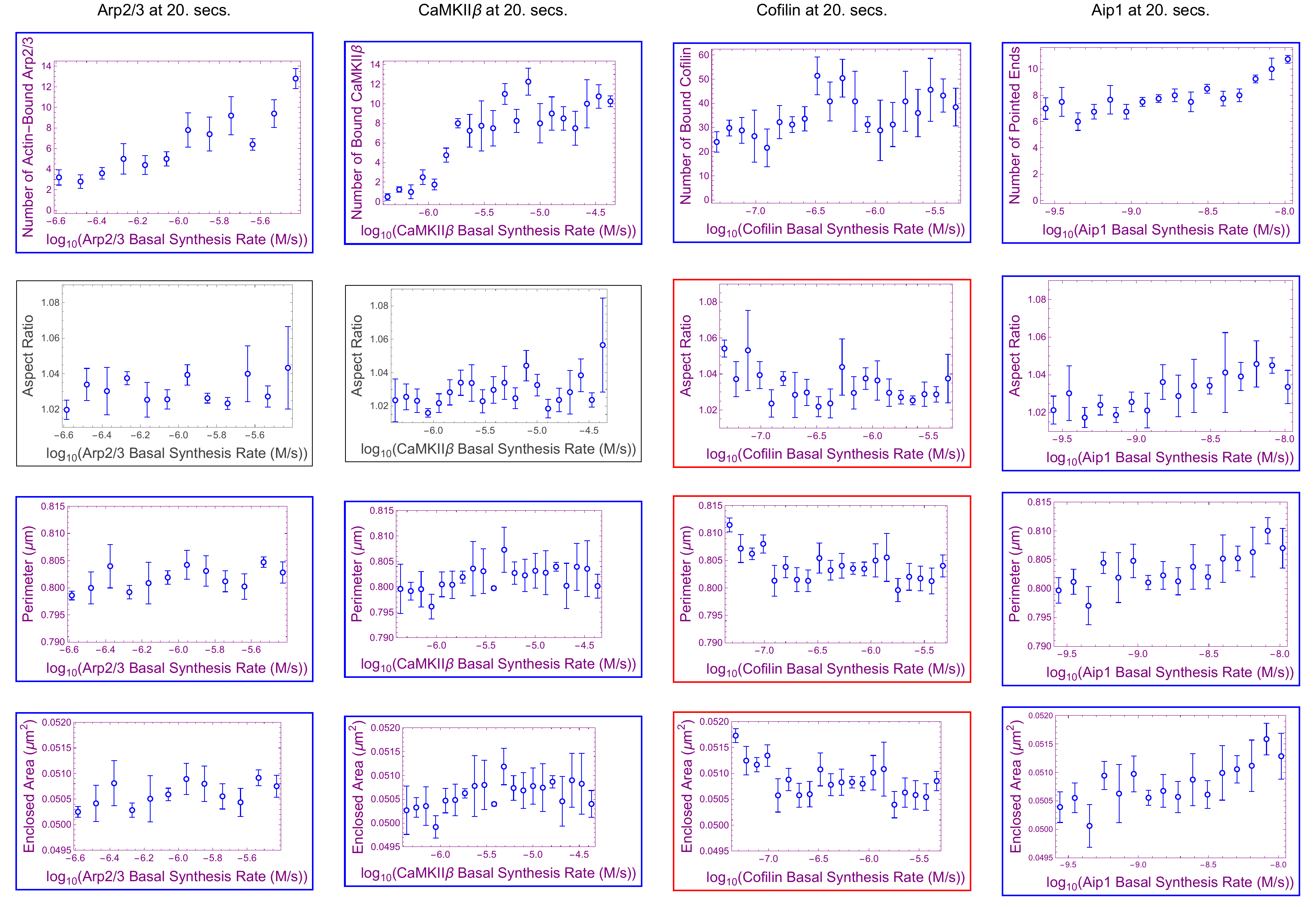}
\caption{Dependence of membrane properties on synthesis rates of four actin-binding regulatory proteins demonstrated in parameter sweeps. The basal synthesis rate provided in Section~\ref{section::discussion} Table~\ref{table:parameters} is the middle value for cofilin and CaMKII$\beta$ horizontal-axes. Error bars show standard error of the mean (SEM) with Bessel's correction, for a number of stochastic simulations each. We calculate membrane aspect ratio, membrane perimeter, and membrane enclosed area for each simulation snapshot. The number of simulations is 5 for Arp2/3 and cofilin and 4 for CaMKII$\beta$ and Aip1, corresponding to each mean value with error bar in the plots. Indigo graphs are significant under p-value threshold 0.05. For the frames surrounding significant individual plots, blue is a significant increasing trend, and red is a significant decreasing trend as determined by Table~\ref{table:stats}. The range on log-scale for the parameter sweep consists of $10\times$ smaller to at most $10\times$ larger than the estimated, basal synthesis rates for CaMKII$\beta$ and cofilin, with range truncated at the upper end for Arp2/3 and Aip1. Rows represent membrane properties across four ABPs and columns represent a cytoskeletal network property followed by three membrane properties for each ABP. 
}
\label{figure:parametersweepstubby}
\end{figure}

In the parameter sweep shown in Fig.~\ref{figure:parametersweepstubby} for stubby spine heads, we see that as the synthesis rates of cofilin, Arp2/3, and CaMKII$\beta$ concentration increase,
they increase (Arp2/3 with p = 0.0074; CaMKII$\beta$ with p = 0.00030; Aip1 with p = 0.00048; in Table~\ref{table:stats}) and decrease (cofilin with p = 0.033; in Table~\ref{table:stats}) spine head size. In all cases we consider a statistical test p-value of $< 0.05$ to be ``significant''.

\begin{table}[h]
\centering
\tiny
\begin{tabular}{||l|l|l|l|l||}
\hline
 \multicolumn{5}{||c||}{Stubby spine head at 20.0 seconds}\\
 \hline
 Membrane property / value type & Arp2/3 (n=5) & CaMKII$\beta$ (n=4) & cofilin (n=5) & Aip1 (n=4) \\ 
 \hline
 Aspect ratio / Jonckheere-Terpstra p-value & 0.26 & 0.077 & \textbf{0.015} & \textbf{0.00030} \\
 -- / Pearson's correlation p-value & 0.68 & 0.074 & \textbf{0.015} & \textbf{0.00020} \\
 -- / Pearson's correlation r-value & -0.045 & 0.23 & -0.20 & 0.41 \\
 \hline
 Perimeter / -- & \textbf{0.010} & \textbf{0.039} & \textbf{0.00050} & \textbf{0.00071} \\
 -- / -- & 0.062 & 0.074 & \textbf{0.0013} & \textbf{0.00020} \\
 -- / -- & 0.20 & 0.20 & -0.32 & 0.45 \\
 \hline
Enclosed area / -- & \textbf{0.0074} & \textbf{0.033} & \textbf{0.00030} & \textbf{0.00048} \\
 -- / -- & \textbf{0.044} & \textbf{0.050} & \textbf{0.00074} & \textbf{0.00018} \\
 -- / -- & 0.22 & 0.22 & -0.33 & 0.45 \\
 \hline
\end{tabular}
\caption{Statistical values for monotonicity and correlation in parameter sweeps of stubby spine heads. Jonckheere-Terpstra p-values are calculated using ``jonckheereTest'' in the ``PMCRPlus'' package of the R programming language. Stubby statistical values match the format of the parameter sweep of Fig.~\ref{figure:parametersweepstubby}. Trend tests for monotonicity, p-values $<0.05$ are bold under the Jonckheere 1-sided test, with sidedness determined by the sign of Pearson's r in a linear fit.}
\label{table:stats}
\end{table}

\subsection{Epistasis and antagonism between Arp2/3 and CaMKII$\beta$} \label{subsection::epistasis}
In our stubby spine head simulations, interaction
occurs between CaMKII$\beta$ and Arp2/3. Our implementation of mutually exclusive binding at actin nodes for CaMKII$\beta$ and Arp2/3 is observed in biological experiment (\cite{Kim2015}). Since we are measuring
several somatic ``phenotypes''--which are membrane aspect ratio, membrane perimeter, and membrane enclosed area--
that could be exposed to selection, a genetic analysis approach 
(genotype to phenotype, skipping the complex system and model in between) such as epistasis is appropriate. The approach to epistasis in this paper is to elevate or lower synthesis rate of an ABP, e.g. CaMKII$\beta$, by a factor $\eta^q$, $q\in\{1,2,4\}$, and concomitantly increase Arp2/3 by a factor of $\eta$.

In our computational model, we include the actin filament bundling protein CaMKII$\beta$. During a situation like LTP, CaMKII$\beta$ is known to unbind from the actin cytoskeleton because of an influx of calcium ions into the synaptic spine head that bind the Calmodulin (CaM) protein and reduce binding affinity of CaMKII$\beta$ for actin filaments (\cite{Wang2019}). Fig.~\ref{fig:paramrates} shows the decrease in concentration of CaMKII$\beta$ during a biological model of LTP by glutamate uncaging as done in \cite{Bosch2014}. We model this change to CaMKII$\beta$ by fitting an ODE provided in \cite{Quintana-Rangamani2024} to data in \cite{Bosch2014}. From the fit, we arrive at a value for CaMKII$\beta$ synthesis change during a model of LTP that is negative (Fig.~\ref{fig:paramrates} and $k_\text{influx, CaMKII$\beta$}$ in Table~\ref{table:parameters}). The negative value augmentation to CaMKII$\beta$ synthesis influences the binding rate of CaMKII$\beta$ by having less free CaMKII$\beta$ available during simulation. Based on this phenomenon for CaMKII$\beta$, our subsequent simulation experiments will be under assumption that Arp2/3 synthesis increases and CaMKII$\beta$ synthesis decreases during the model of LTP.

\begin{figure}[h]
\centering
\includegraphics[width=\textwidth]
{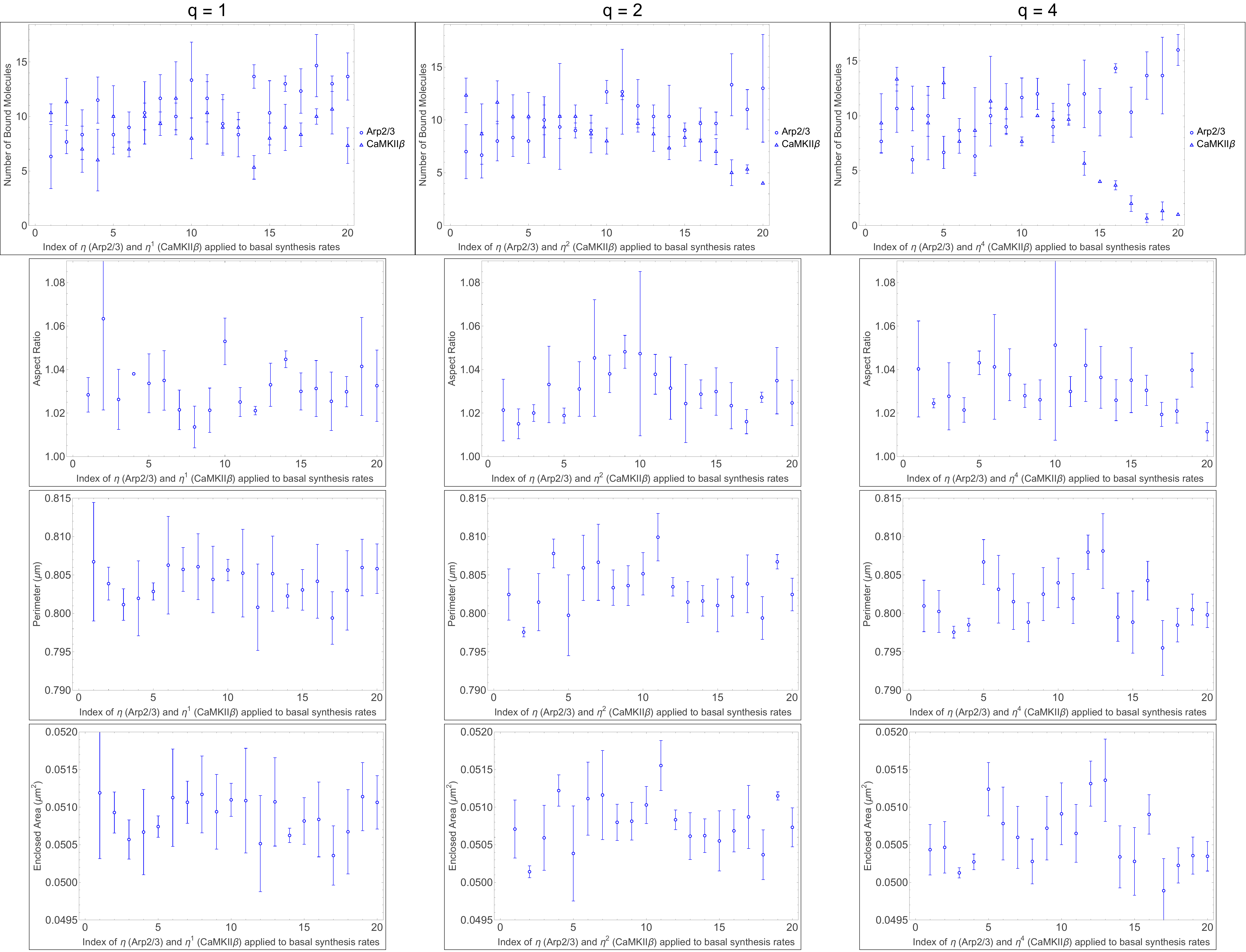}
\caption{Epistatic masking of Arp2/3 on CaMKII$\beta$ in the stubby spine head: Dependence of membrane properties on synthesis rates of CaMKII$\beta$ and Arp2/3 demonstrated in parameter sweeps. $\eta$ is directly applied to Arp2/3 synthesis, and $\eta$ (first column), $\eta^2$ (middle column), and $\eta^4$ (last column) are applied to CaMKII$\beta$ synthesis in reverse to test for epistatic nullification. CaMKII$\beta$ synthesis is decreased by succesive applications of 1/$\eta^q$ and Arp2/3 synthesis is increased by successive applciations of $\eta$. For a chosen multiplicative value of $\eta$, it is sequentially multiplied or divided to the original, basal synthesis rates, but at different powers of $\eta$ for CaMKII$\beta$ or Arp2/3, informed by the statistical values in Table~\ref{table:stats}. The basal synthesis rates provided in Table~\ref{table:parameters} are the middle values of the horizontal-axes. Error bars show standard error of the mean (SEM) with Bessel's correction for n = 3 simulations per data point.
}
\label{figure:parametersweepstubbyeta}
\end{figure}

\begin{figure}[h]
\centering
\includegraphics[width=\textwidth]
{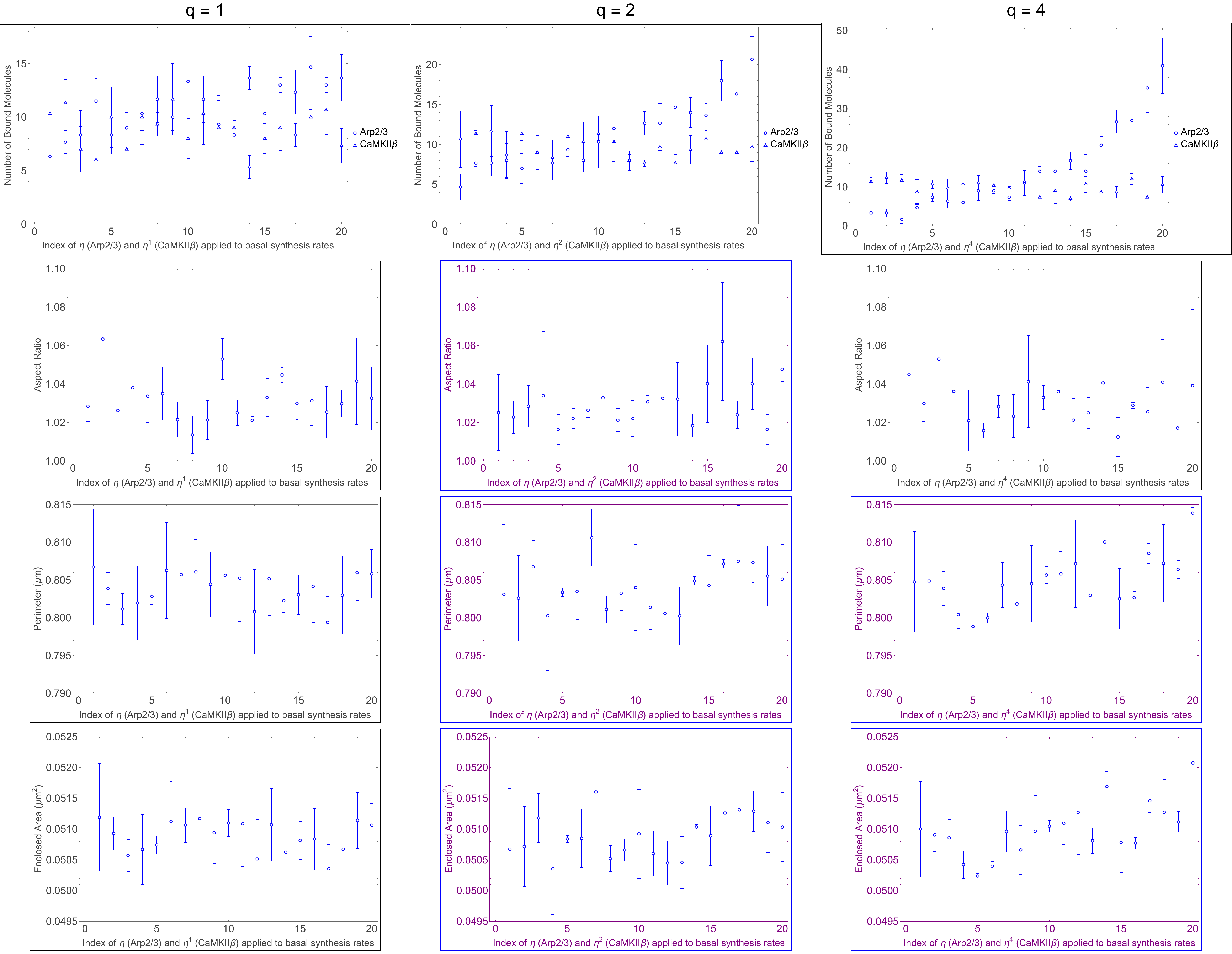}
\caption{Masking effect's uni-directionality of Arp2/3 on CaMKII$\beta$ in the stubby spine head: Dependence of membrane properties on synthesis rates of CaMKII$\beta$ and Arp2/3 demonstrated in parameter sweeps. CaMKII$\beta$ synthesis is decreased by succesive applications of 1/$\eta$ and Arp2/3 synthesis is increased by successive applciations of $\eta^q$. $\eta$ is applied in reverse to CaMKII$\beta$ synthesis, and $\eta$ (first column), $\eta^2$ (middle column), and $\eta^4$ (last column) are directly applied to Arp2/3 synthesis to test for epistatic direction. For a chosen multiplicative value of $\eta$, it is sequentially multiplied or divided to the original, basal synthesis rates, but at different powers of $\eta$ for CaMKII$\beta$ or Arp2/3, informed by the statistical values in Table~\ref{table:stats}. The basal synthesis rates provided in Table~\ref{table:parameters} are the middle values of the horizontal-axes. Error bars show standard error of the mean (SEM) with Bessel's correction for n = 3 simulations per data point.
}
\label{figure:parametersweepstubbyetaunmasked}
\end{figure}

 To clearly observe a phenotype, we have activated changes to synthesis rates in our model pursuant to glutamate uncaging and to long-term potentiation (LTP) induction, including decreasing CaMKII$\beta$ synthesis rate and increasing Arp2/3 synthesis rate by 10$\times$ that decrease to CaMKII$\beta$. This ratio is estimated from fitting ODEs designed in \cite{Quintana-Rangamani2024} to published data in \cite{Bosch2014}. Now, as we apply the method of epistatic analysis, presented in \cite{AveryandWasserman1992}, we assume that there is an interaction between Arp2/3 and CaMKII$\beta$ based on our implementation of mutually exclusive binding. In a double parameter sweep (Fig.~\ref{figure:parametersweepstubbyeta}), only Arp2/3 and CaMKII$\beta$ determine the change to spine head morphology. Cofilin and Aip1 are under control conditions for LTP in the parameter sweeps for Arp2/3 and CaMKII$\beta$ in Fig.~\ref{figure:parametersweepstubby} and in those in Fig.~\ref{figure:parametersweepstubbyeta}. We will provide evidence from observations, in Fig.~\ref{figure:parametersweepstubbyeta} and Fig.~\ref{figure:parametersweepstubbyetaunmasked} parameter sweeps, that Arp2/3 masks the effect of spine head growth by CaMKII$\beta$ observed in second column of Fig.~\ref{figure:parametersweepstubby} (r = 0.22 and p = 0.033; in Table~\ref{table:stats}). By continuing to argue that this observation of masking is one-way and directional, we will show that Arp2/3 stops the effect of CaMKII$\beta$ synthesis on spine head morphology through an epistasis of Arp2/3 on CaMKII$\beta$, during growth similar to LTP.

 Here, we will show that Arp2/3 masks CaMKII$\beta$ to show a condition in masking that is necessary for epistasis, but not yet totally sufficient, using data for nullification summarized in Table~\ref{table:statseta} and presented in Fig.~\ref{figure:parametersweepstubbyeta}. To be a sufficient explanation for epistasis, the interpretation should conclude both masking, shown in this paragraph, and uni-directionality. As shown in column one of Fig.~\ref{figure:parametersweepstubbyeta}, which corresponds to $q=1$, Arp2/3 and CaMKII$\beta$ can nullify each other's effect on spine head area (r = -0.065 and p = 0.47; in Table~\ref{table:statseta}). This null relationship is roughly expected at $q$ value of one by its implied setting of synthesis rates at a nullification 1:1 log-ratio (Arp2/3 increased in steps of $\eta^1$ and CaMKII$\beta$ in steps of 1/$\eta^1$) and by data showing that CaMKII$\beta$ affects spine head size in an approximately 1:1 log-ratio to Arp2/3 (r = 0.22 for Arp2/3 and r = 0.22 for CaMKII$\beta$ area; r = 0.20 for Arp2/3 and r = 0.20 for CaMKII$\beta$ perimeter; with p-values $<0.05$; in Table~\ref{table:stats}). Arp2/3 nullifies CaMKII$\beta$ not only at $q=1$, but also at $q=2$ and $q = 4$ (p = 0.38 for power of two; p = 0.33 for power of four; for area in Table~\ref{table:statseta} with $\eta$ applied to CaMKII$\beta$). The masking effect is measured by our simulation parameter sweeps starting by assuming an additive nullifiction effect. The starting 1:1 log-ratio is implied by  data summarized in Table~\ref{table:stats} from sweeps of Arp2/3 and CaMKII$\beta$ alone (Fig.~\ref{figure:parametersweepstubby}). As the log-ratio favors CaMKII$\beta$ synthesis value in ratios 1:2 ($q=2$) and 1:4 ($q=4$), the correlation r-value between Arp2/3 and CaMKII$\beta$, stays near-zero (r = 0.032 for $q=2$ and r = -0.077 for $q=4$) as it does for $q=1$ (r = -0.065). As no value for $q$ tested by applying to CaMKII$\beta$ penetrates the nullification phenotype between Arp2/3 and CaMKII$\beta$, we can interpret that Arp2/3 masks phenotypic spine head growth effort by CaMKII$\beta$ existing in column two of Fig.~\ref{figure:parametersweepstubby} (r = 0.22, p = 0.033; in Table~\ref{table:stats}). In stubby spine head simulations, we show that Arp2/3 can mask CaMKII$\beta$ mediated effect on spine heads. 

 Next, we will show that Arp2/3 uni-directionally masks CaMKII$\beta$, implied by data showing that Arp2/3 can penetrate the nullification in spine head growth of the log-ratio 1:1 between Arp2/3 and CaMKII$\beta$. Arp2/3 synthesis changes spine head size regardless of the increase or decrease applied to CaMKII$\beta$ synthesis, different from the masking effect we observed onto CaMKII$\beta$ that is explained in the previous paragraph. We observe this evidence for one-way directionality in masking. The supporting data is the significant correlation between Arp2/3 or CaMKII$\beta$ synthesis and their effects on spine head growth in Fig.~\ref{figure:parametersweepstubbyetaunmasked} for higher gradations of change to Arp2/3 synthesis than to CaMKII$\beta$. Past the previously observed nullification value of $q = 1$, Arp2/3 significantly increases spine head size at the higher gradation values of $q\in\{2,4\}$ (r = 0.17 and p = 0.031 for $\eta^2$; r = 0.41 and p = 0.00069 for $\eta^4$; in Table~\ref{table:statseta}). The presence of an effect for synthesis multiplier $\eta^q$ favoring gradation of Arp2/3 synthesis demonstrates that Arp2/3 successfully penetrates the phenotypic nullification for concomitant and opposite gradation of CaMKII$\beta$ synthesis. The penetrance effect by Arp2/3 indicates that CaMKII$\beta$ does not mask phenotypic effect in spine head size by Arp2/3 synthesis change. Only Arp2/3 masks CaMKII$\beta$ and not CaMKII$\beta$ for Arp2/3, which indicates Arp2/3 is epistatic to CaMKII$\beta$ during stubby spine head growth phase, in a model implemented for similarity to LTP.

 Earlier in section~\ref{subsection::epistasis}, we show that because of epistasis, we can mask a CaMKII$\beta$ effect on spine head growth by an opposite effect by Arp2/3. CaMKII$\beta$ synthesis change alone would otherwise effect spine head size, an observation supported by biological experiment in \cite{Pi2010}. Some studies report that bound amount of CaMKII$\beta$ correlates with spine head size (\cite{Asrican2007,Zernov2022}). This relationship is consistent with what we observe about CaMKII$\beta$ in Fig.~\ref{figure:parametersweepstubby} column two.

An alternative hypothesis is that the difference in binding energies of Arp2/3 and CaMKII$\beta$, represented by their dissociation constants ($K_D$), accounts for the competition, where $k_{\text{off}}$ for Arp2/3 is not the unbinding rate of unitary Arp2/3, but the debranching rate of daughter filaments. Here, 
\(
K_{D, \text{ CaMKII$\beta$}} = \frac{k_{\text{off, CaMKII$\beta$}}}{k_{\text{on, CaMKII$\beta$}}}
=\frac{0.23 \text{ (1/s)}}{0.5\times10^6 \text{ (1/(M s))}} \approx .46 \text{ $\mu$M}
\) and
\(
K_{D, \text{ Arp2/3}} = \frac{k_{\text{off, Arp2/3}}}{k_{\text{on, Arp2/3}}}
=\frac{2\times10^{-3} \text{ (1/s)}}{3000 \text{ (1/(M s))}}
\approx .67 \text{ $\mu$M}.
\)
By these $K_D$ values, Arp2/3 and CaMKII$\beta$ have similar $K_D$s, which rules out binding energy as an explanation for our observation of epistasis.

Below, we show mathematical reasoning for double parameter sweeps as evidence for epistasis in addition to antagonism. Let $\alpha=\log_{10}(\text{Arp2/3 synthesis rate})$ and $\gamma=\log_{10}(\text{CaMKII$\beta$ synthesis rate})$ be the change in system variables by taking the logarithms of Arp2/3 and CaMKII$\beta$ synthesis rates. If 

\begin{equation}
    \frac{d \text{ phenotype}}{d \alpha}/\frac{d \text{ phenotype}}{d \gamma} \approx -q,
    \label{A}
\end{equation}
    
\noindent we can introduce a new variable $\mu$, to be varied, such that in the case of antagonism 

\begin{equation}
    \frac{d\alpha}{d \mu}/\frac{d\gamma}{d \mu} = (1/2)/(q/2) = 1/q.
    \label{B}
\end{equation}
A reversal of the powers of $\eta$ applied to $\alpha$ and $\gamma$ leads to the same method for change of variables. We find that $q=-1$ by substituting 0.22 and 0.22 from Table~\ref{table:stats} for the numerator and denominator in Eq. (\ref{A}). $q=-1$ represents the nullification observed between Arp2/3 and CaMKII$\beta$. In Table~\ref{table:statseta}, we have that higher values of $q$ show masking that indicates more than nullification at those higher values of $|q|$. CaMKII$\beta$ is unable to raise an effect due to rising Arp2/3 at the same steps of applying $\eta$ to Arp2/3 and $1/\eta^q$ to CaMKII$\beta$; hence, the masking in this case is epistasis.

The reverse situation, with a reverse 2D change of variables, is different. $\alpha$ dominates at all higher $q$ so $\gamma$ cannot be epistatic to it.
    
\begin{figure}[h]
\centering
\includegraphics[width=\textwidth]
{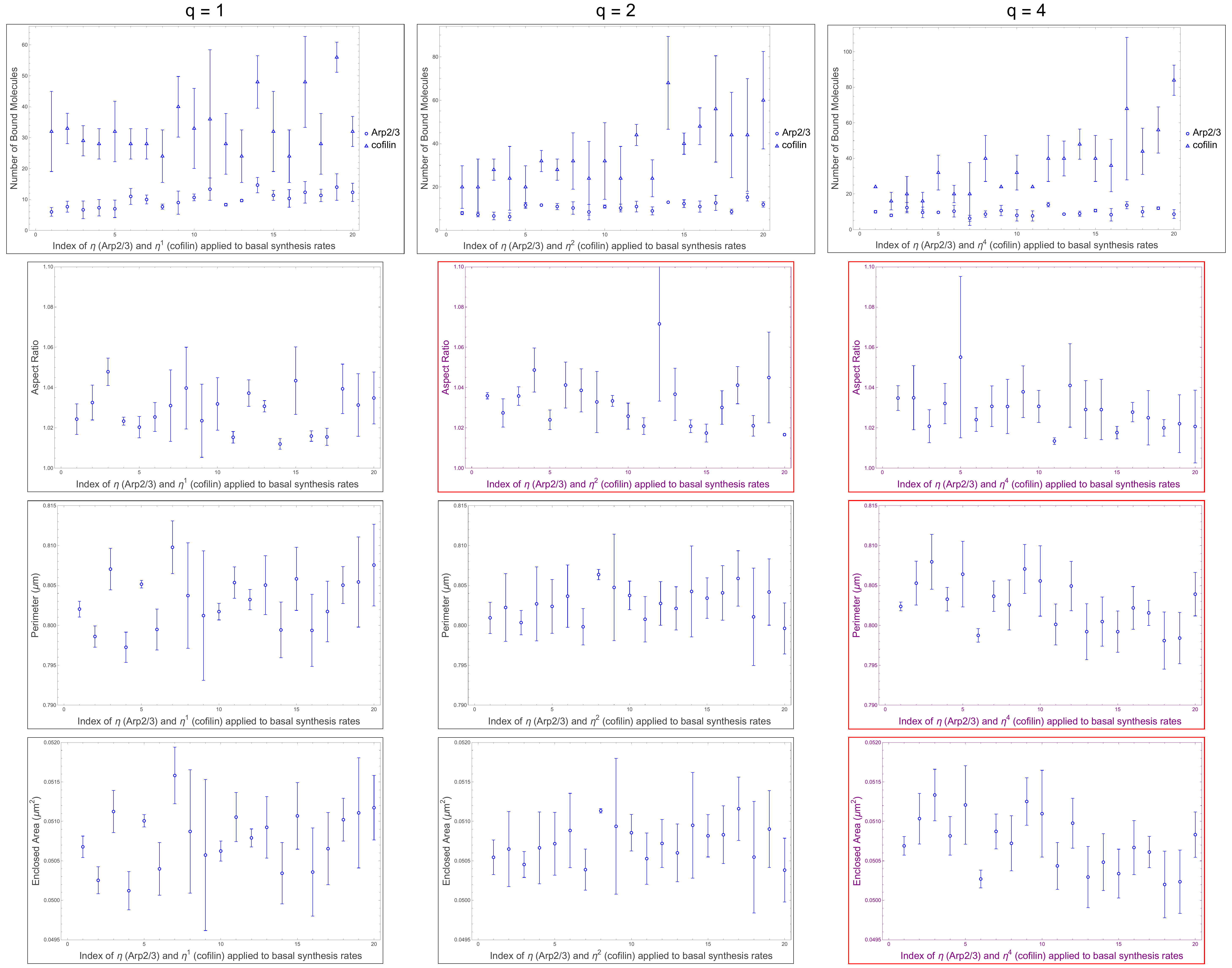}
\caption{Interaction between Arp2/3 and cofilin in the stubby spine head: Dependence of membrane properties on synthesis rates of Arp2/3 and cofilin demonstrated in parameter sweeps. For a chosen value of $\eta$, it is sequentially multiplied or divided to the original, basal synthesis rates, but at different powers of $\eta$ for Arp2/3 or cofilin, informed by the statistical values in Table~\ref{table:stats}. $\eta$ is directly applied to Arp2/3 synthesis, and $\eta$ (first column), $\eta^2$ (middle column), and $\eta^4$ (last column) are applied to cofilin synthesis. The basal synthesis rates provided in Table~\ref{table:parameters} are the middle values of the horizontal-axes. Error bars show standard error of the mean (SEM) with Bessel's correction for n = 3 simulations per data point. Indigo colored graphs are significant under p-value threshold 0.05. For the frames surrounding individual plots, blue is increasing trend, and red is decreasing trend.
}
\label{figure:parametersweepstubbyetacof}
\end{figure}

Here, $\mu$ is the variable that is varied in the $\eta$ parameter sweeps, i.e. in Fig.~\ref{figure:parametersweepstubbyeta}--~\ref{figure:parametersweepstubbyetacof}. 
If Equations (\ref{A}) and (\ref{B}) are satisfied,
for a set of axes in which $\mu$ is varied, 
then

\begin{equation}
    \label{eq::maskedvar}
    \frac{d \text{ phenotype}}{d \alpha}\frac{d\alpha}{d \mu}+\frac{d \text{ phenotype}}{d \gamma}\frac{d\gamma}{d \mu}=0,
\end{equation}
which demonstrates 
nullification of the $\gamma$ species by the $\alpha$ species
(Fig.~\ref{figure:parametersweepstubbyeta}) under these conditions. 

To show the existence of such a variable $\mu$, consider the 2D linear change of variables
$\mu = \alpha + \frac{1}{q}\gamma$ and
$\nu=\alpha-\frac{1}{q}\gamma$.
Inverting the change of variables,
$\alpha = (\mu+\nu)/2$ and
$\gamma = (q/2) (\mu-\nu)$.

\begin{table}[h]
\centering
\begin{tabular}{||l|l|l|l||}
\hline
  \multicolumn{4}{||c||}{Stubby spine head at 20.0 seconds: parameter sweep of $\eta^q$ on CaMKII$\beta$ against Arp2/3}\\
 \hline
 Membrane property / value type & $q=1$ (n=3) & $q=2$ (n=3) & $q=4$ (n=3) \\
 \hline
 Aspect ratio / Jonckheere-Terpstra p-value & 0.50 & 0.29 & 0.15 \\
 -- / Pearson's correlation p-value & 0.72 & 0.89 & 0.30 \\
 -- / Pearson's correlation r-value & -0.048 & 0.018 & -0.14 \\
 \hline
 Perimeter / -- & 0.57 & 0.44 & 0.39 \\
 -- / -- & 0.75 & 0.91 & 0.67 \\
 -- / -- & -0.042 & 0.015 & -0.055 \\
 \hline
Enclosed area / -- & 0.47 & 0.38 & 0.33 \\
 -- / -- & 0.62 & 0.81 & 0.56 \\
 -- / -- & -0.065 & 0.032 & -0.077 \\
 \hline
 \multicolumn{4}{||c||}{Stubby spine head at 20.0 seconds: parameter sweep of $\eta^q$ on Arp2/3 against CaMKII$\beta$}\\
 \hline
 Membrane property / value type & $q=1$ (n=3) & $q=2$ (n=3) & $q=4$ (n=3) \\
 \hline
 Aspect ratio / Jonckheere-Terpstra p-value & | & \textbf{0.016} & 0.18 \\
 -- / Pearson's correlation p-value & | & 0.091 & 0.24 \\
 -- / Pearson's correlation r-value & | & 0.22 & -0.16 \\
 \hline
 Perimeter / -- & | & \textbf{0.045} & \textbf{0.0014} \\
 -- / -- & | & 0.30 & \textbf{0.0033} \\
 -- / -- & | & 0.14 & 0.38 \\
 \hline
Enclosed area / -- & | & \textbf{0.031} & \textbf{0.00069} \\
 -- / -- & | & 0.18 & \textbf{0.0016} \\
 -- / -- & | & 0.17 & 0.41 \\
 \hline
 \multicolumn{4}{||c||}{Stubby spine head at 20.0 seconds: parameter sweep of $\eta^q$ on cofilin against Arp2/3}\\
 \hline
 Membrane property / value type & $q=1$ (n=3) & $q=2$ (n=3) & $q=4$ (n=3) \\
 \hline
 Aspect ratio / Jonckheere-Terpstra p-value & 0.36 & \textbf{0.013} & \textbf{0.045} \\
 -- / Pearson's correlation p-value & 0.79 & 0.19 & 0.16 \\
 -- / Pearson's correlation r-value & -0.035 & -0.17 & -0.18 \\
 \hline
 Perimeter / -- & 0.18 & 0.25 & \textbf{0.0050} \\
 -- / -- & 0.27 & 0.41 & \textbf{0.024} \\
 -- / -- & 0.15 & 0.11 & -0.29 \\
 \hline
    Enclosed area / -- & 0.15 & 0.26 & \textbf{0.0056} \\
 -- / -- & 0.26 & 0.39 & \textbf{0.022} \\
 -- / -- & 0.15 & 0.12 & -0.29 \\
 \hline
\end{tabular}
\caption{Statistical values for monotonicity and correlation in double parameter sweeps of CaMKII$\beta$ and cofilin vs. Arp2/3 in stubby spine heads. Jonckheere-Terpstra p-values are calculated using ``jonckheereTest'' in the ``PMCRPlus'' package of the R programming language. Stubby statistical values match the format of the parameter sweep of Fig.~\ref{figure:parametersweepstubbyeta}. Trend tests for monotonicity, p-values $<0.05$ are bold under the Jonckheere 1-sided test, with sidedness determined by the sign of Pearson's r in a linear fit.}
\label{table:statseta}
\end{table}

\subsection{Cofilin can resist spine head growth}
\label{section::cofilin}
In simulations, cofilin has a role in determining spine head size. Cofilin has previously been suggested as a key player in spine head growth (\cite{Paciello2025}). Whether increasing cofilin alone leads to spine size change remained to be seen. In column three of Fig.~\ref{figure:parametersweepstubby}, we find that increasing its levels, within a realistic biological range, in a spine head shrinks the spine head's size in 2D. Simulations here suggest that on the timescale of 20 seconds of a model for LTP, cofilin-dependent remodeling decreases size (p = 0.00030 for area; p = 0.00050 for perimeter; in Table~\ref{table:stats}) for stubby spine heads. A likely reason for the decrease in size is that a filament bound with more cofilin has a weaker angular bending stiffness, which affects its ability to push against the membrane.

 Previously, cofilin has been known to be necessary in order to shrink spine heads during long-term depression (\cite{Pontrello2012}) and its inhibition conducive to spine head growth during long-term potentiation (\cite{Rust2015, Calabrese2014}). We find that it is indeed realistic behavior for cofilin to have such an effect during long-term potentiation in the stubby spine head. Stimulative augmentation of synthesis rates, i.e. from modeling glutamate uncaging, is the implementation of long-term potentiation in our simulation. The simulation result is corroborative with previous research.

 An effect of cofilin observed in simulations is to resist growth provided by Arp2/3 during LTP (p = 0.00030 for enclosed area; in Fig.~\ref{figure:parametersweepstubby}). The effect is not simply independent of Arp2/3 growth nor simply additive as demonstrated in Table~\ref{table:statseta}. Rather, despite cofilin having roughly greater correlation magnitude than Arp2/3 (Table~\ref{table:stats}), i.e. around r = -0.33 for cofilin and around r = 0.22 for Arp2/3, applying $q = 2$ to cofilin, which is twice the nullification value of $q = 1$, does not overcompensate (p = 0.26, r = 0.12; in Table~\ref{table:statseta}) for Arp2/3 growth. Cofilin overcompensates for Arp2/3 growth actually at $q = 4$ (p = 0.0056, r = -0.29; in Table~\ref{table:statseta}). We observe that Arp2/3 does not completely mask the effect of cofilin induced shrinkage of spine head size, i.e. masking occurs only at $q\in\{1,2\}$, but an effect of cofilin on spine head size (r = -0.29, p = 0.0056; in Table~\ref{table:statseta}) is recovered by $q = 4$. Arp2/3 has a conditional nullification on cofilin, based on roughly $q \le 2$, log-ratio threshold between cofilin and Arp2/3 synthesis rates, pertaining to stubby spine head size. The effect from overactivating cofilin in shrinking spine head size is observed in biological experiment (\cite{Raven2023}). The model we present here with Aip1 suggests that, being necessary for Aip1 binding, cofilin supports Aip1 mechanistically, but not synergistically for morphological size (r = -0.33, p = 0.00030 for cofilin; r = 0.45, p = 0.00048 for Aip1; in Table~\ref{table:stats}), during LTP-like spine head growth.

 We also present here a hypothesis for how opposite effects of cofilin and Aip1, both as part of the severing mechanism, on spine head size could work. Cofilin does not allow Arp2/3 to bind to locations where cofillin is bound on an actin filament. This competitive binding limits further action of cofilin-Aip1. Further binding of cofilin on the filaments that do exist weakens those filaments biomechanically by decreasing angular bending stiffness. According to this hypothesis, there exist a lesser number of filaments, and further amounts of cofilin weaken the working ability of actin filaments on growing the spine head membrane. This hypothetical mechanism would out-compete increase in binding locations for Aip1 (Eq.~(\ref{eq::severing_binding_rule})) by quicker binding of cofilin to actin filaments (Eqs.~(\ref{eq::cofilin_2Node_binding_rule}, \ref{eq::cofilin_1Node_accbinding_rule}, \ref{eq::cofilin_single_binding_rule})).

\subsection{Cofilin as only actin filament severing mechanism leads to uncertainty in effects of ABPs}

\label{section::oldmodel}

\begin{figure}[h]
\centering
\includegraphics[width=\textwidth]
{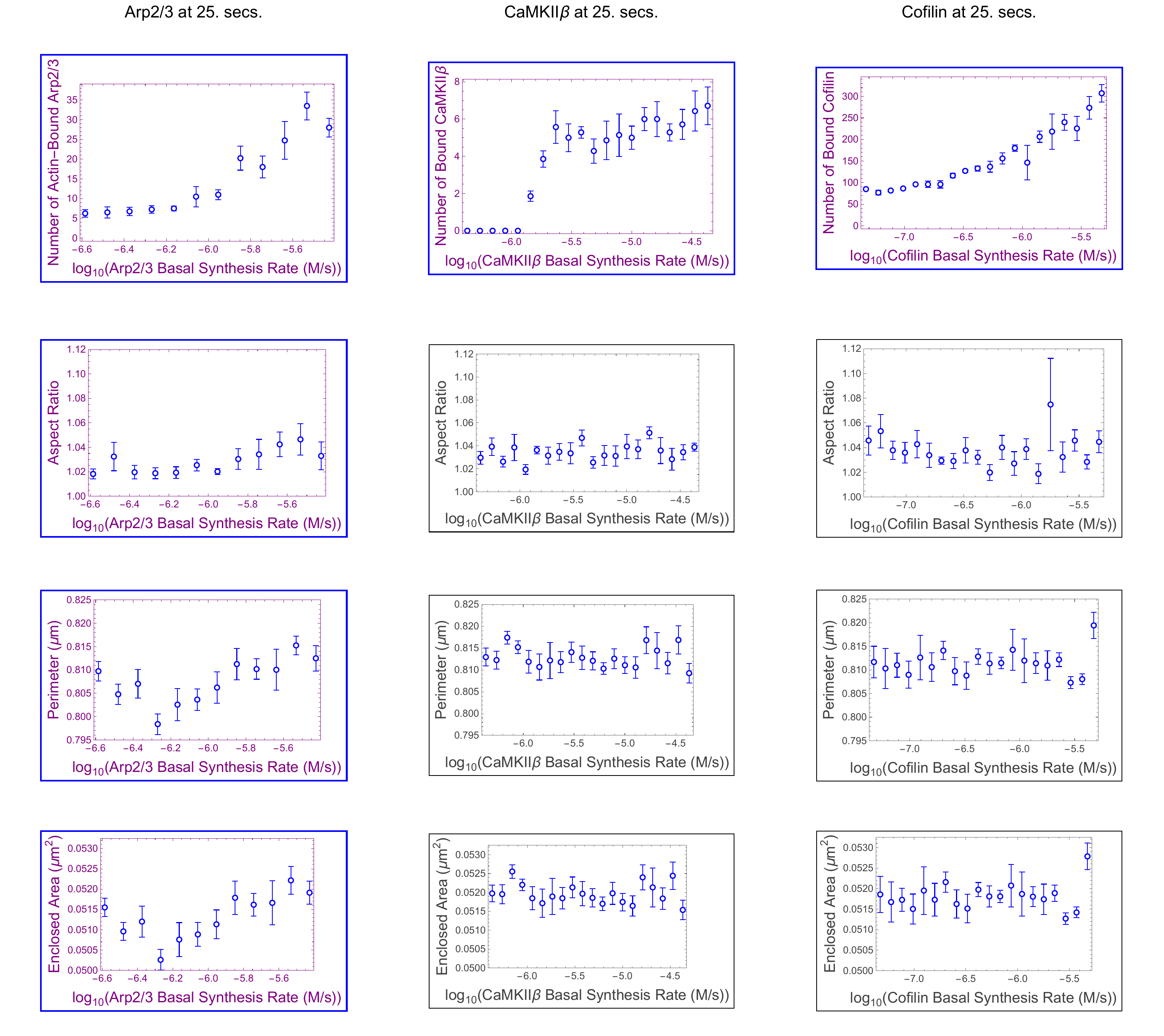}
\caption{Dependence of membrane properties on synthesis rates of three actin-binding regulatory proteins demonstrated in parameter sweeps. This model lacks the ABP Aip1 compared to the model in Fig.~\ref{figure:parametersweepstubby}. The basal synthesis rate provided in Section~\ref{section::discussion} Table~\ref{table:parameters} is the middle value for CaMKII$\beta$ and cofilin horizontal-axes. For Arp2/3 and Aip1, the range is truncated at the upper end of synthesis rate. Error bars show standard error of the mean (SEM) with Bessel's correction, for a number of stochastic simulations each. We calculate membrane aspect ratio, membrane perimeter, and membrane enclosed area for each simulation snapshot. The number of simulations is 4 for Arp2/3 and 5 for cofilin and for CaMKII$\beta$, corresponding to each mean value with error bar in the plots. Indigo graphs are significant under p-value threshold 0.05. For the frames surrounding significant individual plots, blue is a significant increasing trend, and red is a significant decreasing trend as determined by Table~\ref{table:stats}. The range on log-scale for the parameter sweep consists of $10\times$ smaller to at most $10\times$ larger than the estimated, basal synthesis rates for CaMKII$\beta$ and cofilin, with range truncated at the upper end for Arp2/3 and Aip1.
}
\label{figure:parametersweepstubbyold}
\end{figure}

In this paper, we present two models, models with and without Aip1. In this subsection, we focus on the model without Aip1. Cofilin is the only molecule that can sever actin filaments through actin filament breaking (Eq.~\ref{eq::breaking_rule}) in this model. In this model's parameter sweep (Fig.~\ref{figure:parametersweepstubbyold}), Arp2/3 is the only ABP to significantly affect stubby spine head morphology, and it does so with a large r-value (r = 0.43, p = 0.0034; in Table~\ref{table:statsold}) compared to that of other ABPs (r = -0.11 for CaMKII$\beta$ and r = 0.066 for cofilin; in Table~\ref{table:statsold}). In sections~\ref{section::growth}--\ref{section::cofilin}, all four ABPs (including Aip1) have significant effects on stubby spine head morphology. Therefore, Aip1 inclusion in the DGG model leads to less uncertainty about the effect of ABP mechanism on stubby spine head morphology. Inclusion of the Aip1 severing mechanism presented in \cite{Oosterheert2025} can lead to better certainty about other spine head morphology mechanisms. The reason is likely that the addition of the Aip1 binding-and-severing mechanism with a severing event approximately every 0.7 s \cite{Chen2015} for each bound Aip1 protein will generate more actin filaments than through biomechanical breaking alone. This increase in actin network size, will lead to a more biomechanically dynamic actin network that affects membrane morphology.

\begin{table}[h]
\centering
\smaller
\begin{tabular}{||l|l|l|l||}
\hline
 \multicolumn{4}{||c||}{Stubby spine head at 25.0 seconds: parameter sweep of model without Aip1}\\
 \hline
 Membrane property / value type & Arp2/3 (n=4) & CaMKII$\beta$ (n=5) & cofilin (n=5) \\
 \hline
 Aspect ratio / Jonckheere-Terpstra p-value & \textbf{0.0059} & 0.16 & 0.18 \\
 -- / Pearson's correlation p-value & \textbf{0.0048} & 0.30 & 0.86 \\
 -- / Pearson's correlation r-value & 0.41 & 0.10 & -0.018 \\
 \hline
 Perimeter / -- & \textbf{0.0065} & 0.080 & 0.32 \\
 -- / -- & \textbf{0.0043} & 0.35 & 0.47 \\
 -- / -- & 0.41 & -0.095 & 0.073 \\
 \hline
Enclosed area / -- & \textbf{0.0034} & 0.063 & 0.38 \\
 -- / -- & \textbf{0.0026} & 0.29 & 0.51 \\
 -- / -- & 0.43 & -0.11 & 0.066 \\
 \hline
\end{tabular}
\caption{Statistical values for monotonicity and correlation in double parameter sweeps of CaMKII$\beta$ and cofilin vs. Arp2/3 in stubby spine heads. Jonckheere-Terpstra p-values are calculated using ``jonckheereTest'' in the ``PMCRPlus'' package of the R programming language. Stubby statistical values match the format of the parameter sweep of Fig.~\ref{figure:parametersweepstubbyeta}. Trend tests for monotonicity, p-values $<0.05$ are bold under the Jonckheere 1-sided test, with sidedness determined by the sign of Pearson's r in a linear fit.}
\label{table:statsold}
\end{table}

\section{Methods}\label{section::methods}

\label{section:methods}

\subsection{Coarse-graining of multiple actin monomers into single objects}

As 
actin cytoskeleton 
grows or shrinks 
within the synaptic spine 
head, there is a 
continuing 
turnover of F-actin 
governed by
different dynamics at the two different ends of a single filament, the barbed and pointed ends (\cite{Pollard1986}). The main processes that can occur 
in
an actin network 
undergoing
remodeling are elongation and retraction at the barbed end 
with rate parameters
($k_\text{barbed, on}$, $k_\text{barbed, off}$),
elongation and retraction at the pointed end 
with rates ($k_\text{pointed, on}$, $k_\text{pointed, off}$), 
breaking from bending angles
with parameters
($\theta_\text{Break, Actin}, \theta_\text{Break, Cofilactin}, \theta_\text{Break, Boundary}$), severing from Aip1 ($k_\text{sever}$), de-branching from cofilin
($k_\text{de-branch}$),
capping 
($k_\text{cap, on}, k_\text{cap, off}$), 
bundling and unbundling 
($k_\text{CaMKII$\beta$, on}, k_\text{CaMKII$\beta$, off}$), 
and branch nucleation 
($k_\text{branch}$,$k_\text{unbranch}$) (\cite{Hotulainen2010}). 
These rate parameters
are all in
Table~\ref{table:parameters}. We begin with rate constants found in literature and then compute the coarse-grained rate constants -- replacing multiple actin monomers with 2D coordinates at their 
pointed end 
endpoints -- for a more computationally efficient simulation. Using a coarse-grained 
model, we then 
simulate the dynamics of the actin cytoskeleton. 

The barbed end of an actin filament adds monomers faster than the pointed end, which nets removal of monomers faster. Furthermore, the filament near the actin pointed end eventually forms more actin-ADP than near the barbed end, which retains newly added actin-ATP. These dynamics lead to directional pushing onto the membrane predominantly near clusters of barbed ends. To support this mechanism, the membrane-attached ends adaptively step, 
based on Brownian ratchet biophysics (\cite{Peskin1993}), their optimal coarse-graining length.

Coarse-graining is achieved by simulating aggregate rates and storing the
binding numbers of attached molecules and the IDs of linked objects,
for a total of 5--6 parameters per actin object. Below, in 
each DGG rule, 
just a 
relevant
subset of parameters 
are displayed 
inside the 
$\llangle \ldots \rrangle$
parentheses associated with objects.

Coarse-grained rate constants 
for polymerization are found by dividing the coarse-grained number into rates of elongations. The retraction, $k^-_\text{barbed}$, at a coarse-graining number is the rate for which every actin monomer, considering attached molecules, unbinds from the actin filament. Coarse-grained constants for retraction
$k^-_\text{barbed}$
that 
are dependent on the types of actin stored inside the object are computed as a rate divided by the coarse-graining number. Coarse-graining methods 
are approximate but
can significantly address the scalability of an \textit{in silico} model which we use here.

Biophysically, we set the coarse-grained resting length as the unit length of the system. Using the unit length in meters, we scale parameters involving distance units for numerical range in simulation. Furthermore, to define a bond strength parameter of the logitudinal potential, we equate the spring constant, which is the squared-potential curvature, with the equilibrium curvature of an 
anharmonic 
12-6 Lennard-Jones (LJ) potential 
(Eq.~(\ref{LJ_potential})) below.  
$\epsilon_{LJ}$ is a 
parameter for well-depth or equivalently known as the bond dissocation energy. 
Here
$k_{\text{s}}$ is the experimentally measured actin spring constant, scaled as a series of springs from the original experimental measurement, into a single spring of distance between two adjacent monomers
. To yield $k_{\text{s}}$, coarse-graining divides the original spring constant by the number of coarse-grained monomers, i.e. \(k_{\text{s}}/N_{\text{CG}}\), in the equation defining $\epsilon_{LJ}$. After $\epsilon_{LJ}$ for a single spring is estimated, this value is multiplied by $N_{\text{CG}}$ to represent the total energy stored in the series of connections that comprises a single bond between objects in the scaled simulation.
The angular bending constant is the flexural rigidity divided by the equilibrium length of a rod.

\subsection{Actin cytoskeleton remodeling}

Each of the main F-actin
end types, barbed and pointed, 
can elongate or retract based on individual rate constants that also depend on type of bound nucleotide. ADP+Pi and ADP rate constants have been found to be experimentally similar, so for simplicity we equate them in our simulation. 

To provide a better picture of how a rule functions, rules for actin polymerization and retraction are 
shown below. In the ``with'' clauses we have converted the rate constant to a function of species number and membrane area instead of concentration. $R(\theta)$ is a $2 \times 2$ rotation matrix with angle $\theta$.

\begin{center}
\begin{varwidth}{\linewidth}
\textbf{Actin Barbed End Elongation:}
\begin{equation}
\begin{diagram}[size=2em] \label{eq::growth_rule_barbed}
& (
$\Circle$_1 & \rTo & $\CIRCLE$_2 
) &
\\
\longrightarrow &
(
$\Circle$_1 &  \rTo & $\Circle$_2 & \rTo & $\CIRCLE$_3
)
\end{diagram}
\quad
\begin{varwidth}{\linewidth}
$\llangle$
($\text{${ \boldsymbol x}$}_1, \text{${ \theta}$}_1, \text{${ \eta}$}_1$), 
($\text{${ \boldsymbol x}$}_2, \text{${ \theta}$}_2, \text{${ \eta}$}_2$)
$\rrangle$
\\
$\llangle$
($\text{${ \boldsymbol x}$}_1, \text{${ \theta}$}_1, \text{${ \eta}$}_1$), 
($\text{${ \boldsymbol x}$}_2, \text{${ \theta}$}_2, \text{${ \eta}$}_2$), 
($\text{${ \boldsymbol x}$}_3, \text{${ \theta}$}_3, \text{${ \eta}$}_3$)
$\rrangle$
\end{varwidth}
\end{equation}
\quad \quad \text{\boldmath $\mathbf{with}$} \ \
$\frac{k_{\text{barbed, on, $\text{${ \eta}$}_3$}}}{N_\text{CG}}$ $N_{\text{Actin, free}}$
\\ \par \quad \quad \ \ 
$\begin{cases}
\theta_2 \sim \mathcal{N}(0, \sigma_\theta=\sqrt{2/L_p}) \; ; \\
\text{${ \boldsymbol x}$}_3 = \text{${ \boldsymbol x}$}_2 + R(\theta_2)\cdot(\text{${ \boldsymbol x}$}_2-\text{${ \boldsymbol x}$}_1)\\
\text{$\text{${ \eta}$}_3$$\in$ \{ATP, ADP\}} 
\end{cases}$
\end{varwidth}
\end{center}

\begin{figure}
\label{figure::diagram}
\centering
\begin{subfigure}{.45\linewidth}
\includegraphics[width=\linewidth]{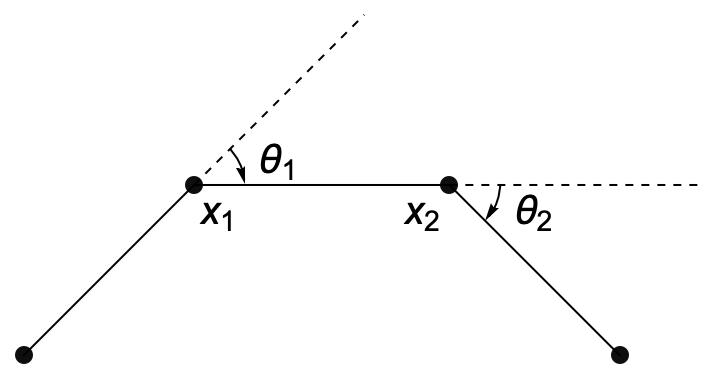}
\caption{}\label{fig:seqangles}
\end{subfigure}
\begin{subfigure}{.45\linewidth}
\includegraphics[width=\linewidth]{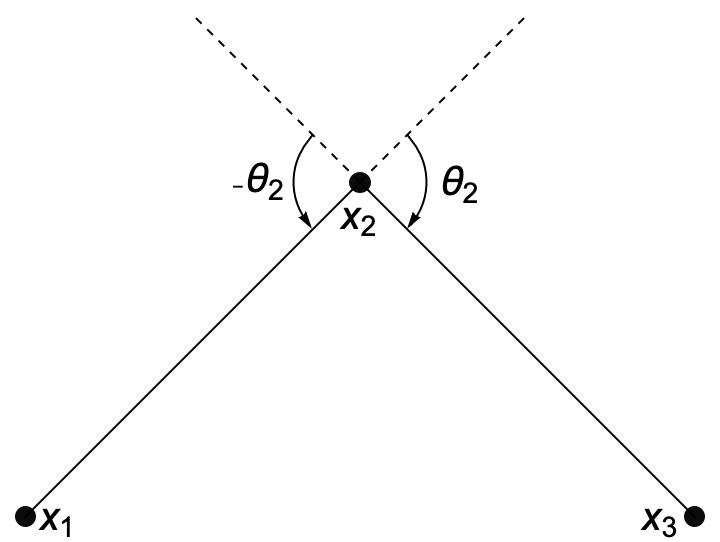}
\caption{}\label{fig:oppangles}
\end{subfigure}
\raisebox{0.15\textheight}{\begin{subfigure}{.45\linewidth}
\includegraphics[width=\linewidth]{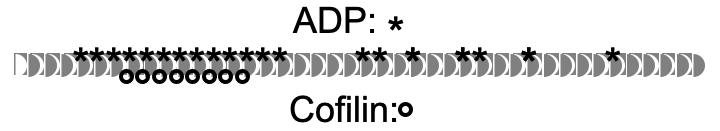}
\caption{}\label{fig:actinfil}
\end{subfigure}}
\begin{subfigure}{.45\linewidth}
\includegraphics[width=\linewidth]{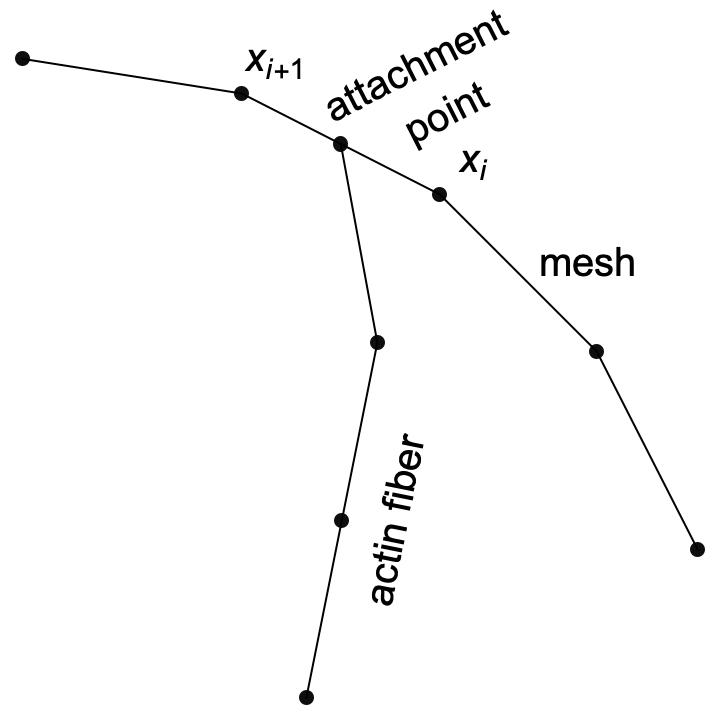}
\caption{}\label{fig:membraneviz}
\end{subfigure}
\caption{Diagrammatic visualization of the physical system with (a) illustration of angle definitions, (b) providing opposite and equal angles with vectors, (c) visualizing an actin filament with ADP bound near cofilin boundaries, more toward the pointed end, as well as cofilin binding sequentially, and (d) depicting an actin fiber attached to a membrane mesh.}
\end{figure}

Here 
\(\theta\) represents the middle angle between consecutive vectors of three nodes, which is zero for end nodes. 
The ``;'' represents sequential execution of 
two successive parameter assignments, 
so that $\theta_2$ is sampled before use in setting $\text{${ \boldsymbol x}$}_3$. 
The sampling of  \(\theta_2\) is centered around zero as its mean,
with a standard deviation 
$\sigma_\theta$
taking into
account the persistence length $L_p$.
The topology and geometric parameters of an actin fiber are 
illustrated in {Figure \ref{fig:seqangles}}. 
 As shown in the ``where'' clause in
Eq.~(\ref{eq::growth_rule_barbed}),
\(\eta\) represents the type of nucleoside phosphate 
(one of ATP, ADP, or ADP+Pi) 
attached to the next new actin object, where we let ADP stand for both ADP and ADP+Pi due to similar rates of elongation and retraction.
${ \boldsymbol x}, \theta$, and $\eta$,
there are other actin object parameters to be 
introduced below, suppressed here for readability. 

Node type symbols
$\Circle$ and $\CIRCLE$ represent interior and end segments of an actin filament,
each segment being one or several actin units long depending on an adjustable
coarse-graining 
parameter. 
Additional node type symbols include
$\resizebox{2.0ex}{!}{$\triangle$}$ for actin network {branch point} junction segments,
$\resizebox{2.0ex}{!}{\rotatebox[origin=c]{30}{$\blacktriangle$}}$ for end-capped actin segments,
$\resizebox{1.5ex}{!}{$\filledsquare$}$ for Arp2/3 proteins at branch points at the end of a new fiber,
$\resizebox{1.5ex}{!}{$\square$}$ for Arp2/3 proteins at branch points not at the end of a new fiber,
and symbols 
$\halfcirc[0.7ex] \in \{\Circle, \CIRCLE, \rotatebox[origin=c]{30}{$\blacktriangle$}\}$,
$\boxbox\in \{\resizebox{1.5ex}{!}{$\filledsquare$} ,
$\resizebox{1.5ex}{!}{$\square$}$ \}$,
$\resizebox{2.0ex}{!}{$\filledstar$} \in \{\Circle, \CIRCLE, 
\resizebox{1.5ex}{!}{$\square$},
\resizebox{1.5ex}{!}{$\filledsquare$} , \resizebox{2.0ex}{!}{$\triangle$}, \rotatebox[origin=c]{30}{$\blacktriangle$}\}$
that each represent a choice among these basic symbols. We consider graph-local rewrite rules for cytoskeleton networks, and show how to derive DGG rules from a mutually exclusive and exhaustive collection of particular graph-local neighborhood types for the left hand sides of the
rules, all from a global energy function that is
a sum of graph-local terms. 


The format of rule 
Eq.~(\ref{eq::growth_rule_barbed}) 
above is similar to how it is implemented in Plenum with objects stored connected to one another in a graph, data stored within the objects such as position, angle, and nucleotide type, and clauses following the keyword ``with'' and ``solving'' to denote the propensity rule firing rate or differential equation respectively.
The 
foregoing
rule is presented as a mathematically idealized and ``prettyprinted'' version of its 
computational implementation in any specific piece of software.
Executable rules in this paper were implemented in the Plenum package (\cite{Yosiphon2009})
for the Mathematica computer algebra and problem-solving environment.
In Plenum the foregoing rule looks like 
Figure \ref{figure:prettyprint}. 

\begin{figure}[h]
\begin{center}
\begin{minipage}{\textwidth}
\begin{verbatim}
(*Rule for elongating a two node graph from free actin-ATP*)
{aP == actin[ID, coords, IDNext, ang, nucleotide], 
  barbedEnd[IDNext, coordsNext, nucleotideNext, nullPointer, dist], 
  newID[NID],
  actinATPCount[numActin],
  spineHeadArea[area], s1 == spine[sID, sCoord, sIDNext], 
  s2 == spine[sIDNext, sCoordNext, sIDNextNext]} 
  -> 
  {actinATPCount[numActin - nCG],
  newID[NID + 1], aP,
  spineHeadArea[area],
  (* make the replacement node for the new INT: *)
  actin[IDNext, coordsNext, NID, randAng, nucleotideNext],
  (* make the newly created end node: *)
  barbedEnd[NID, 
   If[pointSegmentDistance[coordsNext, sCoord, sCoordNext] < 
     overgrowthL, 
    raySegmentIntersection[coordsNext, coordsNext - coords, sCoord, 
     sCoordNext], 
    coordsNext + 
     actinObjectRise newPos[coordsNext - coords, randAng]/
       Norm[coordsNext - coords]], ATP, 
   If[pointSegmentDistance[coordsNext, sCoord, sCoordNext] < 
     overgrowthL, sID, nullPointer], actinObjectRise], s1, s2
  },
with[kPlusBarbedT/nCG numActin* 
  grammarPDF[NormalDistribution[0.0, \[Theta]p], randAng] Boole[
   numActin > nCG] Boole[
   rayIntersectsSegmentQ[coordsNext, coordsNext - coords, sCoord, 
    sCoordNext]]]
\end{verbatim}
\end{minipage}
\end{center}
\caption{Actual Plenum rule text corresponding to
the rule of Eq.~(\ref{eq::growth_rule_barbed}). Here, we make use of the sub-types to an ``actin'' general type, which includes the ``barbedEnd[]'' and ``actin[]'' headers. Sub-dividing types allow for a corpus of rules in the DGG ruleset that are compatible with a cognitive learning algorithm called ``chunking'' (\cite{Laird1984}). Chunking involves sub-dividing a problem and storing solutions for future use. The casework leads to, in total, hundreds of DGG rules that efficiently contribute to simulation in the Plenum modeling language.}
\label{figure:prettyprint}
\end{figure}

From left to right, the parameters inside actin objects are 
an integer-valued
object ID, spatial coordinates, 
an object ID ``pointer'' to
the next actin 
object (for non-barbed end actin objects),
the angle formed by the three objects, the angle formed with the branch (for junction objects), the
type of bound molecule ATP, ADP, ADPPi, cofilin, CaMKII$\beta$
bound to the actin object, the spine vertex to which the end is bound, else a null pointer, (for end objects), and equilibrium length for the end object. The function ``newpos'' rotates a vector around the origin by an angle that is sampled by another function ``grammarPDF'',
as in the additional clause of Eq.~(\ref{eq::growth_rule_barbed}). 

Similarly,

\begin{center}
\begin{varwidth}{\linewidth}
\textbf{Actin Barbed End Retraction:}
\begin{equation}
\label{eq::retraction_rule_barbed}
\begin{diagram}[size=2em] 
& (
$\Circle$_1 &  \rTo & $\Circle$_2 & \rTo & $\CIRCLE$_3
) &
\\
\longrightarrow &
(
$\Circle$_1 & \rTo & $\CIRCLE$_2
)
\end{diagram}
\quad
\begin{varwidth}{\linewidth}
$\llangle$
($\text{${ \boldsymbol x}$}_1$, $\text{${ \theta}$}_1$, $\text{${ \eta}$}_1$), 
($\text{${ \boldsymbol x}$}_2$, $\text{${  \theta}$}_2$, $\text{${ \eta}$}_2$), 
($\text{${ \boldsymbol x}$}_3$,$ \text{${  \theta}$}_3$, $\text{${  \eta}$}_3$)
$\rrangle$
\\
$\llangle$
($\text{${ \boldsymbol x}$}_1$, $\text{${ \theta}$}_1$, $\text{${ \eta}$}_1$), 
($\text{${ \boldsymbol x}$}_2$, $\text{${ \theta}$}_2$, $\text{${ \eta}$}_2$)
$\rrangle$
\end{varwidth}
\end{equation}
\quad \quad \text{\boldmath 
$\mathbf{with}$} \ \
$\frac{k_{\text{barbed, off, $\text{${ \eta}$}_3$}}}{N_\text{CG}}$ $N_{\text{Actin, free}}$
\end{varwidth}
\end{center}

In addition there are two further cases of the
foregoing two rules with different context nodes
( $\square_1$ and $\triangle_1$
in addition to $\Circle_1$), i.e. 
nodes that remain unchanged in rule firing, like enzymes in a catalysed chemical reaction.
Pairing the foregoing two rules together with
the $\rightleftharpoons$ bidirectional arrow,
and omitting the propensity functions,
these cases are:

\begin{center}
\begin{varwidth}{\linewidth}
\textbf{Actin Barbed End Elongation:}
\begin{equation}
\begin{diagram}[size=2em] \label{eq::growth_rule_barbed_2}
& (
\resizebox{1.5ex}{!}{$\square$}_1 & \rTo & $\CIRCLE$_2 
) &
\\
\rightleftharpoons &
(
\resizebox{1.5ex}{!}{$\square$}_1 &  \rTo & $\Circle$_2 & \rTo & $\CIRCLE$_3
)
\end{diagram}
\quad
\begin{varwidth}{\linewidth}
$\llangle$
($\text{${ \boldsymbol x}$}_1, \text{${ \theta}$}_1, \text{${ \eta}$}_1$), 
($\text{${ \boldsymbol x}$}_2, \text{${ \theta}$}_2, \text{${ \eta}$}_2$)
$\rrangle$
\\
$\llangle$
($\text{${ \boldsymbol x}$}_1, \text{${ \theta}$}_1, \text{${ \eta}$}_1$), 
($\text{${ \boldsymbol x}$}_2, \text{${ \theta}$}_2, \text{${ \eta}$}_2$), 
($\text{${ \boldsymbol x}$}_3, \text{${ \theta}$}_3, \text{${ \eta}$}_3$)
$\rrangle$
\end{varwidth}
\end{equation}
\end{varwidth}
\end{center}

and

\begin{center}
\begin{varwidth}{\linewidth}
\textbf{Actin Barbed End Elongation:}
\begin{equation}
\label{eq::growth_rule_barbed_3}
\begin{diagram}[size=2em] 
& (
\resizebox{2.0ex}{!}{$\triangle$}_1 & \rTo & 
\resizebox{1.5ex}{!}{$\filledsquare$}_2
) &
\\
\rightleftharpoons &
(
\resizebox{2.0ex}{!}{$\triangle$}_1 & \rTo & 
\resizebox{1.5ex}{!}{$\square$}_2 & \rTo & $\CIRCLE$_3
)
\end{diagram}
\quad
\begin{varwidth}{\linewidth}
$\llangle$
($\text{${ \boldsymbol x}$}_1, \text{${ \theta}$}_1, \text{${ \eta}$}_1$), 
($\text{${ \boldsymbol x}$}_2, \text{${ \theta}$}_2, \text{${ \eta}$}_2$)
$\rrangle$
\\
$\llangle$
($\text{${ \boldsymbol x}$}_1, \text{${ \theta}$}_1, \text{${ \eta}$}_1$), 
($\text{${ \boldsymbol x}$}_2, \text{${ \theta}$}_2, \text{${ \eta}$}_2$), 
($\text{${ \boldsymbol x}$}_3, \text{${ \theta}$}_3, \text{${ \eta}$}_3$)
$\rrangle$
\end{varwidth}
\end{equation}
\end{varwidth}
\end{center}

For each of these barbed end elongation/retraction rules
there is a corresponding pointed end rule.
The first two are shown below.

\begin{center}
\begin{varwidth}{\linewidth}
\textbf{Actin Pointed End Elongation:}
\begin{equation}
\label{eq::growth_rule_pointed}
\begin{diagram}[size=2em] 
& (
$\CIRCLE$_2 & \rTo & $\Circle$_3
) &
\\
\rightleftharpoons &
(
$\CIRCLE$_1 &  \rTo & $\Circle$_2 & \rTo & $\Circle$_3
)
\end{diagram}
\quad
\begin{varwidth}{\linewidth}
$\llangle$
($\text{${ \boldsymbol x}$}_2, \text{${ \theta}$}_2, \text{${ \eta}$}_2$), 
($\text{${ \boldsymbol x}$}_3, \text{${ \theta}$}_3, \text{${ \eta}$}_3$)
$\rrangle$
\\
$\llangle$
($\text{${ \boldsymbol x}$}_1, \text{${ \theta}$}_1, \text{${ \eta}$}_1$), 
($\text{${ \boldsymbol x}$}_2, \text{${ \theta}$}_2, \text{${ \eta}$}_2$), 
($\text{${ \boldsymbol x}$}_3, \text{${ \theta}$}_3, \text{${ \eta}$}_3$)
$\rrangle$
\end{varwidth}
\end{equation}
\quad \quad \text{\boldmath $\mathbf{with}$} \ \
$\frac{k_{\text{pointed,$\text{${ \eta}$}_3$}}}{N_\text{CG}}$ $N_{\text{Actin, free}}$ 
\par
\quad \quad \ \ 
$\begin{cases}
\theta_2 \sim \mathcal{N}(0, \sigma_\theta) \; ;\\
\text{${ \boldsymbol x}$}_1: = \text{${ \boldsymbol x}$}_2 + R(-\theta_2)\cdot(\text{${ \boldsymbol x}$}_2-\text{${ \boldsymbol x}$}_3)\\
\text{$\text{${ \eta}$}_1$$\in$ \{ATP, ADP\}} 
\end{cases}$
\end{varwidth}
\end{center}


\begin{center}
\begin{varwidth}{\linewidth}
\textbf{Actin Pointed End Retraction:}
\begin{equation}
\label{eq::retraction_rule_pointed}
\begin{diagram}[size=2em] 
& (
$\CIRCLE$_1 &  \rTo & $\Circle$_2 & \rTo & $\Circle$_3
) &
\\
\rightleftharpoons &
(
$\CIRCLE$_2 & \rTo & $\Circle$_3
)
\end{diagram}
\quad
\begin{varwidth}{\linewidth}
$\llangle$
($\text{${ \boldsymbol x}$}_1, \text{${ \theta}$}_1, \text{${ \eta}$}_1)$, 
($\text{${ \boldsymbol x}$}_2, \text{${ \theta}$}_2, \text{${ \eta}$}_2)$, 
($\text{${ \boldsymbol x}$}_3, \text{${ \theta}$}_3, \text{${ \eta}$}_3$)
$\rrangle$
\\
$\llangle$
($\text{${ \boldsymbol x}$}_2, \text{${ \theta}$}_2, \text{${ \eta}$}_2$), 
($\text{${ \boldsymbol x}$}_3, \text{${ \theta}$}_3, \text{${ \eta}$}_3$)
$\rrangle$
\end{varwidth}
\end{equation}
\quad \quad \text{\boldmath $\mathbf{with}$} \ \
$\frac{k_{\text{pointed,$\text{${ \eta}$}_1$}}}{N_\text{CG}}$ $N_{\text{Actin, free}}$
\end{varwidth}
\end{center}


We again have that 
$\sigma_{\theta} = \sqrt{\frac{2}{L_p}}$,
where $L_p$ is persistence length transformed to unit length of an actin filament segment, according to the statistics of a semiflexible polymer
and a random walk in small angles. 
Note that angle $\theta_2$ is being used
in the reverse direction of traversal along the fiber,
and hence has its sign reversed, as shown in 
Fig.~\ref{fig:oppangles}. 
This specification is also used to update the angles in the direct neighborhood of the center node, after moving it. Thus, angles are accessed for potential severing and not recomputed at each simulation iteration.

In addition to end elongation and retraction, there exist several more rule types for network remodeling such as:
\begin{enumerate}
    \item Ordinary differential equation (ODE) molecule rules for synthesis and degradation

    Basal and stimulated (e.g., influx) synthesis rates for simulated proteins are estimated from experimental data as done previously (\cite{Quintana-Rangamani2024}). As in \cite{Quintana-Rangamani2024}, the following differential equation estimates the experimental system which provides the data:
    \begin{align}
        \frac{dM}{dt} &= I_{S, M} + I_M - k_M M,
        \label{eq:M_ODE}
    \end{align}
    where $M$ is the molecule's normalized fluorescence measurement, $I$ represents the influxes from stimuli ($I_{S,M}$) and basal rates ($I_M$) respectively, and $k_M$ is the degradation. For each molecule of interest, we fit the curves in \cite{Bosch2014} to differential equation Eq.~(\ref{eq:M_ODE})
    above. The results for CaMKII$\beta$ (bundling) and Arp2/3 are shown 
    in Figure~\ref{fig:paramrates} 
    comparing the experimental data trend with the parameter fitting. The 
    resulting
    estimated rates are provided in Table 1. We set the initial number of proteins at the steady-state concentrations provided by basal synthesis rates. The combined basal and stimulation rates create actin and ABPs (cofilin, Arp2/3, CaMKII$\beta$, Aip1), simulating LTP induction, as in \cite{Bosch2014} and \cite{Quintana-Rangamani2024}.

    \begin{figure}[H]
    \centering
    \includegraphics[width=0.65\textwidth]{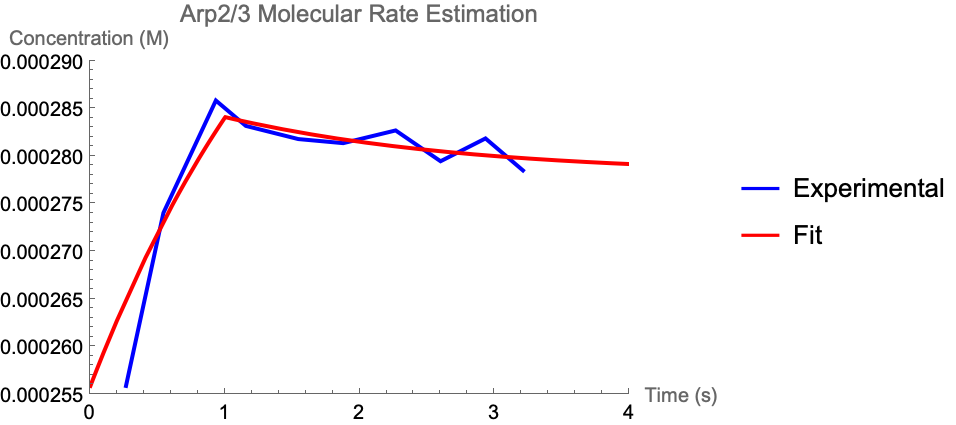}
    \includegraphics[width=0.65\textwidth]{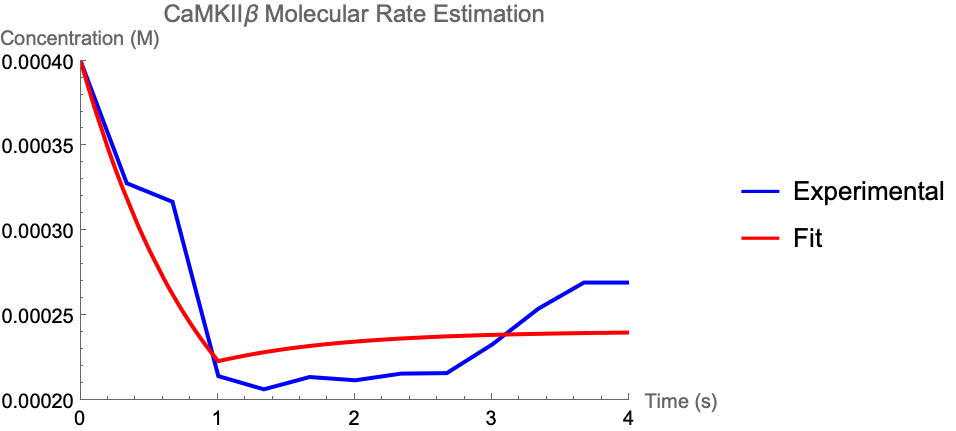}
    \caption{In the top plot, the estimation of Arp2/3 synthesis, influx, and degradation rates from experimental data (\cite{Bosch2014}) are shown, and in the bottom plot, those for CaMKII$\beta$, a bundling protein, are shown. Standard errors in estimated rates are shown in Table~\ref{table:parameters}.}
    \label{fig:paramrates}
    \end{figure}
    
    \item Stochastic rules for post-polymerization phosphate release
    
    The stochastic rules for phosphate release follow actin-attached nucleotide state transitions that regulate binding and unbinding of cofilin in addition to actin polymerizing dynamics. Thus, these stochastic rules regulate cytoskeletal remodeling propensities like synthesis and degradation ODE rules. Phosphate release occurs sequentially from ATP to ADP+Pi and from ADP+Pi to ADP. ADP+Pi is an intermediate state, where the inorganic phosphate (Pi) is mainly bound to the actin filament and not to the nucleotide molecule (\cite{Kudryashov2013}). Adjacent cofilin molecules bound to the actin filament accelerate the release of Pi from actin-ADP+Pi to yield actin-ADP.
    
    \item Branching by Arp2/3 nucleation which forms a junction with three objects connected to a single object node with an expected angle of around 70\textdegree $= \frac{70\pi}{180} \text{radians}$.
    \begin{center}
    \begin{varwidth}{\linewidth}
    \textbf{Actin Arp2/3 Branching:}
    \begin{equation}
    \label{eq::branch_rule}
    \begin{varwidth}{\linewidth}
    \quad\space\space$\left(\begin{diagram}[size=1em] 
    & $\Circle$_1 & \rTo & $\Circle$_2 & \rTo & $\Circle$_3 \\ \\
    & &  \resizebox{1.5ex}{!}{$\filledsquare$}_4 & & &
    \end{diagram} \right)$
    \\
    $\longrightarrow$
    $\left(
    \begin{diagram}[size=1em]
    $\Circle$_1 & \rTo &  \resizebox{2.0ex}{!}{$\blacktriangle$}_2 & \rTo & $\Circle$_3 \\
    & & & \rdTo &  \\
    & & & & \resizebox{1.5ex}{!}{$\filledsquare$}_4 
    \end{diagram}
    \right)$
    \end{varwidth}
    \quad
    \begin{varwidth}{\linewidth}
    $\llangle$
    ($\text{${ \boldsymbol x}$}_1$), 
    ($\text{${ \boldsymbol x}$}_2$, $\eta_2$),
    ($\text{${ \boldsymbol x}$}_3$),
    ($\text{${ \boldsymbol x}$}_4$)
    $\rrangle$
    \\
    $\llangle$
    ($\text{${ \boldsymbol x}$}_1$), 
    ($\text{${ \boldsymbol x}$}_2$, $\eta_2$, $\theta$),
    ($\text{${ \boldsymbol x}$}_3$), 
    ($\text{${ \boldsymbol x}$}_4$)
    $\rrangle$
    \end{varwidth}
    \end{equation}
    \quad \text{\boldmath $\mathbf{with}$} \ \
    $k_{\text{branch}}$ $N_{\text{Arp, free}}$ $(1-\delta(\eta_2, \text{cof}))$
    \par
    \quad \ \
    $\begin{cases}
    \text{s} \sim \mathcal{U} (\pm 1) \; ; \\
    \text{$\theta$} \sim \mathcal{N}(\frac{70 \pi}{180} \text{s}, \sigma_{\theta})  \; ;  \\
    \text{${ \boldsymbol x}$}_4 := \text{${ \boldsymbol x}$}_2 + R(\theta_2) \cdot \left(\text{${ \boldsymbol x}$}_2-\text{${ \boldsymbol x}$}_1\right)
    \end{cases}$
    \end{varwidth}
    \end{center}

    Here, the branching mechanism is mutually exclusive with already bound cofilin.

    \begin{center}
    \begin{varwidth}{\linewidth}
    \textbf{Actin Arp2/3 Unbranching:}
    \begin{equation}
    \label{eq::unbranch_rule}
    \begin{varwidth}{\linewidth}
    \quad\space\space$\left(\begin{diagram}[size=1em] 
    $\Circle$_1 & \rTo & \resizebox{2.0ex}{!}{$\blacktriangle$}_2 & \rTo & $\Circle$_3 \\
    & & & \rdTo & \\
    & & & & \resizebox{1.5ex}{!}{$\filledsquare$}_4 
    \end{diagram} \right)$
    \\
    $\longrightarrow$
    $\left(
    \begin{diagram}[size=1em]
    $\Circle$_1 & \rTo & $\Circle$_2 & \rTo & $\Circle$_3
    \end{diagram}
    \right)$
    \end{varwidth}
    \quad
    \begin{varwidth}{\linewidth}
    $\llangle$
    ($\text{${ \boldsymbol x}$}_1, \text{${ \theta}$}_1$), 
    ($\text{${ \boldsymbol x}$}_2, \text{${ \theta}$}_2$),
    ($\text{${ \boldsymbol x}$}_3, \text{${ \theta}$}_3$)
    $\rrangle$
    \\
    $\llangle$
    ($\text{${ \boldsymbol x}$}_1, \text{${ \theta}$}_1$), 
    ($\text{${ \boldsymbol x}$}_2, \text{${ \theta}$}_2$),
    ($\text{${ \boldsymbol x}$}_3, \text{${ \theta}$}_3$),
    ($\text{${ \boldsymbol x}$}_4, \text{${ \theta}$}_4$)
    $\rrangle$
    \end{varwidth}
    \end{equation}
    \quad \text{\boldmath $\mathbf{with}$} \ \
    $k_{\text{unbranch}}$ 
    \end{varwidth}
    \end{center}

    $k_{\text{unbranch}}$ is accelerated by a factor of roughly 50 in the case that cofilin is bound to the junction (black triangle) node. 
    
    \item Capping, which stops the barbed end from elongating or retracting until the cap is removed, and uncapping:

        
    \begin{center}
    \begin{varwidth}{\linewidth}
    \textbf{Actin Barbed End Capping:}
    \begin{equation}
    \label{eq::capping_rule}
    \begin{diagram}[size=2em] 
    & (
    $\CIRCLE$_1
    ) &
    \\
    \longrightarrow &
    (
    \resizebox{1.75ex}{!}{$\blacktriangleright$}_1
    )
    \end{diagram}
    \quad
    \begin{varwidth}{\linewidth}
    $\llangle$
    ($\text{${ \boldsymbol x}$}_1, \text{${ \theta}$}_1$)
    $\rrangle$
    \\
    $\llangle$
    ($\text{${ \boldsymbol x}$}_1, \text{${ \theta}$}_1$)
    $\rrangle$
    \end{varwidth}
    \end{equation}
    \quad \quad \text{\boldmath $\mathbf{with}$} \ \
    $k_{\text{cap, on}}$ $N_{\text{Cap, Free}}$
    \end{varwidth}
    \end{center}

    \begin{center}
    \begin{varwidth}{\linewidth}
    \textbf{Actin Barbed End Uncapping:}
    \begin{equation}
    \label{eq::uncapping_rule}
    \begin{diagram}[size=2em] 
    & (
    \resizebox{1.75ex}{!}{$\blacktriangleright$}_1
    ) &
    \\
    \longrightarrow &
    (
    $\CIRCLE$_1
    )
    \end{diagram}
    \quad
    \begin{varwidth}{\linewidth}
    $\llangle$
    ($\text{${ \boldsymbol x}$}_1, \text{${ \eta}$}_\text{1, Aip1}$)
    $\rrangle$
    \\
    $\llangle$
    ($\text{${ \boldsymbol x}$}_1, \text{${ \eta}$}_\text{1, cof}$)
    $\rrangle$
    \end{varwidth}
    \end{equation}
    \quad \quad \text{\boldmath $\mathbf{with}$} \ \
    $k_{\text{cap, off}}$
    \end{varwidth}
    \end{center}
    
    \item Cofilin binding, which weakens bending stiffness of bound segments called cofilactin, and lowers 
    {the minimal}
    angle for severing at boundaries along the 
    actin filament
    (under the rule of Eq.~(\ref{eq::breaking_rule}) below). 
    These rules require that further actin parameters $\eta_\ast$ be deployed: 

    \begin{center}
    \begin{varwidth}{\linewidth}
    \textbf{Accelerated Cofilin Binding (2-Node):}
    \begin{equation}
    \label{eq::cofilin_2Node_binding_rule}
    \begin{diagram}[size=2em] 
    & (
    \{\resizebox{2.0ex}{!}{$\filledstar$}_1 & \rTo & \{\resizebox{2.0ex}{!}{$\filledstar$} \backslash \boxbox\}_2
    ) &
    \\
    \longrightarrow &
    (
    \{\resizebox{2.0ex}{!}{$\filledstar$}_1 & \rTo & \{\resizebox{2.0ex}{!}{$\filledstar$} \backslash \boxbox\}_2
    )
    \end{diagram}
    \quad
    \begin{varwidth}{\linewidth}
    $\llangle$
    ($\text{${ \boldsymbol x}$}_1, \text{cof}$), 
    ($\text{${ \boldsymbol x}$}_2,\text{ADP}$)
    $\rrangle$
    \\
    $\llangle$
    ($\text{${ \boldsymbol x}$}_1, \text{cof}$), 
    ($\text{${ \boldsymbol x}$}_2, \text{cof}$)
    $\rrangle$
    \end{varwidth}
    \end{equation}
    \quad \quad \text{\boldmath $\mathbf{with}$} \ \
    $\frac{k_{\text{on-edge, cofilin}}}{N_\text{CG}}
    N_{\text{cofilin, Free}}$
    \end{varwidth}
    \end{center}
    Here \text{ADP} indicates
    the nucleotides
    attached to filament segment object $i$ are fully in an ADP state 
    and ``cof'' indicates 
    the attached molecules are in a cofilin bound 
    state, as illustrated in Fig.~\ref{fig:actinfil}.
    
    This rule exists 
    with its flipped version with cofilin binding from the right node to the left node towards the pointed end.

\begin{center}
\begin{varwidth}{\linewidth}
\textbf{Accelerated Cofilin Binding (1-Node):}
\begin{equation}
\label{eq::cofilin_1Node_accbinding_rule}
\begin{diagram}[size=2em] 
& 
(
\{\resizebox{2.0ex}{!}{$\filledstar$} \backslash \boxbox\}_2
) &
\\
\longrightarrow &
(
\{\resizebox{2.0ex}{!}{$\filledstar$} \backslash \boxbox\}_2
)
\end{diagram}
\quad
\begin{varwidth}{\linewidth}
$\llangle$
($\text{${ \boldsymbol x}$}_1, \text{cof-S}$), 
$\rrangle$
\\
$\llangle$
($\text{${ \boldsymbol x}$}_1, \text{cof}$)
$\rrangle$
\end{varwidth}
\end{equation}
\quad \quad \text{\boldmath $\mathbf{with}$} \ \
$\frac{2 k_{\text{on-edge, cofilin}}}{N_{\text{CG}}}
N_{\text{cofilin, Free}}$
\end{varwidth}
\end{center}

This rule starts with matching ``cof-S'' which is a single bound cofilin, as the result of a rare event in Eq.~(\ref{eq::cofilin_2Node_binding_rule}), and converts to ``cof'' to represent the whole node bound with cofilin. In an actin helix, a singly bound cofilin can cooperatively (\cite{Cao2006}) add adjacently to the edge of a cofilin segment in available directions, i.e. left-right leads to $\times$2 in the above ``with'' clause. 


%

    \begin{center}
    \begin{varwidth}{\linewidth}
    \textbf{Bare Filament Cofilin Binding:}
    \begin{equation}
    \label{eq::cofilin_single_binding_rule}
    \left(
    \begin{diagram}[size=2em]
    \{\resizebox{2.0ex}{!}{$\filledstar$} \backslash \boxbox\}_1
    \end{diagram}
    \right)
    \llangle
    (\text{${ \boldsymbol x}$}_1, \text{ADP})
    \rrangle 
    \longrightarrow 
    \left(
    \begin{diagram}[size=2em]
    \{\resizebox{2.0ex}{!}{$\filledstar$} \backslash \boxbox\}_1
    \end{diagram}
    \right)
    \llangle
    (\text{${ \boldsymbol x}$}_1, \text{cof-S})
    \rrangle
    \end{equation}
    $\quad \quad \text{\boldmath $\mathbf{with}$} \ \
    k_{\text{on-single, cofilin}}N_{\text{cofilin,free}}$
    \end{varwidth}
    \end{center}
    
    \begin{center}
    \begin{varwidth}{\linewidth}
    \textbf{Actin Filament Cofilin Unbinding:}
    \begin{equation}
    \label{eq::cofilin_unbinding_rule}
    \left(
    \mbox{\begin{diagram}[size=2em]
    \; 
    \resizebox{2.0ex}{!}{$\filledstar$}_1
    \end{diagram}}
    \right)
    \llangle
    (\text{${ \boldsymbol x}$}_1, \text{cof})
    \rrangle 
    \longrightarrow 
    \left(
    \mbox{\begin{diagram}[size=2em]
    \; 
    \resizebox{2.0ex}{!}{$\filledstar$}_1
    \end{diagram}}
    \right)
    \llangle
    (\text{${ \boldsymbol x}$}_1, \text{ADP})
    \rrangle 
    \end{equation}
    $\quad \quad \text{\boldmath $\mathbf{with}$} \ \
    \frac{k_{\text{off-edge, cofilin}}}{N_\text{CG}}$
    \end{varwidth}
    \end{center}
    
    \item Bundling, which creates another junction type with a CaMKII$\beta$ object connected to two actin objects that are 
    not already connected 
    to each other:


    \begin{center}
    \begin{varwidth}{\linewidth}
    \textbf{Actin Filament CaMKII$\beta$ Binding:}
    \begin{equation}
    \label{eq::cam_bundling_rule}
    \begin{varwidth}{\linewidth}
    \quad\space\space$\left(\begin{diagram}[size=1em] 
    $\Circle$_1 & \rTo & $\Circle$_2 & \rTo & $\Circle$_3\\
    \\ \\ \\
    $\Circle$_4 & \rTo & $\Circle$_5 & \rTo & $\Circle$_6\\
    \end{diagram} \right)$
    \\
    $\longrightarrow$
    $\left(
    \begin{diagram}[size=1em]
    $\Circle$_1 & \rTo & $\Circle$_2 & \rTo & $\Circle$_3\\
    & & \uparrow \\
    & & $\resizebox{2.5ex}{!}{\hexagon}$_7 \\
    & & \downarrow \\
   $\Circle$_4 & \rTo & $\Circle$_5 & \rTo & $\Circle$_6\\
    \end{diagram}
    \right)$
    \end{varwidth}
    \quad
    \begin{varwidth}{\linewidth}
    $\llangle$
    $\text{${ \boldsymbol x}$}_1, 
    \text{${ \boldsymbol x}$}_2, 
    \text{${ \boldsymbol x}$}_3,
    \text{${ \boldsymbol x}$}_4,
    \text{${ \boldsymbol x}$}_5,
    \text{${ \boldsymbol x}$}_6$ 
    $\rrangle$
    \\ \\
    $\llangle$
    $\text{${ \boldsymbol x}$}_1, 
    \text{${ \boldsymbol x}$}_2, 
    \text{${ \boldsymbol x}$}_3,
    \text{${ \boldsymbol x}$}_4,
    \text{${ \boldsymbol x}$}_5,
    \text{${ \boldsymbol x}$}_6,
    \text{${ \boldsymbol x}$}_7$
    $\rrangle$
    \end{varwidth}
    \end{equation}
    \quad \text{\boldmath $\mathbf{with}$} \ \
    $k_{\text{on, Bundling}}  N_{\text{Cam}} \mathcal{H}\left(\mathcal{D}_\mathcal{U} - ||x_2-x_5||  \right) \mathcal{H}\left(\theta_\text{Bundle} 
    -\cos^{-1}
    \left(\frac{(x_3-x_1) \cdot (x_6-x_4)}{||x_3-x_1||||x_6-x_4||}\right)  \right)$
    \end{varwidth}
    \end{center}
    
where $\mathcal{D}_\mathcal{U}$ and $\theta_\text{Bundle}$
are 
maximal 
numerical values of 
distance and angle for coarse-grained binding of CaMKII$\beta$, 
    and where we denote the Heaviside function as 
    
    \begin{equation*}
    \mathcal{H}(x) = 
    \left\{
    \begin{array}{cc}
       & 
        \begin{array}{cc}
          1 & x\geq 0 \\
          0 & x < 0
        \end{array}
    \end{array}
    \right.
    \end{equation*}

The reverse process is given by this rule:
    \begin{center}
    \begin{varwidth}{\linewidth}
    \textbf{Actin Filament CaMKII$\beta$ Unbinding:}
    \begin{equation} \label{eq::cam_unbundling_rule}
    \left(
    \begin{diagram}[size=1em]
    \halfcirc[0.7ex]_1 \\
    \uparrow \\
    \resizebox{2.5ex}{!}{\hexagon}_2 \\
    \downarrow \\
    \halfcirc[0.7ex]_3 \\
    \end{diagram}
    \right)
    \llangle
    \text{${ \boldsymbol x}$}_1, 
    \text{${ \boldsymbol x}$}_2, 
    \text{${ \boldsymbol x}$}_3
    \rrangle 
    \longrightarrow 
    \left(
    \mbox{\begin{diagram}[size=1em]
    \halfcirc[0.7ex]_1 \\
    \\
    \\
    \\
    \halfcirc[0.7ex]_3 \\
    \end{diagram}}
    \right)
    \llangle
    \text{${ \boldsymbol x}$}_1, 
    \text{${ \boldsymbol x}$}_3
    \rrangle 
    \quad \quad \text{\boldmath $\mathbf{with}$} \ \
    k_{\text{off, Bundling}}
    \end{equation}
    \end{varwidth}
    \end{center}

    \item Filament breaking, from the action of cofilin or aip1.

    \begin{center}
    \begin{varwidth}{\linewidth}
    \textbf{Actin Filament De-Branching:} 
    \begin{equation} \label{eq::debranching_rule}
    \begin{array}{l}
    \left(
    \begin{diagram}[size=1em]
    $\halfcirc[0.7ex]$_1 & \rTo & 
    \resizebox{2.0ex}{!}{$\triangle$}_2 & \rTo & $\halfcirc[0.7ex]$_3 & &\\
    & & & \rdTo & & &\\
    & & & & \resizebox{1.5ex}{!}{$\square$}_4 & &\\
    \end{diagram}
    \right)
    \llangle
    \text{${ \boldsymbol x}$}_1, 
    \left(\text{${ \boldsymbol x}$}_2,\eta_2\right),
    \text{${ \boldsymbol x}$}_3,
    \text{${ \boldsymbol x}$}_4
    \rrangle
    \\
    \longrightarrow
    \left(
    \begin{diagram}[size=1em]
    $\halfcirc[0.7ex]$_1 & \rTo & 
    $\Circle$_2 & \rTo & $\halfcirc[0.7ex]$_3 \\
    & & & &  \\
    & & & & \resizebox{1.5ex}{!}{$\filledsquare$}_4
    \end{diagram}
    \right)
    \llangle
    \text{${ \boldsymbol x}$}_1, 
    \text{${ \boldsymbol x}$}_2,
    \text{${ \boldsymbol x}$}_3,
    \text{${ \boldsymbol x}$}_4
    \rrangle
    \end{array}
    \end{equation}
    \\ \quad \quad \text{\boldmath $\mathbf{with}$} \ \
    $k_{\text{debranch}}$ $\exp\left(\delta\left(\eta_2,\text{Cof}\right)\left(\sigma_{\text{comp}}+k_{\text{comp}}\right)\right)$
    \end{varwidth}
    \end{center}
    
    where $k_{\text{debranch}}$ is the rate at which Arp2/3 naturally dissociates from a parent actin filament, $\sigma_{\text{comp}}$ is the rate of increase from cofilin co-occupying a space on the actin filament that Arp2/3 does, and $k_{\text{comp}}$ is the effect of strain and twist accelerating Arp2/3 dissociation due to cofilin binding at near-neighboring spots alonng the actin filament (\cite{Chan2013}). Together, $\sigma_{\text{comp}}$ and  $k_{\text{comp}}$ increase the rate of branching by roughly a factor of 50 (\cite{Chan2013}).

\begin{center}
\begin{varwidth}{\linewidth}
\textbf{Actin Filament Breaking:}
\begin{equation}
\label{eq::breaking_rule}
\begin{diagram}[size=2em] 
& (
$\Circle$_1 & \rTo & 
$\Circle$_2
) &
\\
\longrightarrow &
(
$\CIRCLE$_1 & & $\CIRCLE$_2
)
\end{diagram}
\quad
\begin{varwidth}{\linewidth}
$\llangle$
($\text{${ \boldsymbol x}$}_1, \theta_1, \eta_\text{1, Cofilin}$),
($\text{${ \boldsymbol x}$}_2, \theta_2, \eta_\text{2, Cofilin}$)
$\rrangle$
\\
$\llangle$
($\text{${ \boldsymbol x}$}_1, \theta_1, \eta_\text{1, Cofilin}$),
($\text{${ \boldsymbol x}$}_2, \theta_2, \eta_\text{2, Cofilin}$)
$\rrangle$
\end{varwidth}
\end{equation}
\quad \quad \text{\boldmath $\mathbf{with}$} \ \
$k_{\text{biomech}}$
$\max(\mathcal{H}(|\theta_1|-\theta_\text{Actin})(1-\delta(\eta_1,\text{cof}))$\par \quad \quad \quad \quad \quad
$\mathcal{H}(|\theta_2|-\theta_\text{Actin})(1-\delta(\eta_2,\text{cof}))$,\par \quad \quad \quad \quad \quad
$\mathcal{H}(|\theta_1|-\theta_\text{Boundary})(1-\delta(\eta_1,\text{cof}))$\par \quad \quad \quad \quad \quad
$\mathcal{H}(|\theta_2|-\theta_\text{Boundary})\delta(\eta_2,\text{cof})$,\par \quad \quad \quad \quad \quad
$\mathcal{H}(|\theta_1|-\theta_\text{Boundary})\delta(\eta_1,\text{cof})$\par \quad \quad \quad \quad \quad
$\mathcal{H}(|\theta_2|-\theta_\text{Boundary})(1-\delta(\eta_2,\text{cof})$,\par \quad \quad \quad \quad \quad
$\mathcal{H}(|\theta_1|-\theta_\text{Cofilactin})\delta(\eta_1,\text{cof})$\par \quad \quad \quad \quad \quad
$\mathcal{H}(|\theta_2|-\theta_\text{Cofilactin})\delta(\eta_2,\text{cof}))$\par \quad \quad \quad \quad \quad
\end{varwidth}
\end{center}

This maximum (equivalent to a logical disjunction i.e. ``or'' operation)
of Heaviside functions conditions actin filament breaking on three critical breaking angles dependent on the state of molecules bound to the coarse-grained object's actin monomers. The actin breaking angle is used for 
zero
cofilin states within bound molecules, the cofilactin breaking angle is used for internal actins which have
cofilin bound molecules, and the boundary breaking angle is used elsewhere where adjacent actin objects have different occupancies of cofilin molecules, hence indicating a cofilactin-actin boundary.

\begin{center}
\begin{varwidth}{\linewidth}
\textbf{Aip1 Binding:}
\begin{equation}
\label{eq::severing_binding_rule}
\begin{diagram}[size=2em] 
& (
$\Circle$_1
) &
\\
\longrightarrow &
(
$\Circle$_1
)
\end{diagram}
\quad
\begin{varwidth}{\linewidth}
$\llangle$
($\text{${ \boldsymbol x}$}_1, \theta_1, \eta_\text{1, cof}$),
$\rrangle$
\\
$\llangle$
($\text{${ \boldsymbol x}$}_1, \theta_1, \eta_\text{1, Aip1}$)
$\rrangle$
\end{varwidth}
\end{equation}
\quad \quad \text{\boldmath $\mathbf{with}$} \ \
$k_{\text{on, Aip1}}$
\end{varwidth}
\end{center}

\begin{center}
\begin{varwidth}{\linewidth}
\textbf{Actin Filament Severing:}
\begin{equation}
\label{eq::severing_rule}
\begin{diagram}[size=2em] 
& (
$\Circle$_1 & \rTo & 
$\Circle$_2
) &
\\
\longrightarrow &
(
\resizebox{1.75ex}{!}{$\blacktriangleright$}_1 & & $\CIRCLE$_2
)
\end{diagram}
\quad
\begin{varwidth}{\linewidth}
$\llangle$
($\text{${ \boldsymbol x}$}_1, \theta_1, \eta_\text{1, Aip1}$),
($\text{${ \boldsymbol x}$}_2, \theta_2, \eta_\text{2}$)
$\rrangle$
\\
$\llangle$
($\text{${ \boldsymbol x}$}_1, \theta_1, \eta_\text{1, Aip1}$),
($\text{${ \boldsymbol x}$}_2, \theta_2, \eta_\text{2}$)
$\rrangle$
\end{varwidth}
\end{equation}
\quad \quad \text{\boldmath $\mathbf{with}$} \ \
$k_{\text{sever}}$
\end{varwidth}
\end{center}

\end{enumerate}

Together, these rules lead to a treadmilling effect with turnover of actin monomers when the actin network is sufficiently large. The network preferentially moves outward in the direction of the barbed ends
since
the barbed end elongation rates are faster than the pointed end elongation rates. 
while the nodes toward the pointed ends are eventually depolymerized. In this way it was originally proposed by theorists such as Mogilner and Oster that actin can provide a force to cellular compartments by interacting with their membranes (\cite{Mogilner2003}). Considering 
Newtonian
reactive forces of membrane on fiber and fiber on fiber, actin can also generate other forces within the compartment, leading to internal reorganization and deformation. The mechanism with which the actin network achieves moving the membrane 
along with its own fibers
is the subject of the next section.

\subsection{Biomolecular mechanisms of actin-binding proteins}

Based on the foregoing rules, we summarize in this section the workings of each ABP in the context of actin filament network remodeling and biophysics.

\subsubsection{Actin}

The polymerization and depolymerization of actin are the 
driving 
mechanisms for actin dynamics in this model.
In addition, a filament sub-graph that includes long rods
causes an electrostatic disconnection between two
adjacent 
actin objects as the objects exert only small radial forces on each other.
The adjacency of the actin objects has its connection broken to create two new filaments 
{(Eq. (\ref{eq::breaking_rule}))}.

\subsubsection{Arp2/3}

A junction object 
nucleates 
the Arp2/3 branch as a new filament with a pointed end. Arp2/3 can unbind and de-branch spontaneously, and encouraged by nearby bound cofilin (\cite{Chan2013}).

\subsubsection{Cofilin}

The cofilin mechanism includes a slow binding to an unoccupied cofilin site on an actin protein within a coarse-grained actin object, which is independent of the state of the site's neighbors on an actin filament. There also is a faster binding, accelerated by the state of the site's neighbors (\cite{Cao2006, DeLaCruz2009}). Each bound site is mechanically weakened by a factor of five, as cofilin is known to weaken the bending stiffness of actin filaments (\cite{McCullough2008}). Nearby cofilin-bound sites accelerate the actin-bound 
ATP/ADP
dynamics of neighboring sites.

Cofilin, when present in high numbers, destabilizes the filament by lowering bending stiffness, and causes breaking
(Eq. (\ref{eq::breaking_rule})).
Cofilin binding to an empty site on an actin filament is slow; however, each bound cofilin encourages 
(Eq. (\ref{eq::cofilin_2Node_binding_rule}))
binding of cofilin to adjacent locations along the filament. Cofilin also facilitates the release of ADP+Pi to ADP (\cite{Suarez2011}) when it is present near to the actin monomer and facilitates breaking along interfaces (\cite{DeLaCruz2009}) between cofilin-bound segments and non-bound segments. These mechanisms are implemented in our simulations with bound ATP/ADP and cofilin stored as values in actin object nodes
(Eqs. (\ref{eq::cofilin_2Node_binding_rule}-\ref{eq::cofilin_unbinding_rule})).

Cofilin competes with Arp2/3 for spots on an actin filament and facilitates the dissociation of a daughter branch from a parent filament (\cite{Chan2013}). This is due to steric hindrance of the Arp2/3 molecule and also by altering the biophysics (e.g. twist (\cite{Bibeau2023})) of the actin filament after cofilin binds.

Furthermore, cofilin creates a boundary angle regime for filament breaking, which subsequently occurs at a much smaller angle than cofilin-unbound segments (\cite{McCullough2011}). This is hypothesized to occur, in part, due to cofilin segments having a 5-fold weaker bending modulus than the unbound segments. Correspondingly, the bending modulus is weakened in simulation, by that factor, conditioned on bound cofilin.

\subsubsection{Aip1}

Aip1 has been known to accelerate severing of filaments (\cite{Chen2015}). We implement the mechanism of Aip1 as a cofilactin (cofilin-bound actin) binding molecule with a severing mechanism in addition to that of cofilin (\cite{Oosterheert2025}), with a rate constant from \cite{Chen2015}. After severing occurs, a barbed end capped by Aip1 is created (\cite{Oosterheert2025}), that can dissociate from the cofilactin barbed end to allow barbed end elongation, as well as a new pointed end.

\subsubsection{CaMKII$\beta$}

Filaments that are within a specified distance and angle of each other have a non-zero rate of binding by CaMKII$\beta$ which links them with the 
the actin biophysics described in
section \ref{section:actionbiomechanics} below. 
This includes 
the anisotropic buckling force and 
the bending force, 
together with 
their respective thermal Hessians. 
A biologically realistic equilibrium length between the filaments is kept as the uniform, optimal length until the bundling molecule unbinds.

\subsection{Spine head morphodynamics}

Dynamical Graph Grammars (DGGs) provide a well-defined way to accommodate dynamically changing system structure such as active cytoskeleton represented using dynamic graphs within nonequilibrium statistical physics as defined by the master equation. 
Such structure changes 
operate naturally at a coarser spatial scale than the force exertion of fixed topology objects on one another in 
biophysical kinetics. To obtain multiscale models then it is necessary to unify the two perspectives - grammar-defined discrete state changes and finer-scale biophysics - with specialized DGG kinetics rules that obey biophysical constraints such as Galilean invariance, conservation of momentum, and dissipation of conserved global energy.

In our model the spine head area grows, starting from an area that encloses a starting pair of actin objects. The mechanism of actin-membrane mutual growth is the attachment of actin objects to the membrane mesh as soon as they intersect, allowing propulsive force onto the membrane, 
Newtonian 
reactive force onto the actin network, and pressure onto the actin network. 
The entire model takes place in two dimensions
(viewing the dendritic spine head from the top),
due to our current computational limitations,
although future work will aim at three dimensions. 
The requisite dynamics conditions for the implementation details are provided below.

First, actin rods intersecting through a pseudo-extended actin overgrowth length (\cite{Medwedeff2023}) with a spine head membrane rod have the end attached to the spine head membrane rod through the creation of a new spine head membrane vertex at that Cartesian coordinate. Second, the membrane is allowed to fluctuate from the curvature energy term. The Helfrich energy mean curvature energy is given by:
\[
E_H = \frac{\kappa}{2} \int_S (2H)^2 \, dA
\] \\
To discretize this term for a 2D membrane polygon, we approximate the mean curvature \( H \) at each vertex \( \mathbf{r}_i \). Let \( \mathbf{r}_{i-2} \), \( \mathbf{r}_{i-1} \), \( \mathbf{r}_{i+1} \), \( \mathbf{r}_{i+2} \) be the neighboring vertices of \( \mathbf{r}_i \). 
Using the adjacent neighbors, we take 
the derivative of the unit tangent vector with respect to the arc length $\bar{w}$, which yields 
\begin{equation}
H_{\Gamma(x_a,x_b,x_c)} = \left|\left|\frac{d \mathbf{T}}{d\bar{w}}\right|\right|,
\end{equation}
where
we approximate
\begin{equation}
    \frac{dT^i}{d\bar{w}^i} 
    {
    \simeq}
    \frac{1}{z^i}\left(\frac{x_1^{i+1}-x_1^i}{v^{i+1}}-\frac{x_1^{i}-x_1^{i-1}}{v^{i}},\frac{x_2^{i+1}-x_2^i}{v^{i+1}}-\frac{x_2^{i}-x_2^{i-1}}{v^{i}}\right),
\end{equation}
and  \(v^i=||x^i-x^{i-1}||\) and \(z^i=\frac{v^i+v^{i+1}}{2}\). 
This scheme was presented in \cite{Quintana2020} (Supplementary Material), and when followed, leads to 
the update as follows: 
where $g^i=\left|\left|\frac{x^{i+1}-x^i}{v^i}-\frac{x^i-x^{i-1}}{v^{i-1}}\right|\right|^2$, 
we have that \(H_{\Gamma(x_a,x_b,x_c,x_d,x_e)} = \frac{\sqrt{g^{i-1}}}{z^{i-1}}+\frac{\sqrt{g^i}}{z^i}+\frac{\sqrt{g^{i+1}}}{z^{i+1}}\).

The actin end is allowed to interpolate from the spine head membrane 
after polymerizing and extending beyond the overgrowth threshold length 
for the overgrowth dynamics, which follows the Brownian ratchet polymerization mechanism (\cite{Peskin1993}). An interpolating membrane mesh is maintained that imparts actin-end forces onto the vertices through  
attachment objects connecting actin ends to 
particular membrane rods{as in Figure \ref{fig:membraneviz}.



\subsection{Actin network biophysics}
\label{section:actionbiomechanics}

The 
gradient descent rules below 
together comprise a stochastic first order Euler update for solving the force balance 
system of differential equations. 
It is analogous to stochastic gradient descent. These rules have a parameter $k_{\text{kinetic}}$
which in the infinite limit recovers the ODE exactly. 
Alternatively we could have coded ODEs in the DGG,
but we think that would be less efficient in the
presence of substantial thermal noise.

There are two sub-sectors to our implemented biophysics 
sector: 
(1) pairwise forces due to energy terms coupling two adjacent segments or proteins of a polymer
(and which subserve a fiber buckling process),
and (2) three-node forces that act specifically on fiber bending angles. For each of these kinds of forces,
there are several rules that update position according to the force.
There is also a corresponding energy term whose negative spatial derivative is the force,
and another kind of rule that implements
Hessian Boltzmann sampling which represents thermal noise fluctuations.

To ensure that even dissipative biophysics is compatible with conservation of energy, the 
{DGG kinetic}
rules must all be derivable from a single global energy function.
This function will be a sum of two-object and three-object potential energy functions.
The three-object potential energies will be
a function of the bending angle between three
consecutive graph-connected filament segments,
hence invariant to the Euclidean group
of global translations and rotations.
The two-object potential energies will be
a function of the distance between the
centroids of two graph-connected objects, 
hence also Euclidean-invariant. 

The longitudinal pairwise energy
$U_{\text{sep}}(x_1,x_2) = U_{\text{sep}}(r)$,
where $r$ is the separation distance $r = ||x_1 - x_2||_2$ between
two connected filaments segments
(which could be a small as individual actin proteins),
is standardly assumed to have a minimum-energy
distance $r_{\text{min}}$, 
to increase for larger distances
and asymptote from below to a value of zero energy,
and to increase rapidly to infinity as the 
segment-segment pairwise distance drops
to zero - the repulsive core of the central force law.
For example,
this intermolecular potiential energy
can be taken to be the
LJ potential:

\begin{equation}
    \hat{U}_{\text{LJ}}(x_1,x_2)=\hat{U}_{\text{LJ}}(||x_1-x_2||_2) = \epsilon_{\text{LJ}}\left(\left(\frac{\sigma_\text{LJ}}{||x_1-x_2||}\right)^{12}-\left(\frac{\sigma_\text{LJ}}{||x_1-x_2||}\right)^6\right).
    \label{LJ_potential}
\end{equation}
This potential takes its minimum value at $r = r_{\text{min}}$.
We will use this popular default choice,
though there is no particular reason that actin proteins
should have this potential energy law rather than 
one of many others that have similar properties.


This kind of potential smoothly ``breaks'' the connection
between two segments for large $r$, since an energy constant
with respect to position implies zero force.
Thus two sufficiently distant segments could 
trigger a filament-breaking rule
such as Eq. (\ref{eq::breaking_rule}). 

When 
such a model is applied to
three 
successive
points in a
fiber,
the result is a tri-nodal potential
in which
a buckling effect arises 
whereby under compression
the middle 
filament node
pushes away from the co-linear state. 
The relevant energy combination is:
\begin{equation}
    U_{\text{tri-nodal,sep}}(x_1,x_2,x_3)=\tilde{U}_\text{sep}(x_1,x_2)+\tilde{U}_\text{sep}(x_2,x_3)
\label{trinodalsep}
\end{equation}

For an unbranched interior actin node in a filament at position $x_2$, the gradient of this energy equals the gradient of the globally summed $U_{\text{sep}}$ over all connected pairs. There is a numerical reason to use the trinodal potential rather than the pairwise potential in the main 
DGG rule for buckling. For a nearly straight fiber, the gradients of the two pairwise potentials in which a given
interior (non-boundary) node participates nearly cancel out.
This cancellation is more efficiently achieved
by an exact analytic
vector sum in the gradient calculation of the trinodal potential,
than by the stochastic addition of displacements
resulting from the stochastic firing of a pairwise
DGG rule on randomly chosen nodes which will 
eventually converge to the same stochastic cancellation
on average.

Also for numerical reasons,
we introduce a 
force-clipping two-body potential energy function 
that is used to ensure that the movements of the actin objects smoothly update towards a system minimum. The force-clipping is important because we are using
a potential energy function $\hat{U}(r)$ for interparticle distance $r=||x_1-x_2||$,
such as the
LJ potential $\hat{U}_{LJ}(r)$,
that includes a strong repulsive core.
The singularity of high potential energy at zero distance
reflects the
high repulsive force if the actin objects move close enough to each other such that their electron orbitals repel strongly,
but it also introduces numerical ill-conditioning.

To control this problem, we linearize the potential
at distances $r \le \epsilon$
and match potential value and slope at $r=\epsilon$,
via Taylor's theorem at that point:
\begin{equation}
\hat{U}(x_1,x_2) 
= \hat{U}(r)
\approx \hat{U}_{\epsilon}(r)
\equiv \hat{U}(\epsilon)+\hat{U}'(\epsilon) \times (||x_1-x_2||_2-\epsilon),
\end{equation} 

The resulting potential function
$\tilde{U}_{\epsilon}(r)$ 
is defined piecewise as the original unclipped potential 
for inter-molecular distance above a threshold $\epsilon$; below that distance 
it becomes a linear function without discontinuity that meets the y-intercept:
\begin{equation}
\tilde{U}_{\epsilon}(r) = 
\left\{
\begin{array}{ll}
    \hat{U}(r) & r \ge \epsilon \\
    \hat{U}_{\epsilon}(\epsilon)
        =\hat{U}(\epsilon)+\frac{\partial \hat{U}(r)}{\partial r}\Big|_{r=\epsilon} (r-\epsilon)
            & 0 \le r \le \epsilon
\end{array}
\right. \\ \,
\label{clipped-potential}
\end{equation}
and therefore we have the equation below for the
clipped gradient proportional to the full, clipped force:
\begin{equation}
\frac{\partial \tilde{U}}{\partial r}(r|\epsilon) = 
\left\{
\begin{array}{ll}
    \frac{\partial \hat{U}}{\partial r}(r) & r \ge \epsilon \\
    \frac{\partial \hat{U}(r)}{\partial r}\Big|_{r=\epsilon} & 0 \le r \le \epsilon
\end{array}
\right. \\ \,
\end{equation}
where $r = ||x_1-x_2||_2$.
The force is a well-defined and continuous derivative of the potential 
which acts to 
preserve Newton's laws of motion.

If $G$ is the graph of connected filament segments,
with number of nodes $|G|$ and symmetric adjacency matrix
$[G_{i j}\in{0,1}]$, then the associated global energy is
\begin{equation}
    U_{\text{radial}} = \sum_{i=1}^{|G|} \sum_{j < i}
    G_{i j} U_{\text{sep}}(x_i,x_{j}).
\label{U_sum}
\end{equation}
If we were to ignore branch points in the actin filament network,
the sum of radial separation energies
would decompose over filaments indexed by $F$
and having size $|F|$:
\begin{equation}
    U_{\text{radial}} = \sum_F\sum_{j=1}^{|F|-1}
    U_{\text{sep}}(x_j,x_{j+1})
\label{Ufil_sum}
\end{equation}
Returning to the general graph case, we calculate the gradient:
\begin{equation}  
\begin{split}
    \nabla_{x_l}U_{\text{radial}}
    &=  
    \frac{1}{2} \sum_{i=1}^{|G|} \sum_{j \neq i}
        G_{i j} \nabla_{x_l}  U_{\text{sep}}(x_i,x_{j})
    \\ &=
    \frac{1}{2} \sum_{i=1}^{|G|} \sum_{j \neq i}
        G_{i j} U_{\text{sep}}^{\prime}(x_i,x_{j})
            \nabla_{x_l} ||x_i-x_j||_2 
    \\ &=
    \frac{1}{2} \sum_{i=1}^{|G|} \sum_{j \neq i}
        G_{i j} U_{\text{sep}}^{\prime}(x_i,x_{j})
            ||x_i-x_j||_2^{-1} 
                (x_i-x_j) \cdot \nabla_{x_l} (x_i-x_j)
    \\ &=
     \frac{1}{2} \sum_{i=1}^{|G|} \sum_{j \neq i}
        G_{i j} U_{\text{sep}}^{\prime}(x_i,x_{j})
            ||x_i-x_j||_2^{-1} 
                (x_i-x_j) \cdot I (\delta_{l i} - \delta_{l j})
    \\ &=
     \frac{1}{2} \sum_{j \neq l}
        G_{l j} U_{\text{sep}}^{\prime}(x_l,x_{j})
            ||x_l-x_j||_2^{-1} 
                (x_l-x_j)
        -
      \frac{1}{2} \sum_{i \neq l}
        G_{i l} U_{\text{sep}}^{\prime}(x_i,x_{l})
            ||x_i-x_l||_2^{-1} 
                (x_i-x_l) 
\end{split},
\end{equation}
where \(\delta_{li}\) = 1 if \(l=i\) and 0 otherwise is the Kronecker delta function, so by symmetry, antisymmetry, and then change of index name, followed by collecting terms:
\begin{equation}  
\begin{split}
    \nabla_{x_l}U_{\text{radial}}
    &=
     \frac{1}{2} \sum_{j \neq l}
        G_{l j} U_{\text{sep}}^{\prime}(x_l,x_{j})
            ||x_l-x_j||_2^{-1} 
                (x_l-x_j)
    -
      \frac{1}{2} \sum_{i \neq l}
        G_{l i} U_{\text{sep}}^{\prime}(x_l,x_{i})
            ||x_l-x_i||_2^{-1} 
                (x_i-x_l) 
    \\ &=
     \frac{1}{2} \sum_{j \neq l}
        G_{l j} U_{\text{sep}}^{\prime}(x_l,x_{j})
            ||x_l-x_j||_2^{-1} 
                (x_l-x_j)
    +
      \frac{1}{2} \sum_{i \neq l}
        G_{l i} U_{\text{sep}}^{\prime}(x_l,x_i)
            ||x_l-x_i||_2^{-1} 
                (x_l-x_i) 
    \\ &=
     \frac{1}{2} \sum_{j \neq l}
        G_{l j} U_{\text{sep}}^{\prime}(x_l,x_{j})
            ||x_l-x_j||_2^{-1} 
                (x_l-x_j)
    +
      \frac{1}{2} \sum_{j \neq l}
        G_{l j} U_{\text{sep}}^{\prime}(x_l,x_j)
            ||x_l-x_j||_2^{-1} 
                (x_l-x_j) 
    \\ &=
       \Big[ \sum_{j \neq l}
        G_{l j} U_{\text{sep}}^{\prime}(x_l,x_{j})
            ||x_l-x_j||_2^{-1} \Big]
                x_l
    -
      \Big[ \sum_{j \neq l}
        G_{l j} U_{\text{sep}}^{\prime}(x_l,x_j)
            ||x_l-x_j||_2^{-1} 
                x_j \Big].
\end{split}
\end{equation}

Let
\begin{equation}
\psi(r) = U_{\text{sep}}^{\prime}(r)/r .
\end{equation}
Then finally
\begin{equation}  
\begin{split}
    \nabla_{x_l}U_{\text{radial}}
    =
       \Big[ \sum_{j \neq l}
        G_{l j} \psi(||x_l-x_j||) \Big]
                x_l
     - \Big[ \sum_{j \neq l}
        G_{l j} \psi(||x_l-x_j||) 
                x_j \Big].
\end{split}
\label{Gtransport}
\end{equation}
This is the equation we follow in the local DGG update rules below.

One consequence of Eq.~(\ref{Gtransport}) is
\begin{equation}
    \frac{d {\bf p}}{dt} = \sum_l {\bf F}_l = - \sum_l  {\bf \nabla}_{x_l}U_{\text{radial}} 
    = (1-1) \Big[ \sum_l \sum_{j \neq l}
        G_{l j} \psi(x_l-x_j) x_l \Big] =0,  
\end{equation}
conservation of the total 
system
momentum vector.

In particular for an unbranched filament $F$
of nonzero length $|F|$, 
Eq.~(\ref{Gtransport})
specializes to 

\begin{equation}
\begin{split}
    \nabla_{x_{F \, l}}U_{\text{radial}}               
    &= 
    (1- \delta_{l 1})(1- \delta_{l |F|})
        \Big( \big(
        \psi({||x_l-x_{l+1}||_2})
        + \psi({||x_{l-1}-x_{l}||_2})
        \big) x_l
    \\ & \quad \quad \quad \quad
            - \psi( {||x_l-x_{l+1}||_2} ) x_{l+1}
            - \psi( {||x_{l-1}-x_{l}||_2} ) x_{l-1}
        \Big)
    \\ & \quad +
        \delta_{l 1} \psi({||x_1-x_{2}||_2} )
        (x_1-x_{2}) 
         + \delta_{l |F|} \psi({||x_{|F|}-x_{|F|-1}||_2} )
                (x_{|F|}-x_{|F|-1})  . \\ 
\end{split}
\label{gradU_sum}
\end{equation}

The summands in the sum in Eq.~(\ref{gradU_sum})
correspond to
separate 
grammar rule
diagrams centered on updating
node $l$ which is an interior node
connected to two other nodes of any type
(first term, with $(1- \delta_{l 1})(1- \delta_{l |F|})$ factor), 
vs. an end node at
the first- or last-indexed end
(second and third terms respectively, 
with factors of $ \delta_{l 1}$ and $\delta_{l |F|}$ respectively)  
connected to one other node of any type.

Thus for an unbranched interior filament node:
\begin{equation}
\begin{split}
    \nabla_{x_{F \, l}}U_{\text{radial}}               
    &=\big(
        \psi({||x_l-x_{l+1}||_2})
        + \psi({||x_{l-1}-x_{l}||_2})
        \big) x_l
    \\ & \quad \quad \quad \quad
            - \psi( {||x_l-x_{l+1}||_2}) x_{l+1} 
            - \psi( {||x_{l-1}-x_{l}||_2} ) x_{l-1} .
\end{split}
\label{gradU_unbranched_interior}
\end{equation}

The desired cancellation of longitudinal gradients
along a nearly straight fiber will occur
if $x_l \approx$ the weighted average of
$x_{l+1}$ and $x_{l-1}$,
i.e. if the segment centroids are roughly
colinear and equally spaced.

Under tension,
the quantity $\psi$ tends to be positive so
any deviation from colinearity will generate
a restoring force 
$\propto - \nabla U_\text{radial}$
bringing the middle node
$x_l$ back into line transversally, 
as well as towards equal longitudinal spacing.
Under compression, $\psi$ tends to be negative so
transverse buckling will result.

To model random thermal displacements,
it will also be necessary to calculate the Hessian matrix
$\nabla^2_{F l m}$
of second derivatives.
If $G_{l\, m}=0$ then
the Hessian matrix element is zero,
so the Hessian is a weighted version
of the connection graph but with self-edges added. From 
Eq.~(\ref{Gtransport})
we calculate:

\begin{equation}
\begin{split}
    \nabla^2_{x_l x_m}U_{\text{radial}}               
    &= \delta_{l m} \sum_{j \neq l} G_{l j}
        \psi({||x_l-x_j||_2})
            - (1-\delta_{l m}) G_{l m}
            \psi( {||x_l-x_{m}||_2})
\\ & + \quad 
     \sum_{j \neq l} 
        G_{l j} \psi^{\prime}(||x_l-x_j||_2)
            ||x_l-x_j||_2^{-1} 
                (x_l-x_j) \cdot I (\delta_{m l} - \delta_{m j}) \cdot (x_l-x_j)
    \\  &= \delta_{l m} \sum_{j \neq l} G_{l j}
        \psi({||x_l-x_j||_2})
            - (1-\delta_{l m}) G_{l m}
            \psi( {||x_l-x_{m}||_2})
    \\ & + \quad 
     \sum_{j \neq l} 
        G_{l j} \psi^{\prime}(||x_l-x_j||_2) ||x_l-x_j||_2
         (\delta_{m l} - \delta_{m j}) 
    \\  &= \delta_{l m} \sum_{j \neq l} G_{l j}
        \Big( \psi({||x_l-x_j||_2}) + \psi^\prime({||x_l-x_j||_2}) ({||x_l-x_j||_2}) \Big)
     \\ & \quad \quad \quad \quad
            - (1-\delta_{l m}) G_{l m}
        \Big( \psi( {||x_l-x_{m}||_2}) +  \psi^\prime( {||x_l-x_{m}||_2}) ( {||x_l-x_{m}||_2}) \Big)
\end{split}
\label{HessU_1}
\end{equation}
and finally
\begin{equation}
\begin{split}
    \nabla^2_{x_l x_m} U_{\text{radial}}               
      &= \delta_{l m} \sum_{j \neq l} G_{l j}
         U_\text{sep}^{\prime \prime}({||x_l-x_j||_2})
            - (1-\delta_{l m}) G_{l m}
         U_\text{sep}^{\prime \prime}({||x_l-x_m||_2})
         .
\end{split}
\label{HessU}
\end{equation}
This is a weighted graph Laplacian.
If all $U^{\prime \prime}$ are positive then this matrix
is diagonally dominant with real eigenvalues that are all,
by Gershgorin's theorem, nonnegative.

For a model of small random displacements consistent with a Boltzmann distribution with second-order (only, leaving first order effects to the gradient rules) approximation of energies, we update $x_l$ with the Gaussian.
\begin{equation}
\text{Pr}(x_l^\prime | x_l) 
\propto \exp \big[ - (x^\prime_l - x_l)^T\big( \big|\sum_{j \neq l} G_{l j}
         U_\text{sep}^{\prime \prime}({||x_l-x_j||_2})\big| I
            +\epsilon^2_\text{sep} I \big)(x^\prime_l - x_l) \big]
\label{buckling_Gaussian}
\end{equation}

For first-order gradient dynamics we assume that, after a very brief transient ballistic behavior,
our molecular-scale objects
all come to force balance $F + F_\text{drag}=0$ with a drag force $F_\text{drag}= - \zeta v$, where $v$ is object velocity and $\zeta$ is the friction constant estimated in \cite{Quintana-Rangamani2024} and \cite{Quintana2020} by fitting the synaptic spine head model to experimental measurements in \cite{Bosch2014}.

Thus
%
%
\begin{equation}
    \frac{d x_i}{d t} = \frac{F_i}{\zeta} = - \frac{1}{\zeta} \nabla_i x_i \; .
\label{force-balance}
\end{equation}
Eq.~(\ref{force-balance}) can be a stiff ODE system for inter-molecular potentials, particularly in the highly repulsive regime of the LJ potential. 
Thus, we introduced the clipped potential to formulate force balance (Eq.~(\ref{force-balance}))
while relieving the numerical instability. 


This rule in DGG implementation form is shown below.

\begin{center}
    \begin{varwidth}{\linewidth}
    \textbf{Interior Case Anisotropic Buckling:}
    \begin{equation}
    \label{eq::LJ_buckling_rule}
    \begin{diagram}[size=2em] 
    & (
    \resizebox{2.0ex}{!}{$\filledstar$}_1 & \rTo & 
    \{ $\Circle$, \resizebox{1.5ex}{!}{$\square$} \}_2 
    & \rTo &\resizebox{2.0ex}{!}{$\filledstar$}_3
    ) &
    \\
    \longrightarrow &
    (
    \resizebox{2.0ex}{!}{$\filledstar$}_1 & \rTo & 
    \boxed{\{ $\Circle$, \resizebox{1.5ex}{!}{$\square$} \}_2} 
    & \rTo &\resizebox{2.0ex}{!}{$\filledstar$}_3
    )
    \end{diagram}
    \quad
    \begin{varwidth}{\linewidth}
    $\llangle$  
    $\text{${ \boldsymbol x}$}_1, 
    \text{${ \boldsymbol x}$}_2, 
    \text{${ \boldsymbol x}$}_3$
    $\rrangle$
    \\
    $\llangle$
    $\text{${ \boldsymbol x}$}_1, 
    \text{${ \boldsymbol x}$}_2, 
    \text{${ \boldsymbol x}$}_3$
    $\rrangle$
    \end{varwidth}
    \end{equation}
    \quad \quad \text{\boldmath $\mathbf{with}$} \ \
    $k_{\text{kinetic}}$
\par \quad \quad  \ \ 
$\begin{cases} 
\text{$\Delta_{\text{sep}}(x_1,x_2,x_3)$} :=
 -\frac{1}{\zeta_\text{actin} k_{\text{kinetic}}}\left(\frac{\partial U_\text{trinodal, sep}(x_1,x_2,x_3)}{\partial x_2}\right)
 \\
 = -\frac{1}{\zeta_\text{actin} k_{\text{kinetic}}}\left(\frac{\partial U_\text{radial}([x])}{\partial x_2}\right)
 \\
 =
  -\frac{1}{\zeta_\text{actin} k_{\text{kinetic}}}
 \Big(
        \psi({||x_2-x_{1}||_2})
        + \psi({||x_{2}-x_{3}||_2})
        \big) x_2
        \\ \quad \quad
            - \psi( {||x_2-x_{1}||_2}  ) x_{1}
            - \psi( {||x_{2}-x_{3}||_2}) x_{3}
            \Big)
\end{cases}$
    \end{varwidth}
    \end{center}

{This rule effectively implements a stochastic version
of a forward Euler solver for Eq.~(\ref{force-balance}),
approaching the ODE solution as $k_{\text{kinetic}} \rightarrow \infty$.}
The rules of Eqs.
(\ref{eq::LJ_buckling_rule}), 
(\ref{eq::angle_bending_rule}) below, 
and (\ref{eq::Hessian_Noise_rule}) below
have in common that 
aside from the parameter values
their LHS and RHS graphs are the same,
and that only the parameters
of the central node change,
in a way that is computed systematically
from the global biophysical
energy function. For such rules
we introduce the shorthand diagrammatic
notation:
\begin{equation}
\begin{split}
\mbox{\begin{diagram}[size=2em]
\; 
\Circle_1 & \rLine & \boxed{\Circle_2} & \rLine &\Circle_3
\end{diagram}}
\end{split}
\label{box_notation}
\end{equation}
where the boxing of node 2 is not
part of the graph but serves to
indicate which node has changing
parameter values.

Note that the empty/filled distinction is just a visualization
for a binary parameter in the actual rule objects
that specifies whether the given fiber segment
is an end segment, i.e. has an empty binding site for continued
polymerization, or not. 
$k_{\text{kinetic}}$ is a relative rate of updating biophysical (kinetic) rules;
as it tends to infinity, there are more and more updates of smaller and smaller step sizes each
per unit time, and the limit is an ordinary differential equation system.

Likewise, there is a special buckling rule for a filament branch point.
It is mutually exclusive in domain of applicability with
the interior case above, due to the constraints on actin segment objects in position 2.
Defining similarly to Eq. (\ref{trinodalsep})
\begin{equation}
    U_{\text{quad-nodal,sep}}(x_1,x_2,x_3,x_4)=\tilde{U}_\text{sep}(x_1,x_2)+\tilde{U}_\text{sep}(x_2,x_3) + +\tilde{U}_\text{sep}(x_2,x_4)\, ,
\label{quadnodalsep}
\end{equation}
the rule is:

\begin{center}
    \begin{varwidth}{\linewidth}
    \textbf{Branch Case Anisotropic Buckling:}
    \begin{equation}
    \label{eq::branch_biomech}
    \begin{varwidth}{\linewidth}
    \quad\space\space$\left(\begin{diagram}[size=1em] 
    $\halfcirc[0.7ex]$_1 & \rTo & 
    \resizebox{2.0ex}{!}{$\triangle$}_2 & \rTo & $\halfcirc[0.7ex]$_3 \\
    & & & \rdTo &  \\
    & & & & \resizebox{1.5ex}{!}{$\boxbox$}_4
    \end{diagram} \right)$
    \\
    $\longrightarrow$
    $\left(
    \begin{diagram}[size=1em]
     $\halfcirc[0.7ex]$_1 & \rTo & 
    \resizebox{2.0ex}{!}{$\triangle$}_2 & \rTo & $\halfcirc[0.7ex]$_3 \\
    & & & \rdTo &  \\
    & & & & \resizebox{1.5ex}{!}{$\boxbox$}_4
    \end{diagram}
    \right)$
    \end{varwidth}
    \quad
    \begin{varwidth}{\linewidth}
    $\llangle$
    $\text{${ \boldsymbol x}$}_1, 
    \text{${ \boldsymbol x}$}_2, 
    \text{${ \boldsymbol x}$}_3, 
    \text{${ \boldsymbol x}$}_4$
    $\rrangle$
    \\
    $\llangle$
    $\text{${ \boldsymbol x}$}_1, 
    \text{${ \boldsymbol x}$}_2 + \text{$\Delta_{\text{sep}}(x_1,x_2,x_3,x_4)$}, 
    \text{${ \boldsymbol x}$}_3, 
    \text{${ \boldsymbol x}$}_4$
    $\rrangle$
    \end{varwidth}
    \end{equation}
    \quad \text{\boldmath $\mathbf{with}$} \ \
$k_{\text{kinetic}}$
\par \quad \quad \ \ 
$\begin{cases} 
\text{$\Delta_{\text{sep}}(x_1,x_2,x_3,x_4)$} 
 := -\frac{1}{\zeta_\text{actin} k_{\text{kinetic}}}\left(\frac{\partial U_\text{quad-nodal, sep}(x_1,x_2,x_3,x_4)}{\partial x_2}\right)
 \\
 = -\frac{1}{\zeta_\text{actin} k_{\text{kinetic}}}\left(\frac{\partial U_\text{radial}([x])}{\partial x_2}\right)
 \\
 =
  -\frac{1}{\zeta_\text{actin} k_{\text{kinetic}}}
 \Big(
        \psi({||x_2-x_{1}||_2}) + \psi({||x_2-x_{3}||_2})
        + \psi({||x_{2}-x_{4}||_2})
        \big) x_2
        \\ \quad 
            - \psi( {||x_2-x_{1}||_2}) x_{1} 
            - \psi( {||x_2-x_{3}||_2}) x_{3} 
            - \psi( {||x_{2}-x_{4}||_2}) x_{4}
            \Big)
\end{cases}$
\end{varwidth}
\end{center}

The last mutually exclusive case occurs at any filament end segment.
That rule is shown in Eq. (\ref{eq::LJ_buckling_rule_end}) in Section \ref{boundary_conditions}
on filament boundary conditions.

Edges are constructed by IDs as parameters associated to objects. The forward ID is represented by the arrow, which can emanate from any object including two arrows out of a junction object to point towards an Arp2/3 object (square) as well. 

Another module we add that accounts for more microscopic phenomena than the larger-scale whole ionic interactions between actin monomers is an angle bending energy that accounts for potential alignments of multiple electrostatic interaction residues between the amino acids of actin monomers. This structure imposes a constraint that the actin monomers follow the optimal bending angle created by multiple binding residues between two adjacent proteins. If the positions of three adjacent actin objects are defined as $x_1$, $x_2$, and $x_3$, then the angle bending potential energy is

\begin{equation}
    U_\text{tri-nodal, ang}(x_1,x_2,x_3,\theta_\text{target}) = \frac{k_\text{B}}{2}\left(\cos^{-1}
    \left(\frac{(x_2-x_1) \cdot (x_3-x_2)}{||x_2-x_1||_2||x_3-x_2||_2} \right) - \theta_\text{target}\right)^2.
\end{equation}

We define the scalar quantities
\begin{equation}
    \begin{split}
        a(x_1, x_2, x_3) &= \frac{(x_2-x_1) \cdot (x_3-x_2)}
            {||x_2-x_1||_2||x_3-x_2||_2} \\
        b(x_1, x_2, x_3) &= \frac{||x_3-x_2||_2}
            {||x_2-x_1||_2} \\
        c(x_1, x_2, x_3) &=||x_2-x_1||_2||x_3-x_2||_2 \\
\end{split}
\end{equation}
Note that $a$ is symmetric ($a(x_1, x_2, x_3) = a(x_1, x_2, x_3) $), as is $c$, but $b$ isn't.
Parameter $a = \cos(\theta) \in [-1,1]$.
For three colinear points with $x_2$ in the middle, 
the actual enclosed angle $\theta= 0$
and $a=1$.  
We define the vector quantities
\begin{equation}
    \begin{split}
        L(x_1, x_2, x_3) &= (x_1-2x_1+x_3)\\
        \tilde{L}(x_1, x_2, x_3) &= 
            (b x_1-(b+\frac{1}{b})x_2+\frac{1}{b}x_3).
    \end{split}
\end{equation}
$L$ is symmetric but $\tilde{L}$ is not.
Then we may calculate the gradient of an interior node of three as
\begin{equation}
\nabla_{x_2} U_\text{trinodal, ang} 
    = -k_\text{B} \frac{(\cos^{-1}(a)-\theta_{\text{target}})}{c \sqrt{1-a^2}}
    (L + a \tilde{L})
\end{equation}
and the gradient of an end node of three is
\begin{equation}
\nabla_{x_1} U_\text{trinodal, ang} 
    = -k_\text{B} \frac{(\cos^{-1}(a)-\theta_{\text{target}})}{c \sqrt{1-a^2}}
    ( (x_2-x_3) - a b (x_1 - x_2) ).
\end{equation}
Our target angles are $\theta_{\text{target}}=0$ and 70 degrees.
Angles $\theta=\cos^{-1}(a)$ near 70 degrees are not
a problem in the denominator but those near zero
look like they might be since then $a\simeq 1$.
However, for $\theta_{\text{target}}=0$ 
and $a \simeq 1$ the ratio of terms is actually nonsingular
by l'Hopital's rule.

The corresponding interior-node Hessian as
\begin{equation}
\begin{split} 
    \nabla^2_{x_2 x_2} U_\text{trinodal, ang}
    =& 
    H_\text{interior} 
    + (\cos^{-1}(a)- \theta_{\text{target}}) \hat{H}_\text{interior} \\
    H_\text{interior}=& \frac{k_\text{B}}{c^2(1-a^2)}(L+a \tilde{L}) \otimes (L+a \tilde{L})
    \\ 
\end{split} 
\end{equation}
The $(\cos^{-1}(a)- \theta_{\text{target}})\hat{H}$ term 
has a variable sign depending on the first factor.
We will drop all such matrix terms in order to obtain a
tractable, positive semidefinite approximation for use
in a Gaussian distribution of position parameters
for each modeled object.
{The resulting matrix has one nonnegative eigenvalue
and (in 2D) one zero eigenvalue.
We will further regularize the $H$ matrix to make it
positive definite, by taking $H$ to be the positive eigenvalue
of  $H_\text{interior}$ times the identity matrix,
or equivalently (for a rank-one matrix)
the Frobenius norm or the nuclear norm
of $H_\text{interior}$ times the identity matrix.}

The full angle-bending potential energy is
\begin{equation}
        U_\text{ang}(x, \theta_{\text{target} \; i j k})
    = \frac{k_\text{B}}{2} \sum_i \sum_{j \neq i} \sum_{k \neq j, i}
        G_{i j} G_{j k}
        U_\text{tri-nodal, ang}
        (x_i, x_j, x_k,\theta_{\text{target} \; i j k}) 
\end{equation}
where now $G^2$  can be symmetrized, 
since any antisymmetric component will sum to zero.

We can take its gradient as
\begin{equation}
     \nabla_l  U_\text{ang}(x, \theta_{\text{target} \; i j k})
    = \frac{k_\text{B}}{2} \sum_i \sum_{j \neq i} \sum_{k \neq j, i}
        G_{i j} G_{j k}
       \nabla_l U_{\text{tri} \; ijk}
        (x_i, x_j, x_k,\theta_{\text{target} \; i j k}) 
\end{equation}

\begin{equation}
\begin{split}
     \nabla_l  U_\text{ang}(x, \theta_{\text{target} \; i j k})
    = & -
    \frac{k_\text{B}}{2} \sum_{\langle i j k \rangle_\neq} 
        G_{i j} G_{j k} \Big(
       \nabla_l U_{\text{tri} \; ilk} \delta_{j l}
        + \nabla_l U_{\text{tri} \; ljk} \delta_{i l}
        + \nabla_l U_{\text{tri} \; ijl} \delta_{k l}
        \Big)
    \\ = & 
    \frac{k_\text{B}}{2} \sum_{\langle i j k \rangle_\neq} 
        G_{i j} G_{j k} 
         \frac{(\cos^{-1}(a_{i j k})-\theta_{i j k}a)}{c_{i j k} \sqrt{1-a_{i j k}^2}}
        \Big[
       \delta_{j l} (L_{i j k} + a_{i j k} \tilde{L}_{i j k})
        \\ &
        +  \delta_{i l} ((x_j - x_k) - a_{i j k} b_{i j k}(x_i - x_j))
        + \delta_{k l}((x_j - x_i)- a_{k j i} b_{k j i}(x_k - x_j))
        \Big]
\end{split}
\end{equation}
Defining
\begin{equation}
\begin{split}
        K_{i j k} &= (x_j - x_k) - a_{i j k} b_{i j k}(x_i - x_j) \\
        \tilde{K}_{i j k} &= 
             (x_j - x_i)- \frac{a_{i j k}}{ b_{i j k}}(x_k - x_j) = K_{k j i}
\end{split}
\end{equation}
we have
\begin{equation}
\begin{split}
     \nabla_l  U_\text{ang}(x, \theta_{\text{target} \; i j k})
    = & -
    \frac{k_\text{B}}{2} \sum_{\langle i j k \rangle_\neq} 
        G_{i j} G_{j k} 
         \frac{(\cos^{-1}(a_{i j k})-\theta_{i j k})}{c_{i j k} \sqrt{1-a_{i j k}^2}}
        \\ & \times
        \Big[
        \delta_{i l} K_{i j k}
        + \delta_{k l}K_{k j i}
      - \delta_{j l} (K_{i j k} + K_{k j i})
        \Big].
\end{split}
\end{equation}
$K_{i j k}$ is not symmetric under $(i, j, k)\leftrightarrow(k, j, i)$, but of course $(K_{i j k} + K_{k j i})$ is.
Likewise for the ``usable'' portion of the Hessian,
\begin{equation}
\begin{split}
     \nabla^2_{ l m}  U_\text{ang}(x, \theta_{\text{target} \; i j k})
    = &
    \frac{k_\text{B}}{2} \sum_{\langle i j k \rangle_\neq} 
        G_{i j} G_{j k} 
         \frac{1}{c^2_{i j k} ({1-a_{i j k}^2})}
        \\ & \times
        \Big[
        \delta_{i l} K_{i j k}
        + \delta_{k l}K_{k j i}
      - \delta_{j l} (K_{i j k} + K_{k j i})
        \Big]
        \\ & \times
        \Big[
        \delta_{i m} K_{i j k}
        + \delta_{k m}K_{k j i}
      - \delta_{j m} (K_{i j k} + K_{k j i})
        \Big]
        \\ &
        + O(\cos^{-1}(a_{i j k})-\theta_{\text{target} \; i j k})
        .
\end{split}
\end{equation}
Note that again the sum over $l$ of this expression is zero, because the sum over $l$ removes the Kronecker deltas from inside the first square bracket factor, leaving $K_{i j k}+K_{k j i}-(K_{i j k}+K_{k j i})=0$. In the $l,m$ space this expression is a nonnegatively weighted sum of outer products of the vectors in square brackets, so the matrix is positive semidefinite.

For efficiency in implementation we need to eliminate the Kronecker deltas. So, 
\begin{equation}
\begin{split}
     \nabla_l  U_\text{ang}(x, \theta_{\text{target} \; i j k})
    =  -\frac{k_\text{B}}{2} &
     \Bigg[
        \sum_{\langle j k \rangle_\neq ; \; j,k \neq l} 
        G_{l j} G_{j k} 
         \frac{(\cos^{-1}(a_{l j k})-\theta_{l j k})}{c_{l j k} \sqrt{1-a_{l j k}^2}}
         K_{l j k}
        \\ & +
        \sum_{\langle i j \rangle_\neq ; \; i,j \neq l} 
          G_{i j} G_{j l} 
         \frac{(\cos^{-1}(a_{i j l})-\theta_{i j l})}{c_{i j l} \sqrt{1-a_{i j l}^2}}
         K_{l j i}
        \\ & -
        \sum_{\langle i k \rangle_\neq ; \; i,k \neq l} 
          G_{i l} G_{l k} 
         \frac{(\cos^{-1}(a_{i l k})-\theta_{i l k})}{c_{i l k} \sqrt{1-a_{i l k}^2}}
         (K_{i l k} + K_{k l i})
        \Bigg].
\end{split}
\label{simplified_bending_gradient}
\end{equation}

Using relationships such as
$\delta_{i l} \delta_{i m} = \delta_{l m} \delta_{i l} \delta_{i m}$ and, 
within the $\langle i j k  \rangle_\neq$ sums,
$\delta_{i l} \delta_{k m} = (1- \delta_{l m}) \delta_{i l} \delta_{k m}$,
we find a similarly reduced Hessian expression which however is the sum of
nine terms rather than three. Of these nine terms, three share a factor
of $\delta_{l m} $ i.e. appear on the diagonal of the Hessian, and six
share a factor of $(1- \delta_{l m})$ and are therefore off-diagonal terms.
For example the diagonal terms can be calculated easily:
\begin{equation}
\begin{split}
     \nabla^2_{ l m}  U_\text{ang}(x, \theta_{\text{target} \; i j k})
    =  \frac{k_\text{B}}{2}  \delta_{l m} &
     \Bigg[
        \sum_{\langle j k \rangle_\neq ; \; j,k \neq l} 
        G_{l j} G_{j k} 
         \frac{1}{c^2_{l j k} ({1-a_{l j k}^2})}
         K_{l j k}^2
        \\ & +
        \sum_{\langle i j \rangle_\neq ; \; i,j \neq l} 
          G_{i j} G_{j l} 
         \frac{1}{c^2_{i j l} ({1-a_{i j l}^2})}
         K_{l j i}^2
        \\ & +
        \sum_{\langle i k \rangle_\neq ; \; i,k \neq l} 
          G_{i l} G_{l k} 
         \frac{1}{c^2_{i l k} ({1-a_{i l k}^2})}
         (K_{i l k} + K_{k l i})^2
        \Bigg]
        \\ &
        + (1 - \delta_{l m}) \Big[\ldots \Big]
        + O(\cos^{-1}(a_{i j k})-\theta_{\text{target} \; i j k})
        .
\end{split}
\label{simplified_bending_Hessian}
\end{equation}
The terms diagonal in the $l,m$ space suffice to define a joint Gaussian
model with properly nonnegative eigenvalues,
but more importantly for our purpose, to define a single-$l$ at a time 
Gaussian update formula similar to Eq.~(\ref{buckling_Gaussian}).
\begin{equation}
\text{Pr}(x_l^\prime | x_l) 
\propto \exp \Big[ - (x^\prime_l - x_l)^T
    \Big( \big(\sum_{\langle j k \rangle_\neq } G_{l j} G_{j k}  \ldots +  \ldots +  \ldots \big) I
            +\epsilon^2_\text{ang} I \Big)(x^\prime_l - x_l) \Big]
\label{bending_Gaussian}
\end{equation}

The resulting DGG update rules are:

\begin{center}
    \begin{varwidth}{\linewidth}
    \textbf{Angle Bending 1:} \\
    \begin{equation}
    \label{eq::angle_bending_rule}
    \begin{diagram}[size=2em] 
    & (
    \resizebox{2.0ex}{!}{$\filledstar$}_1 & \rTo & {\resizebox{2.0ex}{!}{$\filledstar$}_2} & \rTo &\resizebox{2.0ex}{!}{$\filledstar$}_3
    ) &
    \\
    \longrightarrow &
    (
    \resizebox{2.0ex}{!}{$\filledstar$}_1 & \rTo & \boxed{\resizebox{2.0ex}{!}{$\filledstar$}_2} & \rTo &\resizebox{2.0ex}{!}{$\filledstar$}_3
    )
    \end{diagram}
    \quad
    \begin{varwidth}{\linewidth}
    $\llangle$
    $\text{${ \boldsymbol x}$}_1, 
    \text{${ \boldsymbol x}$}_2, 
    \text{${ \boldsymbol x}$}_3$
    $\rrangle$
    \\
    $\llangle$
    $\text{${ \boldsymbol x}$}_1, 
    \text{${ \boldsymbol x}$}_2 + \text{$\Delta_{\text{ang}}(x_1,x_2,x_3)$}, 
    \text{${ \boldsymbol x}$}_3$
    $\rrangle$
    \end{varwidth}
    \end{equation}
    \quad \quad \text{\boldmath $\mathbf{with}$} \ \
    $k_{\text{kinetic}}$
    \par \quad \quad \ \ 
    $\begin{cases}
\text{$\Delta_{\text{ang}}(x_1,x_2,x_3)$} &:= 
    -\frac{1}{\zeta_\text{actin} k_{\text{kinetic}}}
        \frac{\partial\text{$U_{\text{tri-nodal,ang}}(x_1,x_2,x_3,\theta(\tau_1,\tau_2,\tau_3))$}}{\partial x_2}
        \\
        & = - \frac{k_\text{B}}{\zeta_\text{actin} k_{\text{kinetic}}}
                \frac{(\cos^{-1}(a_{1 2 3})-\theta(\tau_1,\tau_2,\tau_3))}{c_{1 2 3} \sqrt{1-a_{1 2 3}^2}} (K_{1 2 3} + K_{3 2 1})
    \end{cases}$
    \end{varwidth}
\end{center} \

\begin{center}
    \begin{varwidth}{\linewidth}
    \textbf{Angle Bending 2:} \\
    \begin{equation}
    \label{eq::angle_bending_rule_2}
    \begin{diagram}[size=2em] 
    & (
    \resizebox{2.0ex}{!}{$\filledstar$}_1 & \rTo & {\resizebox{2.0ex}{!}{$\filledstar$}_2} & \rTo &\resizebox{2.0ex}{!}{$\filledstar$}_3
    ) &
    \\
    \longrightarrow &
    (
    \boxed{\resizebox{2.0ex}{!}{$\filledstar$}_1} & \rTo & \resizebox{2.0ex}{!}{$\filledstar$}_2 & \rTo &\resizebox{2.0ex}{!}{$\filledstar$}_3
    )
    \end{diagram}
    \quad
    \begin{varwidth}{\linewidth}
    $\llangle$
    $\text{${ \boldsymbol x}$}_1, 
    \text{${ \boldsymbol x}$}_2, 
    \text{${ \boldsymbol x}$}_3$
    $\rrangle$
    \\
    $\llangle$
    $\text{${ \boldsymbol x}$}_1 + \text{$\Delta_{\text{ang}}(x_1,x_2,x_3)$}, 
    \text{${ \boldsymbol x}$}_2, 
    \text{${ \boldsymbol x}$}_3$
    $\rrangle$
    \end{varwidth}
    \end{equation}
    \quad \quad \text{\boldmath $\mathbf{with}$} \ \
    $k_{\text{kinetic}}$
    \par \quad \quad \ \ 
    $\begin{cases}
\text{$\Delta_{\text{ang}}(x_1,x_2,x_3)$} &:=
    -\frac{1}{\zeta_\text{actin} k_{\text{kinetic}}}\frac{\partial\text{$U_{\text{tri-nodal,ang}}(x_1,x_2,x_3,\theta(\tau_1,\tau_2,\tau_3))$}}{\partial x_1}
        \\
        & = - \frac{k_\text{B}}{\zeta_\text{actin} k_{\text{kinetic}}}
                \frac{(\cos^{-1}(a_{1 2 3})-\theta(\tau_1,\tau_2,\tau_3))}{c_{1 2 3} \sqrt{1-a_{1 2 3}^2}} K_{3 2 1}
\end{cases}$
    \end{varwidth}
\end{center}

... and likewise a third rule (with the directed edges reversed) for $x_3$.
Note that the second angle-bending rule is not exclusive to filament end cases,
though the first angle-bending rule will not apply to those cases.
Here $\theta(\tau_1,\tau_2,\tau_3)= \theta(\tau_3,\tau_2,\tau_1)$ is the preferred (minimal-energy) angle
for three successive nodes as a function of their type data (including end-node status) $\tau$,
which we have visualized with different icons $\Circle, \CIRCLE$ etc..

The other module we add for biophysics is aimed at providing some thermal noise for spatial fluctuations near-equilibrium which describes continual movement of the actin filaments within the context of minimizing the Hessian of the chosen potential. 

Hessian Boltzmann sampling is derived from the Taylor expansion of a potential
\begin{equation}
    U(x+u)\approx U(x)+U'(x)^T \cdot u+\frac{1}{2}u^T \cdot H|_x \cdot u+...,
\label{Taylor_U}
\end{equation}
where $x$ is the 
current state,
$u$ is a displacement vector in the state,
and $H  = \nabla^2 U$ is the Hessian matrix of second derivatives of $U$. 

We extend the ``heat-bath" thermal simulation algorithm to a Metropolis-Hastings version by fulfilling the following Bayesian equation for detailed balance and probability flow between two states:
\begin{equation}
    \frac{A(x'|x)}{A(x|x')} = \frac{p(x') q(x | x')}{p(x) q(x' | x)},
\end{equation}
where $x'$ is the new state and $x$ is the old or current state, which we henceforth equate with position. $q$ is the probability distribution for proposing a change from the old state to the new state. $p$ is known as the target steady-state distribution to which we want the Markov Chain Monte Carlo (MCMC) algorithm to converge. For the Metropolis-Hastings algorithm the displacement proposal distributions are not homogeneous, and we analogously extend the ``heat-bath" algorithm to a novel thermal noise algorithm that assigns displacement update probabilities according to the current state. The covariance of this multivariate Gaussian is the inverse of the Hessian multiplied by thermodynamic $k_\text{Boltzmann} T$ (\cite{Das2019}). We let the mean of the sampling distribution be zero. The linear term of the Taylor expansion -- assumed to be near-zero -- is accounted for instead in the viscous gradient dynamics.

In this algorithm we derive a heat-bath acceptance probability for thermal noise by considering respective forward and reverse versions of the target state probabilities $p(x)$ and proposal probabilities 
$q(x'|x)$, and imposing the standard constraint of detailed balance. 
We consider the Boltzmann probabilities of the thermal state of a system to rely on the pure 
separation
potential at that state. For proposal probabilities conditioned on the current state, which is our normal distribution, we take them to be proportional to the Boltzmann probability with Hessian energy term in the Taylor expansion of the current state. When the acceptance probabilities are algebraically manipulated into heat-bath form,  we find that the acceptance probability takes the form
\begin{equation}
    A(x'|x) = \frac{e^{-\frac{\Delta U - \Delta U_q}{k_\text{Boltzmann} T}}\sqrt{|H(x)/H(x^{\prime})|}}
    {1 + e^{-\frac{\Delta U - \Delta U_q}{k_\text{Boltzmann} T}}\sqrt{|H(x)/H(x^{\prime})|}}\,
\end{equation}
where $U$ is the global mechanical energy, $\Delta U$ is the change in $U$ due to a local move from $x$ to $x^\prime$,
$U_q$
is just the quadratic (Hessian) part of the local Taylor expansion (Eq. (\ref{Taylor_U})) of $U$, $|H(x)|$ is the determinant of the Hessian $H(x)$ at $x$,
and $\Delta U_q(x^\prime|x) = U_q(x^\prime|x) - U_q(x|x^\prime)$.
The constant term of course drops out under the difference $\Delta$ operation.
The factors of $\sqrt{|H(x^\prime)|}$ arise from a Gaussian integral approximation
to the partition function normalization $\int dx \exp(-\beta U_q(x|x^\prime))$ 
for the heat bath move proposal $q(x|x^\prime)$,
which depends on move starting position $x^\prime$.

If we instead used both the linear and quadratic terms as in Eq. (4.7) of \cite{Lavenda1991},
then the approximation of $\Delta U$ would be better except near the local energy minimum,
leading to good acceptance ratios $A$. However, we opt to separate thermal noise from
active drive processes in the rule set since they are physically distinct.
The two options are equivalent near the local energy minimum, at thermal equilibrium or quasi-equilibrium.
The probability of a ``proposal'' step under $\Delta U_q$ will be a zero-mean Gaussian
or Normal distribution with precision matrix (inverse covariance matrix) $H/k_\text{Boltzmann} T$.
The acceptance of such a step will depend strongly on its alignment 
with the gradient where the gradient is high,
but will be suitably thermal near energy minima in the manner of an Ornstein-Uhlenbeck process.

We have calculated above the relevant Hessians for both buckling and bending energies.
These Hessians can be summed together or, if one is expected to be substantially smaller
in a matrix norm sense than the other, only the larger one can be kept.
Since the central LJ-style potentials can have high derivative values at the cores,
we retain just the buckling Hessian. Then, for a change in one particular $\overrightarrow{x_i}$ alone, 

\begin{equation*}
    \Delta_i U([x]) = \Delta_i U_\text{sep}(\overrightarrow{x_i}|\text{nbd}(x_i)),
\end{equation*}
i.e. only local summands of $U_\text{sep}$ are involved.

The DGG rule is shown below.

\begin{center}
    \begin{varwidth}{\linewidth}
    \textbf{Hessian Thermal Noise:} \\
    \begin{equation}
    \label{eq::Hessian_Noise_rule}
    \begin{diagram}[size=2em] 
    & (
    \resizebox{2.0ex}{!}{$\filledstar$}_1 & \rTo & {\resizebox{2.0ex}{!}{$\filledstar$}_2} & \rTo &\resizebox{2.0ex}{!}{$\filledstar$}_3
    ) &
    \\
    \longrightarrow &
    (
    \resizebox{2.0ex}{!}{$\filledstar$}_1 & \rTo & \boxed{\resizebox{2.0ex}{!}{$\filledstar$}_2} & \rTo &\resizebox{2.0ex}{!}{$\filledstar$}_3
    )
    \end{diagram}
    \quad
    \begin{varwidth}{\linewidth}
    $\llangle$
    $\text{${ \boldsymbol x}$}_1, 
    \text{${ \boldsymbol x}$}_2, 
    \text{${ \boldsymbol x}$}_3$
    $\rrangle$
    \\
    $\llangle$
    $\text{${ \boldsymbol x}$}_1, 
    \text{${ \boldsymbol x}$}_2 + \text{$\Delta_{H}$}, 
    \text{${ \boldsymbol x}$}_3$
    $\rrangle$
    \end{varwidth}
    \end{equation}
    \quad \quad \text{\boldmath $\mathbf{with}$} \ \
    $k_{\text{kinetic}}p(\Delta U_H(x_1,x_2,x_3)) \mathcal{N}(\Delta_H; 0, \Sigma_{\text{therm}})$
    \par \quad \quad  \ \ 
$\begin{cases}
p(\Delta U_H(x_1,x_2,x_3)):=\frac{e^{-\frac{\Delta U(x_1,x_2,x_3)-\Delta U_q(x_1,x_2,x_3)}{k_\text{Boltzmann} T}}\sqrt{|H(x_2)/H(x_2+\Delta_H)|}}{1+e^{-\frac{\Delta U(x_1,x_2,x_3)-\Delta U_q(x_1,x_2,x_3)}{k_\text{Boltzmann} T}}\sqrt{|H(x_2)/H(x_2+\Delta_H)|}} \\
\Sigma_{\text{therm}} := 2 \frac{k_\text{Boltzmann} T}{\zeta_\text{actin} k_\text{kinetic}} H^{-1}(x_2)
\end{cases}$
    \end{varwidth}
\end{center}

For consistency we also need to include the actin fiber end case.
It is in section \ref{boundary_conditions} below.

Next, we explain how the dynamics of the membrane can be implemented in the DGG simulation also in the form of stochastic rules. There are three terms which comprise the total membrane energy: membrane area energy, membrane length energy, and membrane Helfrich mean curvature energy for the spine head membrane (\cite{Quintana2020}). The total energy is

\begin{equation}
    U_{\text{mem}} = 2 \kappa \int_{\Gamma} (H^2) d \bar{w}.
\end{equation}

Here, $\kappa$ the bending modulus, $\Gamma$ the 1D representation of the manifold that is the 2D spine head membrane surface, $H$ the mean curvature, and $\bar{w}$ the arc length (\cite{Quintana2021}). Mean curvature is approximated using forward finite differences.

Parameters used in our simulations are provided in Table 1.

\subsection{Efficient implementation}

We have approached the 
issue of computational efficiency in our simulation in multiple ways including
rule-writing, and biophysical theory.

In rule-writing, we sample tri-nodal forces so that the net update is weaker in magnitude than pairwise forces, and anisotropic towards a perpendicular axis. 
In theory of biophysical kinetics, we derived
a clipping potential ({Eq.~(\ref{clipped-potential})})
that sets the force to a constant value past a multiplicative factor below the optimal length. Also, the bending energy's gradient
and Hessian are
simplified to a few terms (Eqs.~(\ref{simplified_bending_gradient}) 
and (\ref{simplified_bending_Hessian})).

For the filament breaking rule (Eq.~(\ref{eq::breaking_rule})), we store an angle parameter for each actin node, such that all internal angles are not re-computed at each step, but the angle of each node is updated with the firing of a stochastic spatial rule that samples it. Then, they are accessed by Eq.~(\ref{eq::breaking_rule}) by its ``with'' clause. 

\subsection{Boundary conditions}
\label{boundary_conditions}

The first boundary condition rule is for buckling.
It is mutually exclusive in domain of applicability with
the previous two (interior and junction node) cases.

\begin{center}
    \begin{varwidth}{\linewidth}
    \textbf{End Case Anistropic Buckling:} \\
    \begin{equation}
    \label{eq::LJ_buckling_rule_end}
    \begin{diagram}[size=2em] 
    & (
    \resizebox{2.0ex}{!}{$\filledstar$}_1 & \rLine & 
    \{ \CIRCLE, \resizebox{1.5ex}{!}{$\filledsquare$} \}_2 
    ) &
    \\
    \longrightarrow &
    (
    \resizebox{2.0ex}{!}{$\filledstar$}_1 & \rLine & 
    \boxed{\{ \CIRCLE, \resizebox{1.5ex}{!}{$\filledsquare$} \}_2} 
    )
    \end{diagram}
    \quad
    \begin{varwidth}{\linewidth}
    $\llangle$
    $\text{${ \boldsymbol x}$}_1, 
    \text{${ \boldsymbol x}$}_2$
    $\rrangle$
    \\
    $\llangle$
    $\text{${ \boldsymbol x}$}_1, 
    \text{${ \boldsymbol x}$}_2 + \text{$\Delta_{\text{sep}}(x_1,x_2)$}$
    $\rrangle$
    \end{varwidth}
    \end{equation}
    \quad \quad \text{\boldmath $\mathbf{with}$} \ \
    $k_{\text{kinetic}}$
    \par \quad \quad \ \ 
$\begin{cases} 
\text{$\Delta_{\text{sep}}(x_1,x_2)$} :=
 -\frac{1}{\zeta_\text{actin} k_{\text{kinetic}}}\left(\frac{\partial U_\text{sep}(x_1,x_2)}{\partial x_2}\right)
 \\
 = -\frac{1}{\zeta_\text{actin} k_{\text{kinetic}}}\left(\frac{\partial U_\text{radial}([x])}{\partial x_2}\right)
 \\
 =
  -\frac{1}{\zeta_\text{actin} k_{\text{kinetic}}}
        \psi({||x_2-x_1||_2}) (x_2-x_1)
\end{cases}$
    \end{varwidth}
\end{center}

Likewise, the Hessian Thermal Noise end case is:

\begin{center}
    \begin{varwidth}{\linewidth}
    \textbf{End Case Hessian Thermal Noise:} \\
    \begin{equation}
    \label{eq::Hessian_Noise_end_rule}
    \begin{diagram}[size=2em] 
    & (
    \{$\CIRCLE$, \resizebox{1.5ex}{!}{$\filledsquare$} \}_1 & \rLine & {\resizebox{2.0ex}{!}{$\filledstar$}_2}
    ) &
    \\
    \longrightarrow &
    (
    \boxed{\{$\CIRCLE$, \resizebox{1.5ex}{!}{$\filledsquare$} \}_1} & \rLine & {\resizebox{2.0ex}{!}{$\filledstar$}_2}
    )
    \end{diagram}
    \quad
    \begin{varwidth}{\linewidth}
    $\llangle$
    $\text{${ \boldsymbol x}$}_1, 
    \text{${ \boldsymbol x}$}_2$
    $\rrangle$
    \\
    $\llangle$
    $\text{${ \boldsymbol x}$}_1 + \text{$\Delta_{H}$}, 
    \text{${ \boldsymbol x}$}_2$
    $\rrangle$
    \end{varwidth}
    \end{equation}
    \quad \quad \text{\boldmath $\mathbf{with}$} \ \
        $k_{\text{kinetic}}p(\Delta U_H(x_1,x_2)) \mathcal{N}(\Delta_H; 0, \Sigma_{\text{therm}})$
\par \quad \quad \ \ 
$\begin{cases}
p(\Delta U_H(x_1,x_2)):=\frac{e^{-\frac{\Delta U(x_1,x_2)-\Delta U_q(x_1,x_2)}{k_\text{Boltzmann} T}}\sqrt{|H(x_1)/H(x_1+\Delta_H)|}}{1+e^{-\frac{\Delta U(x_1,x_2)-\Delta U_q(x_1,x_2)}{k_\text{Boltzmann} T}}\sqrt{|H(x_1)/H(x_1+\Delta_H)|}} \\
\Sigma_{\text{therm}} := 2 \frac{k_\text{Boltzmann} T}{\zeta_\text{actin} k_\text{kinetic}} H^{-1}(x_2)
\end{cases}$
\end{varwidth}
\end{center}

In each of these rules the undirected edge on both LHS and RHS is a shortand notation
for a pair of directed-edge rules. In the first of the pair, both undirected
edges are replaced with edges directed in the same left-to-right direction in the picture;
and in the second of the pair, 
both undirected
edges are replaced with edges directed in the same right-to-left direction.

\begin{table}
\centering
\tiny
\begin{tabular}{||l|l|l||}
 \hline
 Parameter & Value & Description \\ 
 \hline\hline
 $k_{\text{kinetic}}$ & 0.35 $\frac{1}{\text{s}}$ & Kinetics propensity \\
 $r_{\text{spinehead, init}}$ & 0.125 $\mu$m & Starting spine head radius \\
 $N_{\text{CG}}$ & 12 monomers& Number of actin monomers in a coarse-grained object \\
 T & 310 K & Temperature \\ 
 $k_{\text{Boltzmann}}$ & $1.38\times10^{-23} \frac{\text{J}}{\text{K}}$ & Boltzmann's constant \\
 $\epsilon_\text{Clip}$ & 0.75 & Clipping factor for LJ potential\\
 $\theta_\text{Arp}$ & $70^{\circ}$ & Arp2/3 branching angle (\cite{Pollard1998}) \\
 $k_\text{s}$ & $15.9 \frac{\text{N}}{\text{m}}$ & Spring constant of adjacent actin monomers (\cite{Mogilner2003}) \\
 $k_\text{B}$ & $4.0\times10^{-26}$ N$m^2$ & Bending stiffness of actin filament (\cite{Isambert1995}) \\
 $k_\text{B, Arp}$ & $k_\text{B}$ N$m^2$ & Bending stiffness of Arp2/3 branch\\
 $\epsilon_{\text{LJ}}$ & $7.36\times10^{-16}$ J & Dissociation energy for LJ potential (\cite{Hoyer2022}) \\
 $\mathcal{D}_\mathcal{U}$ & 0.0175 $\mu$m & Binding distance constraint for CaMKII$\beta$ \\
 $\theta_\text{Bundle}$ & $15^\circ$ & Bundling angle constraint for CaMKII$\beta$ (\cite{Kim2007}) \\
  $k_{\text{s, Arp2/3}}$ & $20 \frac{N}{m}$ & Spring constant for branch bond to F-actin (\cite{Kim2007}) \\
  $k_{\text{s, CaMKII$\beta$}}$ & $20 \frac{N}{m}$ & Spring constant for bundling bond to F-actin (\cite{Kim2007}) \\
 $k_\text{B, CaMKII$\beta$}$ & $k_{\text{s}, \text{ Cam}}$ & Bending stiffness for CaMKII$\beta$ bundling segment \\
 $\zeta_{\text{actin}}$ & $500 \frac{\text{N}}{\text{m/s}}$ & Friction coefficient of actin network (\cite{Quintana2020})\\
  $\zeta_\text{mem}$ & $500 \frac{\text{N}}{\text{m/s}}$& Friction coefficient of membrane vertex(\cite{Quintana2020})\\
 $L_p$ & $17.7\times 10^{-6}$ m & Persistence length of actin filament (\cite{Gittes1993}) \\
 $\theta_{\text{break, Actin}}$ & $57^{\circ}$ & Breaking angle of bare actin (\cite{McCullough2011}) \\
 $\theta_{\text{break, Cofilactin}}$ & $73^{\circ}$ & Breaking angle of cofilactin (\cite{McCullough2011}) \\
 $\theta_{\text{break, Boundary}}$ & $31^{\circ}$ & Breaking angle of actin-cofilactin boundary (\cite{McCullough2011}) \\
 $k_\text{barbed, on, ATP}$ & $11.6\times 10^6 \frac{1}{\text{M}\cdot\text{s}}$ & Barbed end elongation rate constant (ATP) (\cite{Pollard1986})\\
 $k_\text{barbed, off, ATP}$ & $1.4 \frac{1}{\text{s}}$ & Barbed end retraction rate constant (ATP) (\cite{Pollard1986})\\
 $k_\text{pointed, on, ATP}$ & $1.3\times 10^6 \frac{1}{\text{M}\cdot\text{s}}$ & Pointed end elongation rate constant (ATP) (\cite{Pollard1986})\\
 $k_\text{pointed, off, ATP}$ & $0.81 \frac{1}{\text{s}}$ & Pointed end retraction rate constant (ATP) (\cite{Pollard1986})\\
 $k_\text{barbed, on, ADP}$ & $3.8\times 10^6 \frac{1}{\text{M}\cdot\text{s}}$ & Barbed end elongation rate constant (ADP) (\cite{Pollard1986})\\
 $k_\text{barbed, off, ADP}$ & $7.2 \frac{1}{\text{s}}$ & Barbed end retraction rate constant (ADP) (\cite{Pollard1986})\\
 $k_\text{pointed, on, ADP}$ & $0.16\times 10^6 \frac{1}{\text{M}\cdot\text{s}}$ & Pointed end elongation rate constant (ADP) (\cite{Pollard1986}) \\
 $k_\text{pointed, off, ADP}$ & $0.27 \frac{1}{\text{s}}$ & Pointed end retraction rate constant (ADP) (\cite{Pollard1986})\\
$k_
\text{branch}$ & $3000. \frac{1}{\text{M}\cdot\text{s}}$ & Arp2/3 nucleation rate (\cite{Smith2013})\\
 $k_{\text{unbranch}}$ & $0.47 \frac{1}{\text{s}}$ & Unbinding of Arp2/3 (\cite{Smith2013})\\
 $k_\text{cap, on}$ & $6.3 \times 10^6 \frac{1}{\text{M}\cdot\text{s}}$ & Barbed end capping on rate constant (\cite{Wear2003}) \\
 $k_\text{cap, off}$ & $9.5 \times 10^{-4} \frac{1}{\text{s}}$ & Aip1 end-capping off rate constant (\cite{Hayakawa2019}) \\
 $k_\text{CaMKII$\beta$, on}$ & $0.5 \times 10^6 \frac{1}{\text{M}\cdot\text{s}}$ & Binding rate constant of CaMKII$\beta$ (\cite{Khan2016})\\
 $k_\text{CaMKII$\beta$, off}$ & $0.23 \frac{1}{\text{s}}$ & Unbinding rate constant of CaMKII$\beta$ (\cite{Khan2016})\\
 $k_\text{ATP, Hydrolysis}$ & $0.35 \frac{1}{\text{s}}$ & ATP hydrolysis rate constant on actin filaments (\cite{Roland2008})\\
 $k_\text{cof, Pi}$ & $0.035 \frac{1}{\text{s}}$ & Release of Pi from actin filament with nearby cofilin (\cite{Roland2008})\\
 $k_\text{Pi}$ & $0.006 \frac{1}{\text{s}}$ & Release of Pi from actin filament (\cite{Roland2008})\\
 $k_\text{cof-on, edge, ADP}$ & $17 \times 10^6 \frac{1}{\text{M}\cdot\text{s}}$ & Recruitment of cofilin to bound cofilin in ADP presence (\cite{Wioland2017})\\
 $k_\text{cof-off}$ & $0.7 \frac{1}{\text{s}}$ & Unbinding of cofilin from actin filament (\cite{Wioland2017})\\
 $k_\text{single, on, Cof}$ & $10^4 \frac{1}{\text{M}\cdot\text{s}}$ & On rate constant of isolated Cof. (\cite{DeLaCruz2009})\\
 $k_\text{sever}$ & $1.43 \frac{1}{s}$ & Rate of severing by Aip1 (\cite{Chen2015})\\
 $k_\text{on, Aip1}$ & $112\times10^6 \frac{1}{\text{M}\cdot\text{s}}$ & On rate of Aip1 to cofilactin (\cite{Hayakawa2019}) \\
 $k_\text{debranch}$ & $2 \times 10^{-3}$ & De-branching rate from parent filament (\cite{Chan2013}) \\
 $\sigma_\text{comp}$ & $4.0$ & Competition constant for cofilin against Arp2/3 (\cite{Chan2013}) \\
 $k_\text{synth, actin}$ & $19.5 \times 10^{-6} \frac{\text{M}}{\text{s}}$ & Basal synthesis rate of actin(\cite{Bosch2014,Quintana-Rangamani2024,Helm2021}) \\
 $k_\text{influx, actin}$ & $18.8\times 10^{-6} \frac{\text{M}}{\text{s}}$ & Stim. synthesis rate of actin (\cite{Bosch2014,Quintana-Rangamani2024,Helm2021}) \\
 $k_\text{deg, actin}$ & $0.096 \frac{\text{1}}{\text{s}}$ & Degradation rate of actin (\cite{Bosch2014,Quintana-Rangamani2024}) \\
 $k_\text{synth, arp2/3}$ & $2.6 \times 10^{-6} \frac{\text{M}}{\text{s}}$ & Basal synthesis rate of Arp2/3 (\cite{Bosch2014,Quintana-Rangamani2024})\\
 $k_\text{influx, arp2/3}$ & $0.41 \times 10^{-6} \frac{\text{M}}{\text{s}}$ & Stim. synthesis rate of Arp2/3 (\cite{Bosch2014,Quintana-Rangamani2024})\\
 $k_\text{deg, arp2/3}$ & $0.053 \frac{1}{\text{s}}$ & Degradation rate of Arp2/3 (\cite{Bosch2014,Quintana-Rangamani2024}) \\
 $k_\text{synth, cof}$ & $0.47 \times 10^{-6} \frac{\text{M}}{\text{s}}$ & Basal synthesis rate of cof. (\cite{Bosch2014,Quintana-Rangamani2024}) \\
 $k_\text{influx, cof}$ & $0.78 \times 10^{-6} \frac{\text{M}}{\text{s}}$ & Stim. synthesis rate of cof. (\cite{Bosch2014,Quintana-Rangamani2024}) \\
 $k_\text{deg, cof}$ & $0.057 \frac{1}{\text{s}}$ & Degradation rate of cof. (\cite{Bosch2014,Quintana-Rangamani2024}) \\
  $k_\text{synth, CaMKII$\beta$}$ & $4.29 \times 10^{-6} \frac{\text{M}}{\text{s}}$ & Basal synthesis rate of CaMKII$\beta$(\cite{Bosch2014,Quintana-Rangamani2024,Helm2021})\\
   $k_\text{influx, CaMKII$\beta$}$ & $-1.96 \times 10^{-6} \frac{\text{M}}{\text{s}}$ & Stim. synthesis rate of CaMKII$\beta$(\cite{Bosch2014,Quintana-Rangamani2024,Helm2021})\\
 $k_\text{deg, CaMKII$\beta$}$ & $ 0.052\frac{1}{\text{s}}$ & Degradation rate of CaMKII$\beta$ (\cite{Bosch2014,Quintana-Rangamani2024}) \\
 $k_\text{synth, cap}$ & $0.0046 \times 10^{-6} \frac{\text{M}}{\text{s}}$ & Basal synthesis rate of end-capping protein (\cite{Bosch2014,Quintana-Rangamani2024}) \\
  $k_\text{influx, cap}$ & $0.0014 \times 10^{-6} \frac{\text{M}}{\text{s}}$ & Stim. synthesis rate of end-capping protein (\cite{Bosch2014,Quintana-Rangamani2024}) \\
 $k_\text{deg, cap}$ & $0.052 \frac{1}{\text{s}}$ & Degradation rate of end-capping protein (\cite{Bosch2014,Quintana-Rangamani2024})\\
 $\kappa$ & $0.0005$ $\text{pN} \mu\text{m}$ & Membrane curvature strength (\cite{Quintana2020}) \\
 [1ex] 
 \hline
\end{tabular}
\caption{Table of parameters used in DGG simulations.}
\label{table:parameters}
\end{table}

\section{Discussion}\label{section::discussion}

We have demonstrated how to use our Dynamical Graph Grammar (DGG) simulation to reveal trends of actin cytoskeletal elements on dendritic spine head membrane expansion in 2D. 
The results support the existence of biophysical effects from the number of ABPs bound to actin filament networks.
In particular CaMKII$\beta$ molecules are affected by Arp2/3 and are hypostatic in growth of spine head during the growth phase of a model for long-term potentiation. Cofilin, in large numbers, counteracts Arp2/3 based growth. There can be net effects from combined mechanisms, e.g. mechanical bending relaxation and filament breaking, as shown by cofilin's impact on membrane size. The addition of Aip1 as a filament severing molecule to the model provided greater certainty about the role of other ABPs than through cofilin as the sole severing molecule. The DGG modeling language easily accommodated model variations and elaborations pertaining to cytoskeleton in dendritic spine.

An important advancement of our model is the method of biophysics modules in a DGG modeling language. The symbolic expressiveness of our stochastic simulator allows non-linear representations for velocity gradient and thermal noise Hessian based on an anharmonic potential compared to an integrator requiring linearization of the instantaneous force like Cytosim (\cite{Nedelec2007}). Implementation of biophysics in a stochastic grammar simulator that is event-based rather than looping over objects is more efficient than computation of inter-object potential (Eq.~(\ref{LJ_potential})) gradient for every interaction, which would additionally require more calibration of time step for effective wall clock simulation run time. The DGG stochastic algorithm handles time-stepping efficiently in a way that adapts to computational burden from the increasing size of pool of objects. We can update the combinatorial casework and physical update equations in biophysical rules in a DGG grammar whereas simulators like Cytosim are a pre-determined numerical integration algorithm that is difficult to update according to physical theory.

Our model can reveal epistatic relationships between cytoskeleton binding proteins. For the transient growth phase of a model of stubby spine head long-term potentiation, we find that Arp2/3 is epistatic to CaMKII$\beta$. Such a relationship can be used to determine which protein masks another protein's effect under control conditions of the rest of the ABPs in the model. We observe benefits of the Aip1 extension on the relation of ABPs to stubby spine head morphology. In the model without Aip1, there is greater uncertainty for the role of each of three ABPs (Arp2/3, CaMKII$\beta$, cofilin), and Arp2/3, the network branching protein, dominates the contribution to spine head morphology. The new model is clear in the role of each ABP included in it (sections~\ref{section::growth}--\ref{section::cofilin}), with a reduction in the role of Arp2/3.

Consistent with previous experimental studies discussed in Section~\ref{section::cofilin}, we find that cofilin reduces spine head size. Our modeling result shows that, under our modeling conditions, actin filament severing by Aip1 increases spine head size. A mutually exclusive implementation of competitive binding to actin filaments between cofilin and Arp2/3 as well as between cofilin and CaMKII$\beta$ in our model is a reason for decreased spine head size from increased cofilin synthesis. We hypothesize that the competitive binding between cofilin and Arp2/3 means increased number of bound cofilin limits the extent the actin network can grow by limiting branching events. The severing events that occur under this increasingly limiting factor are also further limited by CaMKII$\beta$ binding events to the newly polymerized locations of actin filaments, aligned in parallel by the severing events that do not depend on a breaking angle, that disallow recruitment of cofilin to neighboring unbound locations on actin filaments. The role of cofilin that does occur under inclusion of Aip1 is weakening of angular bending stiffness, which decreases the ability of filaments to push the membrane to grow in size. The role of Aip1 in severing actin filaments by binding to cofilactin has been determined by biological experiment recently (\cite{Oosterheert2025}), explaining  previous speculation about Aip1 affecting actin filament severing rate (\cite{Chen2015}), and we include it as a mechanistic extension in our model with the addition of just a few simpler rules to the pre-existing several hundred rules. The DGG modeling language has proven to be highly expressive in representing biological, biochemical, and biophysical knowledge in an extensible way in this domain, as we scaled up the number of rules and the amount of detail they express.

There are many avenues for future work to extend our model. Additional modules can be incorporated by design of DGGs for new cytoskeletal elements. Simulations can reach bigger scales of the spine head receiving timed signals. Implementation in the 
more numerically powerful, accelerated DGGML simulation package 
(\cite{Medwedeff2023, Medwedeff2024, Medwedeff2024b}), of DGG in C++, 
currently two-dimensional but with three dimensions under development,
may enable large 3D simulations
and may include model reduction assistance by AI (\cite{Mjolsness2019}). 
Usage of larger coarse-graining numbers would feasibly simulate changes to a synaptic spine head on a long-term scale. The result would be exploration of long-term potentiation using graph-based actin cytoskeletal simulation in synaptic spine heads, also of larger sizes, ultimately to understand mechanisms associated with learning and protect and enhance memory.

Molecular dynamics (MD) simulation, e.g. \cite{Voth2006}, is similar to our simulation system in that it updates positions of simulation objects, but also exhibit differences that make DGGs more suitable for spine head simulation. MD simulations are slow, 
simulating on at most
the nanosecond or even femtosecond scale; DGGs can operate on much coarser spatial and temporal scales, and 
can naturally update the
number or nature of chemical species
or objects in the simulation. So, they work well for remodeling of the actin cytoskeleton. MD simulations use ballistic movement while we implement viscous dynamics to convert forces to velocity in coarse-scale models represented as DGGs, as is appropriate in a viscous medium and spatial scale on all but the fastest time scales. In this way, we provide an efficient, expressive, and suitable simulation for spine head dynamics.

Our graph-based simulation algorithm is able to implement the biophysical model from underlying principles. It is a highly expressive simulation system that incorporates many modules well-suited for changing both network topology and spatial or, more generally, agent-associated parameters. DGGs offer considerable advantages over modern cytoskeletal simulation software, in the ability to offer users rule-based control to implement spatial, non-spatial, and protein-based regulation of the actin cytoskeleton by ABPs based on satisfying constraints. This format is made possible due to the availability of parameterizations associated with objects in both pattern-matching and object-replacements and the convenient, declarative usage of the ``with'' clause in the DGG modeling language.

Four ABPs, out of many more actin regulatory ABPs, have been implemented in the simulation as a demonstration of such a system's interoperability with DGGs. They represent the major cytoskeleton remodeling mechanisms known about actin filament networks. Non-spatial rules associated with regulatory ABPs, will usually have less casework for rule implementation, as the spatial rules are tri-nodal or quad-nodal (junctions) in actin filament sub-graph matching. As a model of learning and memory, we find that simulations of the synaptic spine head not only provide insight into the molecule types of the simulation, but also allow for a new kind of artificial intelligence capability that is spatial and biophysical in a unit of learning. Moreover, as symbolic and graphical artificial intelligence systems (\cite{Mjolsness1997}), DGGs and their rulesets could function as intelligent counterparts of biological systems.

Inherent patterns of the system have been characterized, with several implications: (1) There may also be an advantage for model reduction of this system using artificial neural networks which have nonlinear activation functions to capture spine head behavior. This would result in faster predictions about synaptic spine heads with biological experiments. (2) A memory protein can be a node inside an epistatic relation as demonstrated in this paper. (3) The current  DGG formalism is available for simulation of long-term effects of ABPs.

Such computational observations may in the future be compared to experiment,
perhaps after further modeling such as the extension of the present model to three dimensions.

\section*{Declarations}

\subsection*{Funding}
This work was funded in part by U.S. NIH/NIDA Brain Initiative Grant 1RF1DA055668-01 (all coauthors),
and also by Human Frontiers Science Program Grant HFSP—RGP0023/2018 (EM),
NIH NIMH CRCNS R01-MH129066 (TS, TB), NSF NeuroNex DBI-1707356 (TS, TB), 
and NSF NeuroNex DBI-2014862 (TS, TB).
This work was supported in part by the UC Southern California Hub, with funding from the UC National Laboratories division of the University of California Office of the President (EM). 
\subsection*{Author Contributions}
TB, TS, and EM designed the research.
MH implemented the kinetic rules,
carried out all simulations, and analyzed the data. 
EM provided the DGG theory and the derivation 
of kinetic rules from a global energy function.
TB designed the cytoskeleton structure-altering rules, 
with domain expertise from PR. 
MH, and EM wrote the paper.
\subsection*{Competing interests}
P.R. is a consultant for Simula Research Laboratories in Oslo, Norway and receives income. The terms of this arrangement have been reviewed and approved by the University of California, San Diego in accordance with its conflict-of-interest policies.
E.M. has a Joint Appointment with the Los Alamos National Laboratory Computer, Computational, and Statistical Sciences Division. The terms of this arrangement have been reviewed and approved by the University of California, Irvine, in accordance with its conflict-of-interest policies.
\subsection*{Acknowledgements}
We thank Mayte Bonilla-Quintana for many substantial modeling discussions 
regarding the actin cytoskeleton in dendritic spine head.
We thank Elliot Meyerowitz, Sean Wilson, and Katherine Thompson-Peer for useful discussions.
We thank Arthur York for creating a previous version of the DGG model file
including cytoskeleton structure-altering rules.
\subsection*{Ethics approval and consent to participate}
Not applicable.
\subsection*{Consent for publication}
All co-authors have consented for publication.
\subsection*{Data availability}
Not applicable.
\subsection*{Materials availability}
The Plenum DGG modeling language package written in Mathematica by Guy Yosiphon is available at \url{https://computableplant.ics.uci.edu/theses/guy/Plenum.html} \cite{Yosiphon2009}. Another version of Plenum, with very small modification to run this paper's simulation code, is available at the GitHub repository 
\url{https://github.com/matthewhur836/SpineHead-DGG} under a sub-directory.
\subsection*{Code availability}
The DGG modeling code for this paper, written using the Plenum DGG modeling package, is available at the GitHub repository 
\url{https://github.com/matthewhur836/SpineHead-DGG}.

\printbibliography

\end{document}